\documentclass[%
 reprint,
 amsmath,amssymb,
 aps,
]{revtex4-2}

\usepackage{graphicx}% Include figure files
\usepackage{dcolumn}% Align table columns on decimal point
\usepackage{bm}% bold math
\usepackage{amsmath}       % align, equation, multline, notag, etc.
\usepackage{amssymb}       % \mathbb, \mathcal, etc.
\usepackage{braket}        % \bra{}, \ket{}, \braket{}
\usepackage{xcolor}
\usepackage{booktabs} %% for midrule command

\usepackage[nolist]{acronym}		%only print the acronyms, which we have used
\usepackage{placeins}

\usepackage{xcolor}
\usepackage[normalem]{ulem}   % for \sout (strike-through)

\begin{document}

%%\preprint{APS/123-QED}

\title{Non-equilibrium Instantaneous Approximation and Dipole Forbidden $d-d$ Transitions}
\author{Marco Marino}
 \email{marco2.marino@tu-dortmund.de}
 \affiliation{TU Dortmund University, Otto-Hahn-Straße 4, 44227 Dortmund, Germany}
\author{Lasse Sternemann}
\author{Mirko Cinchetti}%
\author{Frithjof B. Anders}%
 \email{frithjof.anders@tu-dortmund.de}
 \affiliation{TU Dortmund University, Otto-Hahn-Straße 4, 44227 Dortmund, Germany}

\date{August 18, 2026}% It is always \today, today,
\begin{abstract}
    Time- and angle-resolved photoemission spectroscopy provides direct access to pump-induced changes in the electronic structure of correlated materials, but its theoretical description generally requires computationally demanding two-time non-equilibrium calculations. We introduce an instantaneous approximation for pump-driven correlated systems, based on a separation between the rapid decay of the Green's functions in the relative time variable and their slower evolution in the average time. Combined with a one-shot dynamical mean-field construction at the Hubbard I level, the method incorporates the driven dynamics of the local correlated shell into the lattice Green's function while retaining momentum resolution. We also derive an effective coupling for nominally dipole-forbidden $d-d$ excitations. It arises from a dipole-allowed $d-p$ transition followed by $p-d$ hybridization; beyond the instantaneous limit, the same process produces an energy-dependent vertex correction to the optical response. As a proof of principle, we apply the framework to paramagnetic and antiferromagnetic FePS$_3$, using a density-functional theory derived tight-binding model together with a supercell unfolding procedure. The calculated momentum-resolved spectra reproduce the main qualitative features observed after excitation of the first and second $d-d$ transitions in recent time-resolved photoemission experiments.
\end{abstract}

\keywords{Hubbard I, $d-d$ transitions, NEQ-DMFT, Hubbard I-IAP, FePS$_3$, MPS$_3$, NEQ, trARPES, KB equations}

\maketitle

\begin{acronym}[GF]	
\acro{DFT}{density functional theory}
\acro{trARPES}{time resolved angular-resolved photoelectron spectroscopy}
\acro{ARPES}{angular-resolved photoelectron spectroscopy}
\acro{DMFT}{dynamical mean field theory}
\acro{LDA}{local density approximation}
\acro{GF}{Green's function}
\acro{NEQ}{non-equilibrium}
\acro{IAP}{instantaneous approximation}
\acro{PDOS}{projected density of states}
\acro{FDT}{fluctuation-dissipation theorem}
\acro{TB}{tight-binding}
\acro{AF}{anti-ferromagnetic}
\acro{SOC}{spin-orbit coupling}

\end{acronym}

\section{Introduction}
The partially filled $d$ shells of transition-metal compounds often retain a pronounced atomic character. Strong local Coulomb interactions split the $d$-electron configurations in a lattice environment into spin and orbital multiplets and can produce Mott or charge-transfer insulating states that cannot be adequately described within an independent-particle picture. Optical excitations within the $d$ shell provide a sensitive probe of this local many-body structure. Although direct $d-d$ transitions are dipole forbidden, they are observed, for example, in the optical-absorption spectra of transition-metal thiophosphates (M)PS$_3$ (M=$\mathrm{Mn},\mathrm{Fe},\mathrm{Ni}$) \cite{JoyVasudevanFePS3-Optical-1992}. Such excitations can acquire optical weight through ligand-assisted processes in which dipole-allowed $d-p$ transitions are combined with $p-d$ hybridization \cite{optical-conductivity-Hubbard-2010}. Their energies and spectral weights therefore carry information about crystal-field splittings, electronic correlations, and the multiplet structure of the transition-metal ion.

Recent advances in time- and angle-resolved photoemission spectroscopy (trARPES) make it possible to follow the momentum-resolved electronic structure of quantum materials after optical excitation \cite{RevModPhys.96.015003}. In particular, recent measurements on the correlated insulator FePS$_3$ have demonstrated that trARPES can track the ultrafast buildup and decay of local $d-d$ excitations through their signatures in the momentum-resolved electronic structure \cite{NITSCHKE2025100019}. Describing these experiments therefore requires a theoretical framework that connects the pump-driven dynamics of local multiplet states to the transient momentum-resolved photoemission spectrum. This connection is encoded in the non-equilibrium lesser Green's function, which depends on two real-time arguments \cite{FreericksPumpProbe09}. Its calculation must simultaneously account for the local many-body dynamics of the correlated $d$ shell and its embedding into the dispersive electronic structure of the lattice, making a realistic multi-orbital treatment computationally demanding.

Dynamical mean-field theory (DMFT) provides a non-perturbative description of local correlations and has become a standard framework for correlated materials in equilibrium \cite{RevModPhys.68.13,RevModPhys.78.865}. Its extension to non-equilibrium can, in principle, address pump-driven dynamics, but the repeated solution of a time-dependent impurity problem constitutes a major numerical bottleneck. This difficulty is particularly severe when all five 3$d$ orbitals and their associated multiplet states must be retained. A reduced description is therefore desirable when the dominant effect of the pump is the redistribution of population among local $d$-shell states, while the resulting momentum dependence is governed mainly by the underlying lattice Hamiltonian.

In this work, we introduce an instantaneous approximation (IAP) for the non-equilibrium Green's function of a pump-driven correlated system where the lattice symmetry and the single-particle properties are
encoded in the \ac{DFT} derived tight-binding model. The approximation is based on a separation of time scales: the Green's function is assumed to decay rapidly with respect to the relative time, whereas the pump-induced state evolves more slowly with respect to the average time. The electronic structure can then be evaluated at each average time using an instantaneous, frequency-dependent self-energy. We combine this construction with a one-shot DMFT scheme at the Hubbard I level. The pump-driven dynamics of the isolated correlated shell are calculated explicitly, and the resulting local retarded and lesser Green's functions are embedded into the lattice Dyson equation. This procedure retains the local multiplet dynamics and the momentum dependence of the lattice problem without requiring a fully self-consistent non-equilibrium DMFT calculation.

To connect the optical pump directly to the correlated subspace, we derive an effective matrix element for nominally dipole-forbidden $d-d$ transitions. In the microscopic process considered here, the electromagnetic field first drives a dipole-allowed transition between a transition-metal $d$ orbital and a ligand $p$ orbital. A subsequent $p-d$ hopping process restores the charge of the correlated shell while leaving it in an excited multiplet state. When the ligand-mediated charge-transfer process is fast compared with the local dynamics, it reduces to an effective instantaneous coupling within the $d$ shell. If this instantaneous limit is relaxed, the same mechanism generates an energy-dependent vertex correction to the optical response. This derivation provides a microscopic basis for restricting the explicitly driven dynamics to the correlated $d$ shell.

We apply the method to FePS$_3$, a layered van der Waals antiferromagnet whose equilibrium electronic structure and pump-induced $d-d$ excitations have recently been investigated experimentally \cite{nitschke2023valence, NITSCHKE2025100019}. Magnetic van der Waals materials provide a natural setting for studying low-dimensional correlated physics and can be exfoliated and incorporated into heterostructures \cite{castellanos2022van}. Within the broader family of transition-metal van der Waals compounds, the (M)PS$_3$ materials are antiferromagnetic insulators, whereas compounds such as Fe$_n$GeTe$_2$ ($3\le n \le 7$) are ferromagnetic metals \cite{PhysRevB.106.L180409}. FePS$_3$ is particularly suitable for the present study because its local Fe 3$d$ multiplets coexist with a strongly momentum-dependent valence-band structure, and distinct $d-d$ excitations can be selectively addressed by the pump.

In section \ref{sec:theory}, first we introduce the theory related to non-equilibrium electronic problems, with the objective to introduce a separation between the time scales, and averaging over the microscopic one. Then, we show how to justify a $d-d$ transition through a virtual excitation in the $p$ orbitals and back. Lastly, we describe how to relate a real-material problem to a tight-binding problem, and how to map back the tight-binding results to the real-material. In section \ref{sec:III}, we introduce the case study of FePS$_3$, and how we tried to solve the related non-equilibrium problem using the approach introduced in section \ref{sec:theory}. 

In particular, starting from a 
\ac{DFT} derived tight-binding Hamiltonian, we separate the correlated Fe $3$d subspace from the ligand-dominated bands and incorporate the local interactions within the Hubbard I approximation. We also formulate a supercell unfolding procedure that allows the spectra of the antiferromagnetic phase to be represented in the Brillouin zone of the primitive cell and compared directly with experimental momentum maps. The calculated spectra reproduce the principal trARPES features: excitation of the first $d-d$ transition produces momentum maps that remain similar to the equilibrium maps, whereas excitation of the second transition generates substantially stronger changes. These results demonstrate that local multiplet dynamics, embedded into a realistic momentum-dependent electronic structure, can account for essential features of the pump-induced response of FePS$_3$.
\section{\label{sec:theory}Theory}

We introduce the  electronic problem subject to an external time-dependent perturbation and employ the Keldysh-Kadanoff-Baym (KB) formalism \cite{Keldysh65,RevModPhys.58.323}. In order to simplify the equations we introduce the %%Instantaneous Approximation 
\ac{IAP} by assuming that the electronic dynamics happens on a much faster time scale than the pulse duration of 100-200~fs.
In the spirit of the \ac{DMFT}, we restrict ourselves to a local self-energy correction. We avoid to solve the \ac{DMFT} out-of-equilibrium \cite{RevModPhys.86.779} by restricting ourselves to the first \ac{DMFT} iteration
which is equivalent to the Hubbard I approach. In its generalization to a pump-probe, we solve the local decoupled \ac{NEQ} dynamics of the 3$d$ shell exactly using an effective $d-d$ transition matrix and feed the retarded and the lesser \ac{GF} into the lattice Dyson equations. The fundamental material properties are obtained by \ac{DFT} approach providing the realistic tight binding model. In this regard, we introduce a downfolding/upfolding procedure, to isolate the energy window of interest. For magnetic calculations, a supercell must be considered. Therefore, in the final part, we introduce an unfolding scheme that allows us to extract the primitive-cell Brillouin-zone results from the corresponding supercell calculations. We use in the following sections natural units. In particular, we absorb $\hbar$ into the definition of the time, so that time has the unit of inverse energy in the equations.

\subsection{\ac{NEQ}-\ac{GF} theory in the \ac{IAP}}

\label{sec:II.A}
In order to introduce the notation used in this paper, we review the basics on non-equilibrium Green's functions. The general  Hamiltonian for the electronic problem including an external driving term $\hat{H}_{pulse}(t)$ describing the effect of the pump-pulse can be written as
\begin{align}
\label{eq:H-1}
    \hat{H}(t)&=\sum_{\bm{R}\bm{R'}}\sum_{\alpha\beta}{t}_{\alpha\beta}(\bm{R},\bm{R'})\,
    \hat{d}_{\bm{R}\alpha}^{\dagger}\,\hat{{d}}_{\bm{R'}\beta}+\hat{H}_{int}+\hat{H}_{pulse}(t)
\end{align}
where $\hat{d}_{\bm{R}\alpha}$ and $\hat{d}^{\dagger}_{\bm{R}\alpha}$ are the annihilation and creation operators, the index $\bm{R}$ refers to the unit cell position, and the index $\alpha=(i, \zeta, \sigma)$ is a generic index denoting the site position $i$ within the unit cell, site-related orbital index $\zeta$, and spin index $\sigma$. The single particle hopping  matrix element $t_{\alpha\beta}(\bm{R},\bm{R'})$ stems from the kinetic contribution plus the nuclei-electron potential, and $\hat{H}_{int}$ accounts for the electron-electron interaction. Taking into account the translational invariance of the problem, we can write down
\begin{equation}
    t_{\alpha\beta}(\bm{R},\bm{R'})=t_{\alpha\beta}(\bm{R-R'},\bm{0})
\end{equation}
We can then Fourier transform our creation and annihilation operators due to the periodicity of the electronic problem, which takes  us to the Hamiltonian form
\begin{align}
    \hat{H}(t)&=\sum_{\bm{k}}\sum_{\alpha\beta}\tilde{t}_{\alpha\beta}(\bm{k})\,\hat{d}_{\bm{k}\alpha}^{\dagger}\,\hat{{d}}_{\bm{k}\beta}+\hat{H}_{int}+\hat{H}_{pulse}(t)
\end{align}
where
\begin{align}
    t_{ij}^{\alpha\beta}(\bm{R-R'},\bm{0})=\frac{1}{N_{\bm{k}}}\sum_{\bm{k}}e^{i\bm{k}\cdot(\bm{R}-\bm{R'})}\tilde{t}_{ij}^{\alpha\beta}(\bm{k}).
\end{align}
At this point, we introduce the time-ordered two-times Green's function of the electronic problem in momentum space
(we label the time variable with the letter $z$)
\begin{equation}
\label{eq:T-GF-def}
G_{\alpha\beta}(\bm{k}|z,z')=-i\,Tr[\hat{\rho}_0\,\mathcal{T}\{\hat{d}_{H\bm{k}\alpha}(z)\hat{d}_{H\bm{k}\beta}^{\dagger}(z')\}]
\end{equation}
where the density matrix $\hat{\rho}_0 =\rho(z_0=t_0)$ is usually the equilibrium density matrix for the initially prepared system,
$\mathcal{T}$ is the time-ordering operator. The Heisenberg picture representation of the creation and annihilation operators is defined as
\begin{align}
    \hat{d}_{H\bm{k}\alpha}(z)=U(z_0,z)\hat{d}_{\bm{k}\alpha}(z)U(z,z_0).
\end{align}
In the Schroedinger representation the time $z$ is labeling the position of the operator on the time axis having defined the time evolution operator as
\begin{equation}
i\tfrac{d}{dz}U(z,z_0)=\hat{H}(z)U(z,z_0)
\end{equation}
with the initial condition $U(z_0,z_0)=1$, and the formal solution
\begin{equation}
\label{eq:U-z-zp}
    U(z,z')=\mathcal{T}e^{i\int_{z'}^zH(\bar{z})d\bar{z}} .
\end{equation}
from which the adjoint condition   $U^{\dagger}(z,z')=U(z',z)$  is derived. Substituting the two-times time evolution operator $U(z,z')$, Eq.~\ref{eq:U-z-zp}, into the definition Eq.~\ref{eq:T-GF-def}, we obtain
\begin{align*}
&G_{\alpha\beta}(\bm{k}|z,z')\\
&=-i\,Tr[\hat{\rho}\,\mathcal{T}\{U(z_0,z)\hat{d}_{\bm{k}\alpha}(z)U(z,z')\hat{d}_{\bm{k}\beta}^{\dagger}(z')U(z',z_0)\}]\\
&=-i\,Tr[\hat{\rho}\,\mathcal{T}\{e^{i\int_{\gamma}H(\bar{z})d\bar{z}}\hat{d}_{\bm{k}\alpha}(z)\hat{d}_{\bm{k}\beta}^{\dagger}(z')\}]
\end{align*}
where we have introduce the Keldysh time-contour $\gamma$ \cite{RevModPhys.58.323} in the second line and generalized correspondingly the time-ordering operator $\mathcal{T}$
\begin{equation}
\gamma=\gamma_1\oplus\gamma_2=(-\infty,+\infty)\oplus(+\infty,\infty)
\end{equation}
being the system unchanged from $-\infty$ to $z_0$ and from $\max(z,z')$ to $+\infty$. The Kadanoff-Baym (KB) equations
\begin{align}
\label{eq:10}
    \bigg[i\delta_{\alpha\gamma}\tfrac{d}{dz}+\tilde{t}_{\alpha\gamma}(\bm{k})\bigg]G_{\gamma\beta}(\bm{k}|z,z')
    =\delta(z,z')\delta_{\alpha\beta} \nonumber \\
    +\sum_{\gamma}\int_{\gamma}d\bar{z}\,\Sigma_{\alpha\gamma}(\bm{k}|z,\bar{z})G_{\gamma\beta}(\bm{k}|\bar{z},z') 
\end{align} 
are obtained from the dynamics of the Green's function 
\begin{align*}
    i\tfrac{d}{dz}&G_{\alpha\beta}(\bm{k}|z,z')=\\&=\delta(z,z')\delta_{\alpha\beta}+
    Tr[\hat{\rho}\mathcal{T}\{i\tfrac{d}{dz}(\hat{d}_{H\bm{k}\alpha}(z))\hat{d}^{\dagger}_{H\bm{k}\beta}(z')\}]\\
    &=\delta(z,z')\delta_{\alpha\beta}-Tr[\hat{\rho}\mathcal{T}\{[\hat{H}(z),\hat{d}_{H\bm{k}\alpha}(z)]\hat{d}_{H\bm{k}\beta}^{\dagger}(z') \}]
\end{align*}
by separating the quadratic part, from the rest of the Hamiltonian after closing the Martin-Schwinger hierarchy \cite{stefanucci2013nonequilibrium} by assuming that
\begin{align*}
    Tr[\hat{\rho}\mathcal{T}\{e^{i\int_{\gamma}\hat{H}(\bar{z})d\bar{z}}[\hat{H}_{int+pulse}(z),\hat{d}_{\bm{k}\alpha}(z)]\hat{d}^{\dagger}_{\bm{k}\beta} (z')\}]\\
    =\sum_{\gamma}\int_{\gamma}d\bar{z}\,\Sigma_{\alpha\gamma}(\bm{k}|z,\bar{z})G_{\gamma\beta}(\bm{k}|\bar{z},z')
\end{align*}
This new quantity is called self-energy and it is a diagrammatic functional of the Green's function. The different positions of the time-ordered Green's function time variables on the time-contour identify different definitions of Green's function. We can map the time-ordered Green's function to a matrix whose components are equivalent to these different definitions \cite{Kamenev_2011} 
\begin{align}
   G_{\alpha\beta}(\bm{k}|z_1,z_2)&\rightarrow\bm{G}_{\alpha\beta}(\bm{k}|z_1,z_2)\\
   \bm{G}_{\alpha\beta}(\bm{k}|z_1,z_2)&=\begin{pmatrix} G_{11}& G_{12}\\
   G_{21}&G_{22}
   \end{pmatrix}_{\alpha\beta}(\bm{k}|z_1,z_2)
\end{align}
where the $\eta\theta$ dynamical indices pair refers to the Green's function with $z_1$ lying on the branch $\eta=1,2$ and $z_2$ lying on the branch $\theta=1,2$. The matrix structure of the Green's function, and likewise the self-energy in the time domain,
can be used to write the integral over each time branch in a compact way,
\begin{align*}
    \int_{\gamma}d\bar{z}\,\Sigma(z,\bar{z})G(\bar{z},z')\\
    \rightarrow (\bm{\Sigma}\bm{\sigma}_3\bm{G})(z,z')
\end{align*}
where the integral over time is implied. The integrals for intermediate time scales involving the branch $2$ picks up a factor $(-1)$ if we maintain the integration direction from $-\infty$ to $\infty$. This is taken intro account by the Pauli matrix $\bm{\sigma}_3$
\begin{equation}
\bm{\sigma}_3=\begin{pmatrix} 1& 0\\
0&-1
\end{pmatrix}
.
\end{equation}
The off-diagonal elements in the dynamical matrix are known as lesser and greater Green's functions
\begin{align*}
(G_{12})_{\alpha\beta}(z_1,z_2)=G^{<}_{\alpha\beta}(z_1,z_2)
=iTr\{\hat{\rho}\,\hat{d}^{\dagger}_{\beta}(z_2)\hat{d}_{\alpha}(z_1)\}\\
(G_{21})_{\alpha\beta}(z_1,z_2)=G^{>}_{\alpha\beta}(z_1,z_2)
=-iTr\{\hat{\rho}\,\hat{d}_{\alpha}(z_1)\hat{d}^{\dagger}_{\beta}(z_2)] \}\\
\end{align*}
where $G^<$ contains the information on the occupied states of the electronic problem while $G^>$ on the available states. The four components of the dynamical matrix are not independent, and through a Keldysh rotation we can get rid of one of them
\begin{align}
\bm{\sigma}^3\bm{G}&\rightarrow L\bm{\sigma}^3\bm{G}L^{\dagger}\\
    \bm{\sigma}^3\begin{pmatrix}
        G_{11}&G_{12}\\
        G_{21}&G_{22}
    \end{pmatrix}&\rightarrow
     \begin{pmatrix}
        G^{R}&G^{K}\\
        0&G^{A}
    \end{pmatrix}    
\end{align}
with
\begin{equation}
    \bm{L}=\frac{1}{\sqrt{2}}\begin{pmatrix}
        1&-1\\1&1
    \end{pmatrix}
\end{equation}
In particular we obtain
\begin{align}
    G^R=G_{11}-G_{12}=G_{11}-G^<\\
    G^A=G_{11}-G_{21}=G_{11}-G^>\\
    G^K=G_{11}+G_{22}=G_{12}+G_{21}\\
    0=G_{11}+G_{22}-G_{12}-G_{21}
\end{align}
where $G^R$ is known as the retarded Green's function, $G^A$  known as the advance Green's function, while $G^K$ is denoting the kinetic (or Keldysh) Green's function. This representation of the Green's function is called triagonal representation, and its advantage is that the self-energy is also triagonal as a functional of the Green's function. 

Let us write the KB equations taking into account this new notation (the Pauli matrix is now on implied in the product between dynamical matrices)
\begin{align}
\label{eq:20}
\bigg\{\sum_{\beta}\bigg[(\bm{G}_0^{-1})_{\alpha\beta}(\bm{k})-\bm{\Sigma}_{\alpha\beta}(\bm{k})\bigg]\bm{G}_{\beta\gamma}(\bm{k})&\bigg\}(z_1,z_2)\notag\\
=\bm{1}\delta(z_1,z_2)\delta_{\alpha\gamma}&
\end{align}
where $\bm{1}$ is the identity in the dynamical space, and we have introduced the free Green's function
\begin{align}
\sum_{\beta}\bigg[i\delta_{\alpha\beta}-\tilde{t}_{\alpha\beta}(\bm{k})\bigg]&(\bm{G}_0)_{\beta\delta}(\bm{k}|z_1,z_2)\notag\\
   =\bm{1}\delta(z_1,z_2)\delta_{\alpha\delta}&
\end{align}
In non-equilibrium, the Green's function is not translational invariant in time, and the dependency on both times $z_1$ and $z_2$ must be taken into account, so the adjoint KB equations have to be considered, too. 

At this point we introduce the time Wigner transform of the Green's function dynamical component $\eta\theta$
\begin{equation*}
    (G^W_{\eta\theta})_{\alpha\beta}(\bm{k}|T,\omega)=\int_{-\infty}^{\infty}d\tau  \,(G_{\eta\theta})_{\alpha\beta}(\bm{k}|T+\tfrac{\tau}{2},T-\tfrac{\tau}{2})
    e^{i\omega \tau}
\end{equation*}
This allows us to separate the microscopic dynamics on the relative time difference $\tau=z_1-z_2$ and the macroscopic dynamics on the average time $T=(z_1+z_2)/2$. We can use the property of the Wigner transform to write the convolution integral of two quantities implied in Eq.~\eqref{eq:20}, as the product of their respective Wigner transforms
\begin{multline}
\bigg[\sum_{\beta\theta}\int_{-\infty}^{\infty}dt\,(\Sigma_{\eta\theta})_{\alpha\beta}(\bm{k}|z_1,z)
(G_{\theta\lambda})_{\beta\delta}(\bm{k}|z,z_2)\bigg]_W(T,\omega)\\
=\hat{F}^{\Sigma\,G}(\partial_\omega,\partial_T)
\bigg[\sum_{\beta\theta}(\Sigma^W_{\eta\theta})_{\alpha\beta}(\bm{k}|T,\omega)\times\\
\times
\,(G^W_{\theta\lambda})_{\beta\delta}(\bm{k}|T,\omega)\bigg]
\end{multline}
where we introduced the derivative operator
\begin{align*}
    \hat{F}^{\Sigma\,G}=e^{-\frac{i}{2}(\partial_T^{\Sigma}\partial^{G}_{\omega}-\partial_{\omega}^{\Sigma}\partial^{G}_{T})}
\end{align*}
with the derivatives to be considered applied to the self-energy or the Green's function respectively. The time Wigner transform of the KB equations gives us 
\begin{multline}
  \sum_{\beta\theta}\bigg[\hat{F}^{G^{-1}_0G}((G_0^{-1})^{W}_{\eta\theta})_{\alpha\beta}(\bm{k}|T,\omega)\\-\hat{F}^{\Sigma G}(\Sigma_{\eta\theta}^W)_{\alpha\beta}(\bm{k}|T,\omega)\bigg](G_{\theta\lambda}^W)_{\beta\gamma}(\bm{k}|T,\omega)\\
=\delta_{\eta\lambda}\delta_{\alpha\gamma}
\end{multline}
Now, we assume that the microscopic dynamics is much faster than the macroscopic one. This is justified when the electron dynamics is governed by the inverse band width, which is much faster than the typical pulse duration of the pump pulses of $\tau_{p} > 100$fs even in so-called ultrafast spectroscopic experiments. Treating the microscopic system response to macroscopic variations over time $T$ as instantaneous, defines the instantaneous approximation (IAP)
\begin{equation}
 \hat{F}\simeq 1.
\end{equation}
In this limit, the KB equations then become
\begin{multline}
  \sum_{\beta\theta}\bigg[((G_0^{-1})^{W}_{\eta\theta})_{\alpha\beta}(\bm{k}|T,\omega)\\-(\Sigma_{\eta\theta}^W)_{\alpha\beta}(\bm{k}|T,\omega)\bigg](G_{\theta\lambda}^W)_{\beta\gamma}(\bm{k}|T,\omega)\\
=\delta_{\eta\lambda}\delta_{\alpha\gamma}
\end{multline}
where now the adjoint KB equations are equivalent to the not-adjoint ones. In the triagonal representation, the KB equations are then (from now on, the Wigner-transformed quantity will no longer be indicated explicitly, as it can be unambiguously inferred from its arguments)
\begin{multline*}
  \sum_{\beta}\bigg[(G_0^{-1})^{(R,A)}_{\alpha\beta}(\bm{k}|\omega)\\-\Sigma^{(R,A)}_{\alpha\beta}(\bm{k}|T,\omega)\bigg]G^{(R,A)}_{\beta\gamma}(\bm{k}|T,\omega)\\
=\delta_{\eta\lambda}\delta_{\alpha\gamma}
\end{multline*}
\begin{multline*}
    G^{(K)}_{\alpha\beta}(\bm{k}|T,\omega)\\
    = \sum_{\gamma\delta} G^{(R)}_{\alpha\gamma}(\bm{k}|T,\omega)\Sigma^{(K)}_{\gamma\delta}(\bm{k}|T,\omega)G^{(A)}_{\delta\beta}(\bm{k}|T,\omega)\\
    -\sum_{\gamma\delta} G^{(R)}_{\alpha\gamma}(\bm{k}|T,\omega)(G^{-1}_0)^{(K)}_{\gamma\delta}(\bm{k}|\omega)G^{(A)}_{\delta\beta}(\bm{k}|T,\omega)
\end{multline*}
We will be interested mainly in the lesser Green's function, so instead of the kinetic component equation, we will consider the lesser and greater component equations
\begin{multline}
\label{eq:lesser_gf}
G^{(<)}_{\alpha\beta}(\bm{k}|T,\omega)\\
=\sum_{\gamma\delta} G^{(R)}_{\alpha\gamma}(\bm{k}|T,\omega)\Sigma^{(<)}_{\gamma\delta}(\bm{k}|T,\omega)G^{(A)}_{\delta\beta}(\bm{k}|T,\omega)\\
-\sum_{\gamma\delta} G^{(R)}_{\alpha\gamma}(\bm{k}|T,\omega)(G^{-1}_0)^{(<)}_{\gamma\delta}(\bm{k}|\omega)G^{(A)}_{\delta\beta}(\bm{k}|T,\omega)
\end{multline}
\begin{multline}
G^{(>)}_{\alpha\beta}(\bm{k}|T,\omega)\\
= \sum_{\gamma\delta} G^{(R)}_{\alpha\gamma}(\bm{k}|T,\omega)\Sigma^{(>)}_{\gamma\delta}(\bm{k}|T,\omega)G^{(A)}_{\delta\beta}(\bm{k}|T,\omega)\\
-\sum_{\gamma\delta} G^{(R)}_{\alpha\gamma}(\bm{k}|T,\omega)(G^{-1}_0)^{(>)}_{\gamma\delta}(\bm{k}|\omega)G^{(A)}_{\delta\beta}(\bm{k}|T,\omega)
\end{multline}
where in the former equation the lesser component of the inverse of the free Green's function is equal to
\begin{multline}
    (G^{-1}_0)^{(<)}_{\gamma\delta}(\bm{k}|\omega)\\
    =-\sum_{\gamma\delta} (G_0^{(R)})^{-1}_{\alpha\gamma}(\bm{k}|\omega)(G_0^{(<)})_{\gamma\delta}(\bm{k}|\omega)\\
    \times(G_0^{(A)})^{-1}_{\delta\beta}(\bm{k}|\omega)
\end{multline}

The NEQ-IAP equations have analogue structure as their equilibrium versions. We remember that to obtain the NEQ-IAP equations, we have assumed that the microscopic system loses memory fast on the scale of the macroscopic dynamics (we are in the regime of the Markov approximation). If the system is at equilibrium with respect to the external perturbation (we assume that at $T=0$ the external perturbation is zero), then the lesser Green's function is fully determined by the retarded Green's function and the Fermi-Dirac distribution $f(\omega)$ 
\begin{align}
\label{eq:lesser_gf_eq}
 G^{(<)}_{\alpha\beta}(\bm{k}|0,\omega)=-f(\omega)\bigg[G^{(R)}_{\alpha\beta}-G^{(A)}_{\alpha\beta}\bigg](\bm{k}|0,\omega)\notag\\
=-f(\omega)\bigg[G^{(R)}_{\alpha\beta}-(G^{(R)}_{\beta\alpha})^*\bigg](\bm{k}|0,\omega)
\end{align}
This is called \ac{FDT} and in our case is obtained from the equations above assuming that the unperturbed self-energy itself follows the fluctuation-dissipation theorem with a Fermi-Dirac distribution. In the case of the free Green's functions the fluctuation-dissipation theorem is trivially satisfied, this allows us to write down the free contribution in the lesser \ac{GF} as
\begin{multline}
    (G^{-1}_0)^{(<)}_{\alpha\beta}(\bm{k}|\omega)\\
    =-f(\omega)\bigg[
    (G_0^{(R)})^{-1}_{\alpha\beta}-
    (G_0^{(A)})^{-1}_{\alpha\beta}\bigg](\bm{k}|\omega)
\end{multline}
In summary, we have to solve the following equation for the lesser \ac{GF}
\begin{multline}
G^{(<)}_{\alpha\beta}(\bm{k}|T,\omega)=\sum_{\gamma\delta} G^{(R)}_{\alpha\gamma}(\bm{k}|T,\omega)\Bigg[\Sigma^{(<)}_{\gamma\delta}(\bm{k}|T,\omega)\\
+f(\omega)\bigg[
(G_0^{(R)})^{-1}_{\gamma\delta}-
(G_0^{(A)})^{-1}_{\gamma\delta}\bigg](\bm{k}|\omega)\Bigg]G^{(A)}_{\delta\beta}(\bm{k}|T,\omega)
\end{multline}

\subsection{\ac{NEQ}-\ac{IAP} Hubbard I}
\label{sec:II.B}
In the above KB equations, Eq.~\eqref{eq:10}, the external perturbation driving the system is included  together with the electron-electron interactions in the self-energy. In the spirit of the \ac{DMFT} focusing on a local self-energy, we solve the electron-electron interactions trough an Hubbard I approach \cite{PhysRevB.57.6884}. Then, taking into account our \ac{NEQ}-\ac{IAP} approximation, we relate the non-equilibrium dynamics of the system to an instantaneous change of the self-energy. This relation between Hubbard I and \ac{NEQ}-\ac{IAP} approximation is based on the assumption that the external perturbation is mainly affecting the orbitals which relate to the \ac{DMFT} impurity problem.

The Hubbard I approximation can be viewed as the initial iteration step to
obtain a full self-consistent solution for the effective local site. This approximation fails in the metallic phase and intermediate Coulomb interaction: the very narrow quasiparticle bands are not accounted for. The approximation, however, is very good in the Mott insulating phase which is the case in semiconducting or insulating van-der Waals materials. Our playground here, FePS$_3$, lies at the crossover between Mott-Hubbard and charge-transfer insulating behavior, providing a stringent test of our approach.

\subsubsection{Hubbard I in equilibrium conditions}

The assumption of \ac{DMFT}
is that the self-energy is local in space, i.e. momentum independent, 
\begin{equation}
    \Sigma(\bm{k},\omega)\rightarrow\Sigma(\omega).
\end{equation}
This makes the electronic correlation problem manageable, considering a mapping of the lattice problem onto an analogue quantum impurity problem to determine the local self-energy which enters the lattice Green's function. The self-consistency cycle equating the local Green's functions to the k-summed lattice Green's functions defines a feedback loop connecting the local dynamics with the lattice problem. Ignoring the effective hybridization coupling that emerges from this feedback loop, and solving the \ac{DMFT} equations only once, takes the name of Hubbard I approach. 

The local atomic impurity problem is defined as
\begin{equation}
    \hat{H}_{imp}(\vec{R}_i)=\sum_{\alpha}\epsilon_{\alpha}\hat{c}^{\dagger}_{\alpha}\hat{c}_{\alpha}+\frac{1}{2}\sum_{\alpha\beta\gamma\delta}U_{\alpha\beta\gamma\delta}\hat{c}^{\dagger}_{\alpha}\hat{c}^{\dagger}_{\beta}\hat{c}_{\gamma}\hat{c}_{\delta}
\label{eq:H-imp}
\end{equation}
where the indices refer to the local correlated atomic orbitals of a fixed correlated lattice site $\vec{R}_i$ (while in the full lattice problem $\alpha$ denotes the generalized indices including the atomic sites), $U$ is the onsite Coulomb repulsion, and $\epsilon_{\alpha}$ are the atomic energies. Through exact diagonalization (ED) we can extract the related free ($U=0$) and dressed ($U\neq0$) impurity retarded Green's functions and from them obtain the related retarded impurity self-energy
\begin{align*}
   G^{(R)}_{\alpha\beta}(\omega)=\int dt\, e^{i\omega t}G^{(R)}_{\alpha\beta}(t)
    \\\Sigma^{(R)}_{\alpha\beta}(\omega)=(G^{-1}_0)_{\alpha\beta}(\omega)-(G^{(R)})^{-1}_{\alpha\beta}(\omega)
\end{align*}
Note, that the Green's function is independent of $T$ in equilibrium. This self-energy is used as an approximation of the lattice self-energy restricted lattice site $i$ local orbitals (here we expanded the generalized indices of the preceding section, due to the fact that the self-energy correction is applied on certain sites, those related to correlated orbitals)
\begin{equation}
    \hat{c}_{\alpha}\rightarrow\hat{d}_{i\alpha}
\end{equation}
The addition of the local impurity self-energy to the lattice problem, accounts for the dominant correlation effects in the latter. However, these effects are already partially taken into account at the static mean-field level as we will see in the section about \ac{DFT} in \ac{LDA} and a double-counting correction has to be considered \cite{PhysRevLett.87.067205,KAROLAK201011}. We apply an offset $\Delta H= -\mu_{dc} \sum_\alpha n_\alpha$ to the impurity Hamiltonian, and change the double counting parameter $\mu_{dc}$ such that the expected local occupation in the corrected sites is reproduced. We would typically use the \ac{DFT} filling of the orbitals as a guideline in the calculations; in our case \ac{DFT} is not able to reproduce the expected filling. To preserve, instead, the total number of electrons we consider a shift of the chemical potential.

\subsubsection{Hubbard I in non-equilibrium conditions}

Now we extent the approach to an external driving term modelling a pump pulse. We consider an external time-dependent perturbation in the local impurity problem, then through the related self-energy we embed the perturbation in the lattice problem in a simil-Hubbard I approach. The time-dependent isolated local impurity problem is given by
\begin{align}
    \hat{H}_{imp}(t)&=\sum_{\alpha}\epsilon_{\alpha}\hat{c}^{\dagger}_{\alpha}\hat{c}_{\alpha}+\frac{1}{2}\sum_{\alpha\beta\gamma\delta}U_{\alpha\beta\gamma\delta}\hat{c}^{\dagger}_{\alpha}\hat{c}^{\dagger}_{\beta}\hat{c}_{\gamma}\hat{c}_{\delta}+\hat{H}_{pulse}(t) \nonumber \\
    &=\hat{H}_0+\hat{H}_{pulse}(t).
\end{align}
Introducing the interaction representation of the annihilation and creation operators (assuming that the external perturbation is turned on at time $t=t_0$)
\begin{align*}
    c_{I\alpha}(t)=&e^{i\hat{H}_0t}\,c_{\alpha}\,e^{-i\hat{H}_0t} \\
    c_{H\alpha}(t)=&U_{I}(t,t_0)c_{I\alpha}(t)U_{I}(t_0,t)
\end{align*}
where the time evolution operator is defined by the equation
\begin{align}
    \tfrac{d}{dt}U_{I}(t,t_0)=(H_{pulse})_I(t)U_{I}(t,t_0)\notag\\
    =e^{i\hat{H}_0t}\hat{H}_{pulse}(t)e^{-i\hat{H}_0t}U_{I}(t,t_0),
\end{align}
we can write down the lesser and greater Green's functions in the following form
\begin{align}
(G^{(<)})^{\alpha\beta}(t,t') =i\,\mathrm{Tr}\{\hat{\rho}_I(t)\,U_I(t,t')\,\hat{c}^{\dagger}_{I\beta}(t')\times\notag\\\times U_I(t',t)\,\hat{c}_{I\alpha}(t)\}\\
(G^{(>)})^{\alpha\beta}(t,t') =-i\,\mathrm{Tr}\{\hat{\rho}_I(t')\,U_I(t',t)\,\hat{c}_{I\alpha}(t)\times\notag\\\times U_I(t,t')\,\hat{c}^{\dagger}_{I\beta}(t')\}
\end{align}
In order to perform the Wigner transformation we substitute the relative and average times into the lesser and greater Green's functions,
%\addFBA{There is a typo in the arguments of $G^<$}
\begin{align}
    &(G^{(<)})^{\alpha\beta}\!\left(T+\tfrac{\tau}{2},T-\tfrac{\tau}{2}\right) \notag\\
    &=i\,\mathrm{Tr}\{\hat{\rho}_I\!
    \left(T+\tfrac{\tau}{2}\right)
    tc^{\dagger}_{I\beta}\!
    \left(T+\tfrac{\tau}{2},T-\tfrac{\tau}{2} \right)
     c_{I\alpha}\!\left(T+\tfrac{\tau}{2}\right) 
    \}
    \label{e:def-orbital-lesser-gf}
    \\
    &(G^{(>)})^{\alpha\beta}\!\left(T+\tfrac{\tau}{2},T-\tfrac{\tau}{2}\right) \notag\\
    &=-i\,\mathrm{Tr}\{\hat{\rho}_I\!\left(T-\tfrac{\tau}{2}\right)
      tc_{I\alpha}\!\left(T-\tfrac{\tau}{2},T+\tfrac{\tau}{2}\right)
    c^{\dagger}_{I\beta}\!\left(T-\tfrac{\tau}{2}\right) 
  \}
   \label{e:def-orbital-greater-gf}
\end{align}
where we have defined the modified annihilation and creation operators
\begin{align}
    &tc_{I\alpha}(T-\tfrac{\tau}{2},T+\tfrac{\tau}{2}) \\
    &=U_I\!\left(T-\tfrac{\tau}{2},T+\tfrac{\tau}{2}\right)c_{I\alpha}\!\left(T+\tfrac{\tau}{2}\right)
    U_I\!\left(T+\tfrac{\tau}{2},T-\tfrac{\tau}{2}\right)\notag \\
    &tc^{\dagger}_{I\alpha}(T+\tfrac{\tau}{2},T-\tfrac{\tau}{2}) \\&=U_I\!\left(T+\tfrac{\tau}{2},T-\tfrac{\tau}{2}\right)c_{I\alpha}\!\left(T-\tfrac{\tau}{2}\right)
    U_I\!\left(T-\tfrac{\tau}{2},T+\tfrac{\tau}{2}\right)\notag
\end{align}

The expression obtained by integrating over the relative time,
\begin{align}
    G^{(>)}_{\alpha\beta}\!(T,\omega)=\int^{\infty}_{-\infty}d\tau\,e^{i\omega\tau}  G^{(>)}_{\alpha\beta}\!\left(T+\tfrac{\tau}{2},T-\tfrac{\tau}{2}\right),
\end{align}
is used to evaluate the self-energy for the NEQ-IAP Hubbard I approach.

Within the finite dimensional local Fock-space, the two-times Green's functions can be solved exactly by numerically integrating the equation of motion for the creation and annihilation operators, as well as the von-Neumann equation of the density matrix.

The nonequilibrium dynamics is driven by an external laser pulse. We included the effect of the pulse onto the local dynamics by adding the following pulse Hamiltonian,
\begin{align}
\label{eq:pulse_1}
\hat{H}_\text{pulse}(t)=-f(t)\sum_{\alpha\beta}[d_{\alpha\beta}(\omega_0)+d^*_{\beta\alpha}(\omega_0) ]\hat{c}^{\dagger}_{\alpha}\hat{c}_{\beta}
\end{align}
to the equilibrium Hamiltonian. The 
pulse function $f(t)$,
\begin{align}
\label{eq:pulse_2}
f(t)=A\,e^{-\tfrac{(t-t_p)^2}{\sigma^2}}\,\cos(\omega_0(t-t_p)),
\end{align}
is characterized by the pulse
center $t_p$ and the strength  $A$ of the external vector potential. The oscillation time $T=2\pi/\omega_0$ is short compared to the pulse duration $\sigma$. The effective local $d-d$ transition matrix element $d_{\alpha\beta}(\omega_0)$
depends on the laser pump energy $\omega_0$
\footnote{$\hbar$ is absorbed in the definition of the time throughout the paper}
and the orientation of the vector potential $\bm{n}_A$
and is calculated from the DFT results below
in Sec.\ \ref{ddtheory}. A word is in order about this single-particle form for the pump-pulse induced transitions. For matrix elements which conserve
the electron spin, spin forbidden transitions cannot be accounted for in this formulation. For such transition the \ac{SOC} needs to be included.

We can include  dissipative-decoherence channels in the dynamics, adding a Lindblad term ($\mathcal{D}[\cdot]$) \cite{breuer2002theory} to the von-Neumann equation:
\begin{align}
\label{eq:lindblad}
     \frac{d}{dt}\hat{\rho}_I=-i[H,\hat{\rho}_I]+\mathcal{D}[\hat{\rho}_I].
\end{align}
In the following, we considered Lindblad jump (or Kraus) operators for each pair of excited states and ground states, related to the dissipation, thermalization and decoherence of the excited system
\begin{subequations}
 \begin{align}
     \hat{L}_{decay}^{ij}=\sqrt{\gamma_\text{decay}}\ket{g_i}\bra{e_j}\\
     \hat{L}_{therm}^{ij}=\sqrt{\gamma_\text{therm}}\ket{g_i}\bra{g_j}_{j\neq i}\\
     \hat{L}_{deph}^{ij}=\sqrt{\gamma_\text{deph}}\ket{e_i}\bra{e_j}_{j\neq i}
\end{align}
\end{subequations}
where $\ket{e_i}$ are the excited states while $\ket{g_j}$ are the ground states and $\gamma$ denotes the decay rates in these different relaxation channels. In particular, $\gamma_\text{decay}$ describes the decay of the excited states back to the ground state, $\gamma_\text{therm}$ ensures equilibrium-population of
the ground states manifold, while $\gamma_\text{deph}$ has the same function for the excited states. These Kraus operators enter Eq.\ \eqref{eq:lindblad}
\begin{align*}
    \mathcal{D}[\star] =\sum_{a}\sum_{ij}\hat{L}_a^{ij}\star\hat({L}^{ij}_a)^{\dagger}-\tfrac{1}{2}\{\hat{L}^{ij}_a(\hat{L}^{ij}_a)^{\dagger},\star\}
\end{align*} 
where the brackets in the second term refer to an anticommutator, and the index $a$ refers to the decay, the thermalization and the dephasing channel.

The Wigner transforms $G_{\alpha\alpha}(\bm{k}|T,\omega)$ can be negative due to non-locality and quantum
interference effects. This is, however, not observable, since the probabilty of observing a particle is always positive (or zero).  
The laser probe pulse used for detecting the distribution functions \cite{RevModPhys.86.779} always has a finite pulse width.
Hence we need to convolute the Wigner transform
once we make connection to the experimental detection, for instance
via an APRES measurement. In order to preserve the positivity
\begin{align}
\label{eq:convoluted}
    G(\bm{k}|T,\omega)=\int_{T-\Delta_T}^{T+\Delta_T} d\tilde{T}e^{-\tfrac{(T-\tilde{T})^2}{2\sigma_T^2}}\times\notag\\\times\int_{-\Delta_{\omega}}^{+\Delta_{\omega}} d\tilde{\omega}(\sigma_{\omega}/\sqrt{\pi})e^{-\tfrac{\omega^2}{2\sigma_{\omega}^2}}G(\bm{k}|\tilde{T},\tilde{\omega})
\end{align}
where the widths $\sigma_{T,\omega}$ are related to the probe pulse width, and the intervals $I_{T/\omega}=[-\Delta_{T/\omega},\Delta_{T/\omega}]$ are taken large enough for the Gaussian's functions to decay to negligible values within them. This ensures positive definiteness of the diagonal spectral functions \cite{FreericksPumpProbe09}. In particular, we convolute the lattice lesser and retarded Green's function, not the atomic ones, which are used to obtain the relative self-energies.

For completeness, we report the lesser self-energy of the impurity problem in terms of the lesser Green's function, obtained inverting the NEQ-IAP equation related to the lesser Green's function
\begin{multline}
\label{eq:lesser_se}
\Sigma^{(<)}_{\alpha\beta}(T,\omega)=\\
\sum_{\gamma\delta}(G^{(R)})^{-1}_{\alpha\gamma}(T,\omega)G^{(<)}_{\gamma\delta}(T,\omega)(G^{(A)})^{-1}_{\delta\beta}(T,\omega)\\
+\sum_{\gamma\delta} (G_0^{(R)})^{-1}_{\alpha\gamma}(\omega)(G_0^{(<)})_{\gamma\delta}(\omega)(G_0^{(A)})^{-1}_{\delta\beta}(\omega)
\end{multline}
or taking into account the \ac{FDT} for the free Green's functions we obtain 
\begin{multline}
\Sigma^{(<)}_{\alpha\beta}(T,\omega)=\\
\sum_{\gamma\delta}(G^{(R)})^{-1}_{\alpha\gamma}(T,\omega)G^{(<)}_{\gamma\delta}(T,\omega)(G^{(A)})^{-1}_{\delta\beta}(T,\omega)\\
-f(\omega)\bigg[(G_0^{(R)})_{\alpha\beta}^{-1}(\omega)-(G_0^{(A)})_{\alpha\beta}^{-1}(\omega)\bigg]
\end{multline}

\subsubsection{An effective theory for the $d-d$ transitions}
\label{ddtheory}

The optical excitation between two $d$ levels is forbidden in the dipole approximation but experimentally observed. The prototype example is ruby, Cr$^{3+}$ ions embedded in Al$_2$O$_3$, as pointed out by Katsnelson and Lichtenstein \cite{optical-conductivity-Hubbard-2010}. The experimental spectral lines are connected to internal excitations in the 3$d$ subsystem with fixed occupation \cite{FinkelsteinVanVleck1940}. It is well understood by now
that violation of the Laporte rule is connected to the interaction of the 3$d$ subsystem with the surrounding ligands \cite{FinkelsteinVanVleck1940}.

The idea of this section is to introduce an effective theory of the $d-d$ transitions, which is able to capture the nature of these transitions and justify our approximation above. There are different ways to account for the $d-d$ transitions. For the calculation
of the  optical conductivity it is useful to introduce a vertex function \cite{optical-conductivity-Hubbard-2010}. We are, however, interested in deriving an effective $d-d$ transition Hamiltonian. This is based on the separation
of energy scales: the inverse energy scale of the $p-d$ transition matrix element is short compared to the pulse duration.

The dipole forbidden $d-d$ dipole transitions appear through allowed virtual $d-p$ and $p-d$ dipole transitions. The elementary mechanism of the transition is a light induced electron transition to the neighboring sites and then an energy conserving hopping back to the local $d$-shell. Since the pulse duration is long compared to the time associated with the inverse $p-d$ hopping, an effective $d-d$ transition matrix element is generated. This virtual process is weaker than a direct local optical excitation and in line with previous descriptions using ligand field \cite{FinkelsteinVanVleck1940} or vertex functions in the optical conductivity in multi-orbital models \cite{optical-conductivity-Hubbard-2010}.

The best way to obtain an effective theory is passing through the path-integral formalism and integrating out the degrees of freedom we are not interested in. This becomes possible assuming that the neighboring $p$ orbitals are uncorrelated. 

Let us consider the above generic Hamiltonian, Eq.\eqref{eq:H-1}, and express the pulse term explicitly using the coupling matrix element $\chi_{\alpha\beta}(\bm{R},t)$
\begin{align}
    \hat{H}(t)=\sum_{\bm{R}}\sum_{\alpha\beta}\hat{d}_{\bm{0}\alpha}^{\dagger}\,t_{\alpha\beta}(\bm{R})\,\hat{{d}}_{\bm{R}\beta}+\hat{H}_{int}\notag\\
    +\sum_{\bm{R}}\hat{d}_{\bm{0}\alpha}^{\dagger}\,\chi_{\alpha\beta}(\bm{R},t)\,\hat{{d}}_{\bm{R}\beta}
\end{align}
where $\chi_{\alpha\beta}(\bm{R},t)$ is given by the expression
\begin{align}
    \chi_{\alpha\beta}(\bm{R},t)=\int d\bm{r}w_{\bm{0}\alpha}^*(\bm{r})\bigg(\tfrac{-ie(\nabla\cdot\bm{A}(\bm{r},t))}{2m}\times\notag\\
    \times\tfrac{-ie\bm{A}(\bm{r},t)\cdot\nabla+e^2\bm{A}(\bm{r},t)\cdot\bm{A}(\bm{r},t)}{2m}\bigg)w_{\bm{R}\beta}(\bm{r})
\end{align}
with $e$ and $m$ the electron charge and mass, and $w_{\bm{R}\beta}(\bm{r})$ the Wannier function of orbital $\alpha$ in the unit cell $\vec{R}$. Inside $\chi$ we have respectively a dipole term plus a diamagnetic term and a divergence term. The divergence term vanishes in the Coulomb gauge ($\nabla\bm{A}=0$), while the diamagnetic term is negligible in non-metallic systems. In the dipole term, the dependence of the vector potential on the space variable can be omitted, and the vector potential taken out of the integral sign, since the typical  wavelength in experiments of THz up to visible light pump pulses are  much larger than the cell size. Thus, the pulse term can be expressed as
\begin{align}
    \chi_{\alpha\beta}(\bm{R},t)=\bm{A}(t)\cdot\int d\bm{r}w_{\bm{0}\alpha}^*(\bm{r})\bigg(\tfrac{-ie\nabla}{2m}\bigg)w_{\bm{R}\beta}(\bm{r})\notag\\
    =\bm{A}(t)\cdot\bm{\tilde{\chi}}_{\alpha\beta}(\bm{R})
\end{align}
We introduce for simplicity the generic index $i=(\bm{R}\alpha)$ and a related vector notation
\begin{align}
    \hat{H}(t)=\sum_{ij}\hat{d}_{i}^{\dagger}\,(t_{ij}+\bm{A}(t)\cdot\bm{\tilde{\chi}}_{ij})\,\hat{{d}}_{j}+\hat{H}_{int}\notag\\
    =\bm{\hat{d}}^{\dagger}\,(\bm{t}+\bm{A}(t)\cdot\bm{\tilde{\chi}})\,\bm{\hat{d}}+\hat{H}_{int}
\end{align}

Let us introduce now the Green's function generating fermionic function (for a moment we do not consider the source terms, whose derivation gives the different Green's functions)
\begin{align*}
Z=Tr[\hat{\rho}\mathcal{T}\{e^{i\int_{\gamma}H(\bar{z})d\bar{z}}\}]
\end{align*}
where the time-contour $\gamma$ and the correspondent time-ordering operator $\mathcal{T}$ have been already introduced in the first section. To go to a path-integral formulation, we decompose the time-contour into $2N$ time steps ($N$ time steps for each time branch), and between each time step $\Delta \tau$ we insert a coherent state completeness relation, where coherent states are defined by 
\begin{align}
    \hat{d}_{j}^{(i)}\ket{\psi_j^{(i)}}=\psi_j^{(i)}\ket{\psi_j^{(i)}}\\
    \bm{\hat{d}}^{(i)}\ket{\bm{\psi}^{(i)}}=\bm{\psi}^{(i)}\ket{\bm{\psi}^{(i)}}
\end{align}
with the upper index pointing to the respective time step where the completeness relation has been inserted, and the lower index pointing to the generic index above, with the respective vector notation. Note that $\psi_j^{(i)}$ are Grassman variables to maintain the anticommutation relations. In the limit $N\rightarrow\infty$ and $\Delta\tau\rightarrow0$, the generating fermionic function is given by a conventional Grassman variable path integral where we added back the source terms for generating the different Green's functions
\begin{align}
    Z[\bm{\bar{\eta}},\bm{\eta}]=\int d[\bm{\bar{\psi}},\bm{\psi}]e^{iS[\bm{\bar{\psi}},\bm{\psi}]+\int_{\gamma}\bm{\bar{\eta}}(\bar{z})\bm{\psi}(\bar{z})d\bar{z}+\int_{\gamma}\bm{\bar{\psi}}(\bar{z})\bm{\eta}(\bar{z})d\bar{z}}
\end{align}
where the fermionic action has the following form
\begin{align}
    S[\bm{\bar{\psi}},\bm{\psi}]=\int_{\gamma}\bm{\bar{\psi}}(\bar{z})(i\partial_z-\bm{t}+\bm{A}(z)\cdot\bm{\tilde{\chi}} )\bm{\psi}(\bar{z})d\bar{z}\notag\\
    +S_{int}[\bm{\bar{\psi}},\bm{\psi}]
\end{align}
with the interaction part included in $S_{int}$. 

At this point we partition the total set of orbitals into the Wannier functions that are related to $d$ orbitals, and we indicate the rest as $p$ orbitals
\begin{equation}
    \bm{\psi}=(\bm{\psi}^p,\bm{\psi}^d)
\end{equation}
Moreover, we consider a mean-field approximation by absorbing the factorized
contribution of $S_{int}$ into the single particle properties and set $S_{int}=0$ for the moment. The idea is to define a material specific action  $S[\bm{\bar{\psi}},\bm{\psi}]$ based on the DFT (LDA) single particle properties and address the remaining two-particle fluctuations $S_{int}$ at a later stage. After partitioning the orbitals as outlined above, the generating function becomes
\begin{align}
Z[\bm{\bar{\eta}},\bm{\eta}]=\int d[\bm{\bar{\psi}}^p,\bm{\psi}^p]d[\bm{\bar{\psi}}^d,\bm{\psi}^d]e^{i(S^{pp}+S^{dd}+S^{pd})[\bm{\bar{\psi}},\bm{\psi}] }\notag\\
^{+\int_{\gamma}\bm{\bar{\eta}}(\bar{z})\bm{\psi}(\bar{z})d\bar{z}+\int_{\gamma}\bm{\bar{\psi}}(\bar{z})\bm{\eta}(\bar{z})d\bar{z}}
\end{align}
 where
\begin{align}
    S^{pp}=\int_{\gamma}\int_{\gamma}\bm{\bar{\psi}}^p(\bar{z})((G^{pp}_0)^{-1}(\bar{z},\bar{z}')\notag\\
    +\bm{A}(z)\cdot\bm{\tilde{\chi}}^{pp}\delta(\bar{z},\bar{z}'))\bm{\psi}^p(\bar{z}')d\bar{z}d\bar{z}'\\
    S^{dd}=\int_{\gamma}\int_{\gamma}\bm{\bar{\psi}}^d(\bar{z})((G^{dd}_0)^{-1}(\bar{z},\bar{z}')\notag\\
    +\bm{A}(z)\cdot\bm{\tilde{\chi}}^{dd}\delta(\bar{z},\bar{z}'))\bm{\psi}^d(\bar{z}')d\bar{z}d\bar{z}'\\
    S^{pd}=\int_{\gamma}\bm{\bar{\psi}}^p(\bar{z})(-\bm{t}^{pd}+\bm{A}(z)\cdot\bm{\tilde{\chi}}^{pd})\bm{\psi}^d(\bar{z})d\bar{z}\notag\\
    +\int_{\gamma}\bm{\bar{\psi}}^d(\bar{z})(-\bm{t}^{dp}+\bm{A}(z)\cdot\bm{\tilde{\chi}}^{dp})\bm{\psi}^p(\bar{z})d\bar{z},
\end{align}
and where the free sub-orbital block matrix Green's functions obey the dynamic equations
\begin{align}
    \int_{\gamma}d\bar{z}''(i\partial_{\bar{z}}-\bm{t}^{pp})\delta(\bar{z},\bar{z}'')G^{pp}_0(\bar{z}'',\bar{z}')=\delta(\bar{z},\bar{z}')\\
    \int_{\gamma}d\bar{z}''(i\partial_{\bar{z}}-\bm{t}^{dd})\delta(\bar{z},\bar{z}'')G^{dd}_0(\bar{z}'',\bar{z}')=\delta(\bar{z},\bar{z}')
\end{align}
At this point, we can integrate the $p$ components, and define an effective action $S^{\rm eff}$ by taking into account only the $d$ components
\begin{equation}
    e^{iS^{\rm eff}}=e^{iS^{dd}}\int d[\bm{\bar{\psi}}^p,\bm{\psi}^p]e^{i(S^{pp}+S^{pd})[\bm{\bar{\psi}},\bm{\psi}] } .
\end{equation}
The integral is a Gaussian Grassmann functional integral that can be easily solved, giving an effective action of the following form
\begin{align}
    S^{\rm eff}=\int_{\gamma}\int_{\gamma}\bm{\bar{\psi}}^d(z)(G^{-1}_{\rm eff}(z,z')-\Sigma(z,z') )\bm{\psi}^d(z)dzdz'
\end{align}
where we have introduced the dressed $d$ orbital Green's function and a self-energy correction including the vector potential term
\begin{align}
    G^{-1}_{\rm eff}(z,z')=(G_0^{dd})^{-1}(z,z')+\bm{t}^{dp}G_0^{pp}(z,z')\bm{t}^{pd}\\
    \Sigma(z,z')=\bm{A}(z)\cdot\bm{\tilde{\chi}}^{dp}G_0^{pp}(z,z')\bm{t}^{pd}\notag\\
    +\bm{t}^{dp}G_0^{pp}(z,z')\bm{\tilde{\chi}}^{pd}\cdot \bm{A}(z')\notag\\
    -\bm{A}(z)\cdot\bm{\tilde{\chi}}^{dp}G_0^{pp}(z,z')\bm{\tilde{\chi}}^{pd}\cdot \bm{A}(z').
    \label{eq:69}
\end{align}
We have expanded the dressed $p$ orbital Green's function to separate the dipole term from the kinetic one, and we have considered the dipole selection rules $\bm{\tilde{\chi}}^{pp}=0$ and $\bm{\tilde{\chi}}^{dd}=0$. The self-energy correction derived with respect to the vector potential provides two new dressed vertices representing the two physical processes contributing to the $d-d$ transitions. In the first process the vector potential induces a dipole transition from $d$ to $p$ orbitals, followed by a propagation in the $p$ orbitals set, and a hopping back to a $d$ orbital. In the second process the electron hops from $d$ to $p$ orbitals first, followed by a propagation in the $p$ orbitals set, and as a final step the vector potential generates a dipole transition from $p$ to $d$ orbitals. This process represents a perturbative correction of the order of $t^{pd}/t^{pp}$, this explains the low experimental intensity of the $d-d$ transitions.

Clearly, the self-energy $\Sigma(z,z')$ is a function of two time arguments and, therefore, contains retardation processes via the $pp$ Green's function matrix $G_0^{pp}(z,z')$. In the Fourier domain these processes contribute to the vertex correction in the optical conductivity. In our approach, we consider the dynamics in the $p$ subset faster than the electronic $d$ dynamics, so we do not have the need to consider retardation effects.

Within the general \ac{NEQ} \ac{GF} approach, the effect
of the pump pulse enters the self-energy, Eq.\ \eqref{eq:69},
as an additional source. This expression can be used to extract the effective $d-d$ transition matrix element $\bm{d}_{\alpha\beta} (\omega)$ entering the Hamiltonian, Eq.\ \eqref{eq:pulse_1}.

We perform a Wigner transform of the first two terms of the self-energy \eqref{eq:69}. Assuming that the pulse duration is longer that the typical time scale of the electronic dynamics, we obtain
\begin{align*}
 \Sigma(T,\omega)|_{1,2} \simeq \bm{A}(T)\cdot\bigg( \bm{\tilde{\chi}}^{dp}G_0^{pp}(\omega)\bm{t}^{pd}+
 \bm{t}^{dp}G_0^{pp}(\omega)\bm{\tilde{\chi}}^{pd}\bigg)
\end{align*}
where $\bm{A}(T)$ denotes the  vector potential, $\bm{\tilde{\chi}}^{dp}$ the $d-p$ dipole transition matrix vector, and $G_0^{pp}(\omega)$ the bare \ac{GF} of
the $p$ orbital subsystem. 
To this end, we obtain the effective $d-d$ transition vector (writing the orbital indices and sums explicitly)
\begin{align}
\label{eq:dipole}
   \bm{d}_{\alpha\beta} (\omega)= \sum_{jk}\bigg( \bm{\tilde{\chi}}^{dp}_{\alpha j}(G_0^{pp})_{jk}(\omega)t^{pd}_{k\beta}\notag\\+
 t^{dp}_{\alpha j}(G_0^{pp})_{jk}(\omega)\bm{\tilde{\chi}}^{pd}_{k\beta}\bigg)
\end{align}
which couples to the vector potential $\bm{A}(T)$ where the $j,k$ indices refer to the $p$ orbitals while the $\alpha,\beta$ indices refer to the $d$ orbitals (the local problem). The coupling vector $\bm{\tilde{\chi}}^{dp}_{\alpha j}$
contains the dipole transition between the $p$ and the $d$ orbitals and is explicitly derived from the DFT results below in Sec.\  \ref{sec:comput}.

%\addFBA{The physics described by this effective term
%is the following.
%First term of  Eq.\ \eqref{eq:dipole}: an electron hops from the $d_\beta$ %orbital to
%the $p_k$ orbital of the p-subsystem via the \ac{TB} matrix element
%$\bm{\tilde{t}}^{pd}_{k\beta}$, propagates within the p-system,
%and then the light field induces a dipole transition of an electron 
%from the $p_j$ orbital to the $d_\alpha$ orbital.
%Alternatively (second term), 
%the light field induces an electron transition from the 
%$\beta$ orbital of d-system to $k$-orbital of the
%the p-system $\bm{\tilde{\chi}}^{pd}_{k\beta}$,
%then the electron propagates in the p-subsystem ($G_0^{pp})_{jk}(\omega)$)
%before it hops back from a $p_j$ orbital to the $d_\alpha$ orbital
%via a \ac{TB} hopping matrix element $ \bm{\tilde{t}}^{dp}_{\alpha j}$.
%Essentially, $\bm{d}_{\alpha\beta} (\omega)$ describes a 
%effective transition between the two d-orbitals $\alpha$ and $\beta$
%mediated via the surrounding $p$ system as a virtual excitation.
%This description is valid if the transition to and from the $p$ subsystem is fast compared to the pulse duration.
%}

\subsection{Density Functional theory and NEQ-IAP Hubbard I}
In order to apply the Hubbard I approach (and the related NEQ-IAP Hubbard I approach) we need a system Hamiltonian in terms of localized orbitals as introduced in Eq.~ \eqref{eq:H-1}. In order to make prediction for real materials, we relate the \ac{DFT} solutions, i.e. Kohn-Sham states, to localized orbital states. 
In this regard, localized Wannier functions represent an optimal choice.

The procedure of extracting a set of Wannier functions from a set of Kohn-Sham states is called Wannierisation. Typically, we are only interested in a subset of the Kohn-Sham states that accounts for low energy active states which are potentially modified by the local Coulomb interaction. A downfolding (DF) procedure projects onto this energy subspace and is combinded with the Wannierisation. Going back from this restricted orbital set to
the Kohn-Sham states picture, an upfolding (UF) transformation is required.

We will underline the basic concepts of the two procedures in the following section, then to introduce an UF procedure in the case of super-cell quantities.
\subsubsection{Downfolding and Upfolding procedures}
Let us introduce our Kohn-Sham Hamiltonian
\begin{equation}
    \hat{H}_{KS}=\sum_{\bm{k}}\sum_{m}\hat{c}_{\bm{k}m}^{\dagger}\epsilon_m(\bm{k})\,\hat{{c}}_{\bm{k}m}
\end{equation}
where the $m$ index refers to the Kohn-Sham states for a given $\bm{k}$ vector, and $c$ and $c^{\dagger}$ are respectively the annihilation and creation operators of Kohn-Sham states. Chosen a certain energy window $I(E)$ of the Kohn-Sham problem, we can partition the Kohn-Sham as $\epsilon_m(\bm{k})\in I(E)$ and  $\epsilon_m(\bm{k})\not\in I(E)$, 
\begin{align}
  \hat{H}_{KS}
  &= \biggl(
       \sum_{\bm{k}m}
       \hat{c}^{KS\dagger}_{\bm{k}m} \,
       \epsilon_{m}(\bm{k}) \,
       \hat{c}^{KS}_{\bm{k}m}
     \biggr)_{\!\mathrm{in}}
  +
  \biggl(
       \sum_{\bm{k}m}
       \hat{c}^{KS\dagger}_{\bm{k}m} \,
       \epsilon_{m}(\bm{k}) \,
       \hat{c}^{KS}_{\bm{k}m}
     \biggr)_{\!\mathrm{out}}
  \notag \\[6pt]
  &= \sum_{\bm{R}}\sum_{ij}\sum_{\alpha\beta}
     \hat{d}_{\bm{0}i\alpha}^{\dagger}\,
     t^{\alpha\beta}_{ij}(\bm{0},\bm{R})\,
     \hat{d}_{\bm{R}j\beta}
  \notag \\[4pt]
  &\quad
  + \biggl(
      \sum_{\bm{k}m}
      \hat{c}^{KS\dagger}_{\bm{k}m} \,
      \epsilon_{m}(\bm{k}) \,
      \hat{c}^{KS}_{\bm{k}m}
    \biggr)_{\!\mathrm{out}}
\end{align}
where 'in' refers to the selected window, while 'out' is related to everything that is outside of the selected window.
The creation and annihilation operators are related by the unitary transformation
\begin{align}
		\hat{c}^{KS}_{\bm{k}m}
		&=
		\sum_{\bm{R}}\sum_{\alpha}
		U_{m\alpha}(\bm{k}) \,
		e^{-i \bm{k} \cdot (\bm{R}+\bm{r}_{\alpha})} \,
		\hat{d}_{\bm{R}\alpha}
		\\[2pt]
		\hat{d}_{\bm{R}\alpha}
		&=
		\frac{V}{(2\pi)^3}
		\sum_{\bm{k}}\sum_{m}
		\big[ U_{m\alpha}(\bm{k}) \big]^{\dagger} \,
		e^{i \bm{k} \cdot (\bm{R}+\bm{r}_{\alpha})} \,
		\hat{c}^{KS}_{\bm{k}m}
\end{align}
with $\bm{r}_{\alpha}$ the center of the $\alpha$ Wannier state and $V$ the unit cell volume. 

We note that the Wannier problem is equivalent to the quadratic part of the electronic problem studied at the beginning of Sec.~\ref{sec:II.A}. 
Note that the Hamiltonian obtained from a \ac{DFT} Kohn-Sham calculation includes already the electron-electron interactions via the mapping of the many-body problem onto an effective single-particle description. Therefore, we need to take into account a double-counting correction \cite{KAROLAK201011} when applying the local Coulomb interaction in the Hubbard I approach.
At this point we use a first-quantization notation, which makes the following results more transparent
\begin{align}
    \ket{\psi_{\bm{k}m}}=(\hat{c}^{KS}_{\bm{k}m})^{\dagger}\ket{0}\\
    \ket{w_{\bm{R}\alpha}}=\hat{d}_{\bm{R}\alpha}^{\dagger}\ket{0} .
\end{align}
The $U$ matrix, $U_{m\alpha}(\bm{k})$, has two components: a disentanglement component $U^{dis}$ and a localization component $U^{loc}$, associated to two different steps of the Wannierisation procedure. The former combines Kohn-Sham states of the inner and outer energy window to produce new generalized Kohn-Sham states inside the selected energy window
\begin{equation}
    \ket{\tilde{\psi}^{KS}_{\bm{k}m}}=\sum_{n\in(in+out)}U^{dis}_{mn}(\bm{k}) \ket{\psi^{KS}_{\bm{k}n}}.
\end{equation}
The latter Fourier transforms the new generalized Kohn-Sham states into Wannier states
\begin{equation}
    \ket{w_{\bm{R}\alpha}}=\tfrac{V}{(2\pi)^3}\sum_{\bm{k}}\sum_{n\in in}U^{loc}_{\alpha n}(\bm{k}) e^{i\bm{k}\cdot\bm{R}}\ket{\tilde{\psi}^{KS}_{\bm{k}n}}.
\end{equation}
A generic quantity can then be transformed between the Kohn-Sham states basis and the Wannier states basis applying the $U$ matrix transformation. In the case of the Green's function the Downfolding and Upfolding procedures are respectively
\begin{align}
  G_{\alpha\beta}(\bm{k},\omega)
  &= \braket{w_{\bm{k}I} | \hat{G} | w_{\bm{k}J}}
  \notag \\
  &= \sum_{nm\in\mathrm{in}}
     \bigl(U_{n \alpha }^{loc}(\bm{k})\bigr)^{\dagger}
     \braket{\tilde{\psi}^{KS}_{\bm{k}n} | \hat{G} | \tilde{\psi}^{KS}_{\bm{k}m}}
     U_{m \beta }^{loc}(\bm{k})
  \notag \\[4pt]
  &= \sum_{\substack{ls\,\in \\ (in+out)}}
     \sum_{nm\in\mathrm{in}}
     \bigl(U_{n \alpha }^{loc}(\bm{k})\bigr)^{\dagger}
     \bigl(U^{dis}_{ln}(\bm{k})\bigr)^{\dagger}
  \notag \\
  &\qquad \times
     G^{KS}_{ls}(\bm{k},\omega)\,
     U^{dis}_{sm}(\bm{k})\,
     U_{m \beta}^{loc}(\bm{k})
  \notag \\[4pt]
  &= \sum_{\substack{ls\,\in \\ (in+out)}}
     \bigl(\tilde{U}_{l \alpha }(\bm{k})\bigr)^{\dagger}
     G_{ls}^{KS}(\bm{k},\omega)\,
     \tilde{U}_{s \beta}(\bm{k})\\
G^{KS}_{ls}(\bm{k},\omega))
&=\sum_{\alpha\beta}\bigl(\tilde{U}^{-1}(\bm{k})\bigr)^{\dagger}_{l\alpha}
     G_{\alpha\beta}(\bm{k},\omega)\,
     (\tilde{U}^{-1}(\bm{k}))_{\beta s}
\end{align}
where the generalized $\tilde{U}$ matrix has been introduced (we remind that these matrices are unitary). 

\subsubsection*{Downfolding and Upfolding procedures in a super-cell}

In this section, we analyze how to upfold a quantity defined in the Brillouin zone (BZ) of a super-cell (SC) to a quantity in the BZ of the related primitive-cell (PC). We remind that defined the primitive vectors of the primitive cell as $\bm{a}^{PC}_i$ where $i=0,1,2$, the primitive vectors of the super-cell are related to the former by the epitaxy matrix $M$ 
\begin{equation}
    \bm{a}^{SC}_i=\sum_jM_{ij}\bm{a}^{PC}_j
\end{equation}
The $\bm{K}$ points in the super-cell BZ are mapped in number $N$ ($=|det(M)|$) $\bm{k}$ points of the primitive cell BZ
\begin{equation}
    \bm{k}=\bm{K}+\bm{G_i}
\end{equation}
where $\bm{G}_i$ is a set of $N$ reciprocal vectors of the super-cell. To each $\bm{k}$ in the primitive BZ corresponds one and only one $\bm{K}$ of the super-cell BZ. The amplitude between Kohn-Sham states of the primitive cell and of the super-cell is (assuming the Wannier states in the super-cell constitute a complete set)
\begin{equation}
  \braket{\psi_{n\bm{k}}^{PC}|\psi_{L\bm{K}}^{SC}}
  = \sum_{\tilde{\bm{K}}\beta}
    \braket{\psi_{n\bm{k}}^{PC}|w_{\beta\tilde{\bm{K}}}^{SC}}
    \braket{w^{SC}_{\beta\tilde{\bm{K}}}|\psi_{L\bm{K}}^{SC}}
\end{equation}
where 
\begin{align}
  \ket{w_{\beta\bm{K}}^{SC}}
    &= \sum_{r\in (in+out)}\tilde{U}_{\beta r}(\bm{K})\ket{\psi_{r\bm{K}}^{SC}} \\
  \ket{w_{\beta\bm{K}}^{SC}}
    &= \sum_{\bm{R}} e^{-i\bm{K}\cdot\bm{R}}\ket{w_{\beta\bm{R}}^{SC}}
\end{align}
The amplitude can then be written as
\begin{equation}
  \braket{\psi_{n\bm{k}}^{PC}|\psi_{L\bm{K}}^{SC}}
  = \sum_{\sigma\in N}\sum_{\beta}
    \braket{\psi_{n\bm{k}}^{PC}|w_{\beta\bm{K}}^{SC}}
    \tilde{U}^{\dagger}_{L\beta}(\bm{K})
\end{equation}
Next, we consider the amplitude between super-cell Wannier states and primitive cell
Kohn-Sham states
\begin{align}
  \braket{\psi_{n\bm{k}}^{PC}|w_{\beta\bm{K}}^{SC}}
  &= \sum_{\alpha}
     \tilde{U}_{n\alpha}^{PC}(\bm{k})\,
     \braket{w_{\alpha\bm{k}}^{PC}|w_{\beta\bm{K}}^{SC}} \notag \\
  &= \sum_{\alpha}
     \tilde{U}_{n\alpha}^{PC}(\bm{k})\,
     F(\alpha,[\beta],\bm{k},\bm{K})
\end{align}
where
\begin{align}
  F(\alpha,[\beta],\bm{k},\bm{K})
  &= \sum_{\bm{r}\bm{R}}
     \braket{w_{\alpha\bm{r}}^{PC}|w_{\beta\bm{R}}^{SC}}
     e^{i\bm{k}\cdot\bm{r}}\,e^{-i\bm{K}\cdot\bm{R}} \notag \\
  &= e^{\,i\sum_j n_j^{\alpha[\beta]}\,\bm{k}\cdot\bm{a}^{PC}_j}
     \sum_{\bm{R}} e^{i(\bm{k}-\bm{K})\cdot\bm{R}} \notag \\
  &= e^{\,i\sum_j n_j^{\alpha[\beta]}\,\bm{k}\cdot\bm{a}^{PC}_j}\,
     \delta_{\bm{k},\,\bm{K}+\sum_l n_l \bm{b}^{SC}_l}
     \label{eq:F}
\end{align}
with $\bm{R}$ a SC lattice vector, $\bm{r}$ a PC lattice vector. The overlap between the two Wannier states in the PC and in the SC is nonzero only when the $\alpha$ PC Wannier state coincides with the $\beta$ SC Wannier state. The required translation is encoded in $n^{[\alpha]\beta}_j$. A nonzero contribution from the sum over $\bm{R}$ further requires 
\begin{equation}
    \bm{k}-\bm{K} = \sum_l n_l \bm{b}_l^{SC}
\end{equation}
with $n_l$ an integer. Given these amplitudes, a SC Brillouin zone quantity can be unfolded to the PC
Brillouin zone
\begin{equation}
  O_{nm}^{PC}(\bm{k},\omega)
  = \sum_{LS\bm{K}}
    \braket{\psi_{n\bm{k}}^{PC}|\psi_{L\bm{K}}^{SC}}\,
    O_{LS}^{SC}(\bm{K},\omega)\,
    \braket{\psi_{S\bm{K}}^{SC}|\psi_{m\bm{k}}^{PC}}.
\end{equation}
For the Green's function, which is diagonal in the Kohn-Sham states, this
simplifies to
\begin{equation}
  O_{n}^{PC}(\bm{k},\omega)
  = \sum_{L\bm{K}}
    \bigl|\braket{\psi_{n\bm{k}}^{PC}|\psi_{L\bm{K}}^{SC}}\bigr|^{2}
    O_{L}^{SC}(\bm{K},\omega).
\end{equation}
Since $\tilde{U}^{PC}$ is not generally known, we take the trace over PC Kohn-Sham states
\begin{equation}
  O^{PC}(\bm{k},\omega)
  = \sum_{L\bm{K}n}
    \bigl|\braket{\psi_{n\bm{k}}^{PC}|\psi_{L\bm{K}}^{SC}}\bigr|^{2}
    O_{L}^{SC}(\bm{K},\omega).
\end{equation}
Expanding the squared modulus we obtain
\begin{align}
  O^{PC}(\bm{k},\omega)
  &= \sum_{L,\bm{K}}
     \sum_{\beta,\gamma}
     \sum_{\alpha,\rho}
     \sum_{n}
     \tilde{U}^{PC}_{n\alpha}(\bm{k})\,
     F(\alpha,[\beta],\bm{k},\bm{K}) \notag \\
  &\quad \times
     \bigl(\tilde{U}^{SC}_{L\beta}(\bm{K})\bigr)^{\dagger}
     \tilde{U}^{SC}_{L\gamma}(\bm{K}) \notag \\
  &\quad \times
     F^{*}(\rho,[\gamma],\bm{k},\bm{K})\,
     \bigl(\tilde{U}^{PC}_{n\rho}(\bm{k})\bigr)^{\dagger} \notag \\
  &\quad \times
     O^{SC}_{L}(\bm{K},\omega).
\end{align}
where $F$ is given by Eq.~\eqref{eq:F}. When the observable is already expressed in the SC Wannier states, the unfolding formula
takes the compact form
\begin{align}
  O_{nm}^{PC}(\bm{k},\omega)
  &= \sum_{\bm{K}\beta,\alpha\rho,\gamma}
     \tilde{U}_{n\gamma}^{PC}(\bm{k})\,
     F(\gamma,[\beta],\bm{k},\bm{K}) \notag \\
  &\quad \times
     O_{\beta\alpha}^{SC}(\bm{K},\omega)\,
     \bigl(\tilde{U}_{m\rho}^{PC}(\bm{k})\bigr)^{\dagger}
     F^{*}(\rho,[\alpha],\bm{k},\bm{K}).
\end{align}
Taking the trace ($O^{PC}=\sum_m O_{mm}^{PC}$) and using the unitarity of $\tilde{U}^{PC}$, the result is
\begin{align}
  O^{PC}(\bm{k},\omega)
  = \sum_{\bm{K}}\sum_{\beta,\alpha}\sum_{\gamma}
    F(\gamma,[\beta],\bm{k},\bm{K})\,
    \nonumber
    \\
   \times O_{\beta\alpha}^{SC}(\bm{K},\omega)\,
    F^{*}(\gamma,[\alpha],\bm{k},\bm{K})
\end{align}
where $\alpha,\beta$ are SC Wannier indices, $\gamma$ is a PC Wannier index.

\section{\label{sec:III}A case study: F\lowercase{e}PS$_3$}

For experimental validation of our NEQ-IAP Hubbard I approach, we consider FePS$_3$, a van der Waals Mott–Hubbard/charge-transfer insulator dominated by 3$d$ states, exhibiting zig-zag antiferromagnetism with a Néel temperature of $T_N\approx $ 120 K.

\begin{figure}[hb!]
\centering
\includegraphics[width=0.4\textwidth]{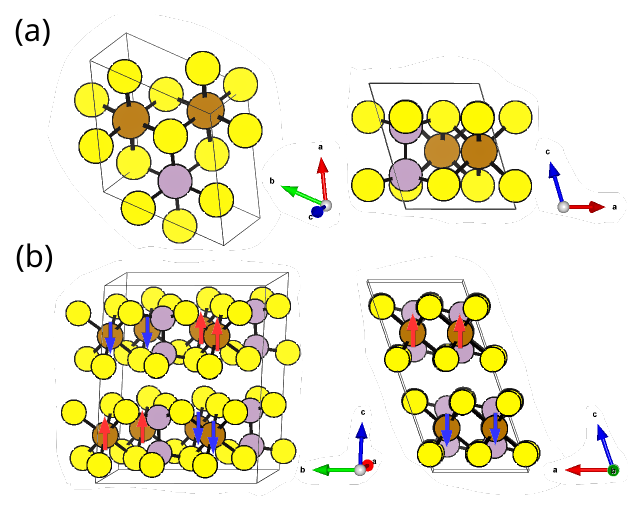}
    \caption{(a) Top view and side-view respectively of the primitve cell of FePS$_3$.(b) Side view of the (zig-zag) antiferromagnetic super-cell of FePS$_3$. Color scheme: brown, Fe; purple, P; yellow, S. Blue and Red arrows in the super-cell of FePS$_3$ indicate the spin orientation of the Fe atoms.}
    \label{fig:feps3_crystal}
\end{figure}

We are interested in both phases: the paramagnetic (PM) and the \ac{AF} phase. The antiferromagnetic phase is described by a super-cell related to the primitive cell by the epitaxy matrix
\begin{equation}
    M=\begin{pmatrix}
        -1&-1&0\\
        1& -1&0\\
        0&0&2
    \end{pmatrix}
\end{equation}
where the local self-energy from a spin-symmetry broken (SB) atomic problem, is applied to the magnetic atoms taking into account different magnetic orientations. The paramagnetic phase instead is described by the primitive cell, where the local self-energy is spin-symmetrized and applied to the magnetic atoms without any differentiation related to a magnetic orientation. A set of super-cells with randomly oriented magnetic atoms, can be used to describe the paramagnetic phase.

\subsection{Computational details}
\label{sec:comput}

We perform density functional theory calculations using the QUANTUM ESPRESSO code \cite{giannozzi2017advanced,giannozzi2009quantum}. We take into account van der Waals dispersion corrections through the vdW-DF-C09 exchange and correlation functional \cite{PhysRevLett.115.136402}. Ultrasoft pseudopotentials, including non-linear core corrections, are used with a plane-wave cutoff of 360 Ry for the charge density and of 90 Ry for the wave function. The atoms are relaxed until the forces reach a maximum of 0.001 Ry/Bohr. The surface Brillouin zone is sampled by a $8\times8\times6$ Monkhorst-Pack \cite{PhysRevB.13.5188} k-point mesh. The lattice parameters for the bulk FePS$_3$ primitive cell are the experimental ones $a =b\simeq 5.9468$~Ang, $c\simeq 6.7222$~Ang, $\alpha\simeq 81.52^{\circ}$, $\beta\simeq 98.48^{\circ}$ and $\gamma\simeq 60.00^{\circ}$ \cite{OUVRARD19851181}. A Wannierisation procedure as implemented in the Wannier90 code \cite{RevModPhys.84.1419} is then applied. The Wannier functions are constructed for the valence and low-lying conduction manifolds using atom-centered projections defined for the atomic $d$ orbitals of the Fe atoms and the atomic $p$ orbitals for the S atoms. The tight-binding Hamiltonian and the unitary transformation matrices thus obtained, are then  manipulated using custom code developed by the authors. This code is employed to compute single-particle properties, many-body Green's functions, self-energies and the Lindblad dynamics, as well as the upfolding and downfolding procedures, as described in the preceding sections. 

It turns out that the resulting local tight-binding Hamiltonian  for the atomic $d$ orbitals requires an additional diagonalization step to identify $t_{2g}$ and $e_{eg}$ orbitals multiplets assuming an $O_h$ local point group symmetry. Note that the real point group symmetry is a $D_{3d}$, and the additional energy difference between the $t_{2g}$ orbitals in LDA is $\simeq 10\%$ of the crystal field splitting and splits the $t_{2g}$ orbitals into an orbital doublet and a singlet.

Once the local Hilbert space has been constructed, the local Hamiltonian  is exactly diagonalized using the TRIQS library \cite{PARCOLLET2015398}. The local Coulomb interaction on the Fe sites is determined by two Racah parameters $U$ and $J$ ($O_h$ symmetry) spanning all Coulomb matrix elements of the $d$ shell and having the values $U=5.50$~eV and $J=0.60$~eV. We comment on the choice of these values below which is guided by the  requirement to obtain an insulator in the paramagentic and in the antiferromagnetic phase, while the energy difference between the $t_{2g}$ and $e_{g}$ orbitals used here is the \ac{DFT} one of $0.67$~eV. Considering an energy difference in agreement with the spectroscopic data of $\simeq1.1$~eV \cite{NITSCHKE2025100019} gives analogue results to the ones presented in this section.

The ability of the Hubbard I method to reproduce a Mott-Hubbard/charge-transfer insulator as found in the experiments \cite{NITSCHKE2025100019} when using the \ac{DFT} constructed tight-binding model depends heavily on the hopping terms $t_{pd}$ and on the orbital energy differences $E_p-E_d$. These parameters of the tight-binding model are initially determined through the \ac{DFT} calculation (i.e. cell strain or atoms relaxation) and through the Wannierisation procedure (centering and spreading of the Wannier functions). In the Wannierisation procedure we minimize spurious hybridization effects between the $p$ and $d$ orbitals, or between these orbitals and orbitals outside of the selected energy window, to reproduce atomic-like Wannier functions, considering a small disentanglement window and few steps in the localization procedure. The obtained Wannier functions for the 3$d$ orbitals are delocalized with
a spread of $\sigma_d\sim 1.79$~Ang$^2$ and $\sigma_p\sim3.20$~Ang$^2$. 
As a consequence, the overlap of the $p$ and the $d$-Wannier functions are large, yielding overly large hopping terms $t_{pd}$. We could reduce the spreading of the single Wannier functions by enlarging the number of Wannier functions and including the P $p$ orbitals. The hopping terms $t_{pd}$, however, remain to be too large for obtaining a paramagnetic insulator. 
We attribute the problem to the missing self-consistency loop of the Hubbard I method believing that this loop would lead to the reduction of the band width and an insulator, as shown by analogous methods \cite{nitschke2023valence,PhysRevB.105.L041108}.

The \ac{DFT} calculation predicts a metal with partially filled $d$ band located around the chemical potential with a filling of about 7e$^-$ in the Fe d orbitals while experimentally an insultator \cite{NITSCHKE2025100019}
is found with an expected valence of Fe$^{2+}$. The task of the many-body correction is to reproduce these experimental findings and use the results to obtain tdARPES data. It is well known \cite{RevModPhys.68.13,Bulla1999} that the Mott transition in the  Hubbard model depends on the ratio $U/t$. In order to obtain a Mott insulator one has two options: (i) increase $U$ or decrease $t$. As mentioned above, the hopping terms $t_{pd}$ are overestimated in our Wannierization due to the extended nature of the 3$d$ Wannier orbitals. We could artificially reduce $t_{pd}$ or increase  $U,J$ above the values used in the literature  \cite{NITSCHKE2025100019} and use the LDA hopping terms $t_{pd}$. We decided for the later one and use the parameters $U,J$ to be ajustable as often in the \ac{DFT}+\ac{DMFT} approach.

We note that the partial overpopulation of the Fe orbitals related to the $d-p$ nature of their $d$ orbitals, is in reality in agreement with a recent experimental report by Wei et al.~\cite{AF-FePS3-2025} claiming that  the X-ray absorption spectroscopy can be consistently interpreted as partially mixed valent state comprising a superposition state of a Fe $|d^6\rangle $ and a Fe $|d^6,L\rangle $ configuration where $L$ denotes a ligand hole. The authors \cite{AF-FePS3-2025} reported a 68.5\% probability of  $|d^6,L\rangle $ configuration which would correspond to an average Fe 3$d$ occupancy of $\approx 6.3$~$e^{-}$. 
Subsequently, we
tune the double-counting correction in the Hubbard I approach such that the local population of the Fe $d$ orbitals is of the order of $6$~$e^-$.

In summary, first, we solve the DFT problem for the primitive cell and extract from it a tight-binding Hamiltonian. In the antiferromagnetic case, this Hamiltonian is extended to the corresponding super-cell. Second, we introduce the Coulomb repulsion
locally on the Fe 3$d$ orbitals (after constraining the local point group symmetry to be $O_h$) and solve the resulting local interacting problem exactly. From its solution we obtain the local interacting and free Green's functions, and from the corresponding local Dyson equation we extract the local $d$ electron self-energy. Third, we obtain the k-dependent \acs{GF} 
%solve the lattice Dyson equation 
using the original DFT tight-binding Hamiltonian
together with this local self-energy.

This scheme is extended to the \ac{NEQ} case. We follow two routes at this point.

In the first simplified route, we treat the effect of the laser pulse by replacing the density matrix by a projector onto those excited states compatible with the laser frequency. Note that the energy difference between non-degenerate local states are always larger then the thermal energy $k_B T$. We calculate the \ac{NEQ} \acp{GF} analog to the equilibrium which yields always  positive-definite spectral functions.

In the second route, we use the effective  pulse  Hamiltonian, Eq.~(\ref{eq:pulse_1}), and the effective transition matrix from Eq.~(\ref{eq:dipole}), to integrate the matrix representation of the density operator as well as the local creation and annihilation operators, needed in the Eqs.\ \eqref{e:def-orbital-lesser-gf} and \eqref{e:def-orbital-greater-gf} in the local $d$ orbital subspace.
The $d-d$ transition matrix is obtained from the DFT tight-binding Hamiltonian and the position operator $\braket{\bm{r}}$ in the Wannier basis. Using the  commutator $\vec{\hat p}/m =i[H,\vec{r}] $ yields the following textbook identity 
\begin{align}
    \braket{w_{\bm{0}\alpha}|\tfrac{-i\bm{\nabla}}{m}|w_{\bm{R}\beta}}\notag=i\braket{w_{\bm{0}\alpha}|[t,\bm{r}]|w_{\bm{R}\beta}}.
\end{align}
Note that we absorbed $\hbar$ in the definition of the time throughout the paper so that frequency and energy are equivalent.
We can write the $p-d$ transition vector as \cite{TomczakEffPeierls2009,PhysRevB.85.205133}
\begin{align}
    \bm{\tilde{\chi}}_{\alpha\beta}(\bm{R})
=&   \frac{1}{N_{\bm{k}}}
\sum_{\bm{k}}e^{i\bm{k} \bm{R}}\bm{\tilde{\chi}}_{\alpha\beta}(\bm{k})\\
    \bm{\tilde{\chi}}_{\alpha\beta}(\bm{k})= &\frac{e}{2}\bm{\nabla}_{\bm{k}}t_{\alpha\beta}(\bm{k})\notag\\
    & +i\tfrac{e}{2}\sum_{\delta}[t_{\alpha\delta}(\bm{k})\braket{\bm{r}}_{\delta\beta}(\bm{k})-\braket{\bm{r}}_{\alpha\delta}t_{\delta\beta}(\bm{k})]
    \label{eqn:100}
\end{align}
The momentum representation $\bm{\tilde{\chi}}_{\alpha\beta}(\bm{k})$ has two contributions: the first term is the direct current $\propto e\bm{v}(\bm{k})$ matrix
for a simultaneous transition between the $d$ and $p$ orbital subsystem. As Tomczak and Biermann pointed out \cite{TomczakEffPeierls2009}, this term is related to hopping processes between different unit cells (encoded in the k-dependency) and corresponds to the usual Peierls substitution \cite{Peierls:1933ec,LuttingerPeierls1951,PhysRevB.101.205140}.
The second term represents the remaining contribution from the commutator after the transformation into the reciprocal space. This term comprises the commutator of the dipole matrix vector $e\braket{\bm{r}}_{\alpha\delta}$ with the \ac{TB} sub-matrix $t_{\delta\beta}(\bm{k})$ connecting both orbital subsystem within the unit cell.
For  channels resonant with the driving frequency of the light field
the term can be connected to a dipole term $\propto e\bm{E} \bm{D}$
using $\bm{E}=-\partial_t \bm{A}$.

To this end, the local $d-d$ transition matrix is then equal to
\begin{align}
\label{eq:effective_dft}
   \bm{d}_{\alpha\beta} (\omega)=\tfrac{1}{N_{\bm{k}}} \sum_{\bm{k}}\sum_{jk}\bigg(\bm{\tilde{\chi}}^{dp}_{\alpha j}(\bm{k})(G_0^{pp})_{jk}(\bm{k},\omega)t^{pd}_{k\beta}(\bm{k})\notag\\+
   t^{dp}_{\alpha j}(\bm{k})(G_0^{pp})_{jk}(\bm{k},\omega)\bm{\tilde{\chi}}^{pd}_{k\beta}(\bm{k})\bigg)
\end{align}
as outlined in Sec.\ \ref{ddtheory}, where we integrate over the first Brillouin zone and the $p$ orbitals Wannier functions contributing to the virtual hopping. The frequency $\omega$ in Eq.\ \eqref{eq:effective_dft} is given by the laser pump frequency, i.e., $\omega\to \omega_0$ \footnote{Technically,
We perform an integration over $\omega$ in a finite interval $\pm\Delta$ around the excited state energy and normalize the result)}.

%\RRchange{However, rather than explicitly propagating the local problem in time during the duration of the light pulse, we make a simplifying assumption: the external pulse drives a population inversion between the ground states and the excited states compatible with the laser frequency. This assumption, lets us evaluate the self-energy at the moment of inversion, simply by replacing the ground-state projector with the excited-state projector in the density matrix.}{} 
%Two justifications underlie \RRchange{this construction }{ these two routes}.

Evaluating the self-energy instant by instant, i.e., through its time propagation, is justified in general by our definition of the \ac{IAP} scheme within the Kadanoff–Baym equations as outline in Sec.~\ref{sec:II.A}.

We have derived an effective description of the dipole-forbidden local $d-d$ transitions in Sec.\ \ref{ddtheory} from a dipole-allowed $d-p$ and $p-d$ transition induced by the external light field in combination with a \ac{TB} hopping matrix element between the $p$ and the $d$ orbital subsystems.

We observed that the lesser self-energy, Eq.~\eqref{eq:lesser_se}, is numerically unstable in proximity of a charge gap, where both the electronic occupational density $\sim Tr\{Im[G^{(<)}]\}$ and the density of states $\sim Tr\{Im[G^{(R)}]\}$ vanish. A way to avoid its direct evaluation is to first introduce an effective distribution $F(T,\omega)$
\begin{align}
    F(T,\omega) &=\frac{N(T,\omega)}{A(T,\omega) }
\end{align}
where $N(T,\omega)= Tr\{Im[G^{(<)}(T,\omega)]\}$, $A(T,\omega)=-2Tr\{Im[G^{(R)}(T,\omega)]\}$,
and reformulate the lesser \ac{GF} in an equilibrium-like form, 
\begin{align}
\label{eq:gf-lesser-effective}
    G^{(<)}_{\alpha\beta}(T,\omega) = -F(T,\omega)\left[G^{(R)}_{\alpha\beta}(T,\omega) - G^{(A)}_{\alpha\beta}(T,\omega)\right].
\end{align}
In a second step, the effective distribution $F(T,\omega)$ is regularized in proximity of the charge gap, 
\begin{align}
    F(T,\omega)= \frac{N(T,\omega) A(T,\omega) + \epsilon^{2} f(\omega)}{A^{2}(T,\omega) + \epsilon^{2}},
\end{align}
interpolating between the ratio $N(T,\omega)/A(T,\omega)$ and the Fermi-Dirac distribution $f(\omega)$, where $\epsilon$ is a small regularization parameter. The regularization of $F(T,\omega)$ is such that (i) in proximity of a pole ($A\gg\epsilon$) the effective distribution  is given by the ratio of the averaged lesser and the spectral function, $F(T,\omega)=N(T,\omega)/A(T,\omega)$, (ii) while far away from any pole ($A\ll\epsilon$), i.e. in a charge gap, it is given by the Fermi-Dirac distribution $F(T,\omega)=f(\omega)$.

We numerically evaluated the difference between the lesser \ac{GF} determined via Eq.\ \eqref{eq:gf-lesser-effective} and the  original lesser Green's function 
calculated by numerical integration, which enters Eq.\ \eqref{eq:gf-lesser-effective} through the definition of $F(T,\omega)$, and obtained an error in the order of $1-2\%$. 
Using Eq.\ \eqref{eq:gf-lesser-effective}, we can express the regularized lesser self-energy as
\begin{align}
    \Sigma^{(<)}_{\alpha\beta}(T,\omega)=-F(T,\omega)\bigg[(G^{(R)})_{\alpha\beta}^{-1}(\omega)-(G^{(A)})_{\alpha\beta}^{-1}(\omega)\bigg]\notag\\
-f(\omega)\bigg[(G_0^{(R)})_{\alpha\beta}^{-1}(\omega)-(G_0^{(A)})_{\alpha\beta}^{-1}(\omega)\bigg].
\end{align}
which avoids the numerical divergences from the inversion of the lesser \ac{GF} in the region of the energy gap.

\subsection{\label{sec:3$d$-dynamics}Local dynamics on the Fe 3$d$ shell}

\begin{figure}[tb!]
    \centering
    \hspace*{0cm}\includegraphics[width=0.4\textwidth]{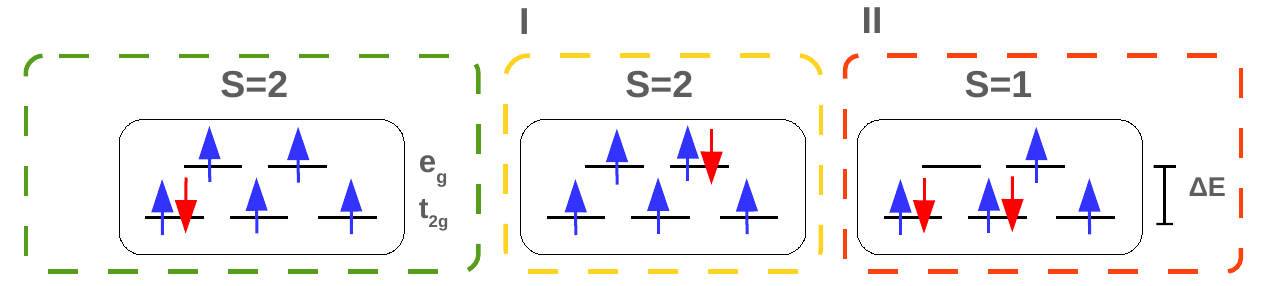}
    \caption{Graph of the main Fe 3$d^6$ excitations. First quadrant: ground-state atomic configuration; Second quadrant (Yellow): first atomic excitation (at $\simeq0.7$~eV). Third quadrant (Orange): second atomic excitation (at $\simeq2.0$~eV). The black horizontal lines represent the atomic orbitals.}
    \label{fig:excitations}
\end{figure}

The correlated equilibrium and non-equilibrium dynamics depend on
the spectrum of $H_\text{imp}$ defined in Eq.~\eqref{eq:H-imp}. In particular, the \ac{DFT} value of the splitting between the $t_{2g}$ and $e_{g}$ orbitals gives the spin allowed  $^5T_{2g}\to$ $^5E_{g}$ transition at an energy of $\simeq 0.7$~eV, while the combination of the Racah parameters $U$ and $J$ selected in order to open the \ac{AF} gap gives a spin forbidden $^5T_{2g}\to$ $^3T_{1g}$ transition at an energy of $2.0$~eV, in proximity to the experimental values of $1.1$~eV and of $1.8$~eV \cite{JoyVasudevanFePS3-Optical-1992}. Meanwhile, the single particle orbital energy is selected such that the ground state  of $H_\text{imp}$ is a d$^6$ configuration.

Fig.~\ref{fig:excitations} shows the schematics of
the lowest excitations in the $N=6$
subspace of the 3$d$ shell. The left panel depicts the fully polarized configuration
of one of the ground state $S=2$ multiplets
which obviously follows Hund's rule: all orbitals are occupied by
one electron with aligned spin and one electron with opposite spin can be added to one
of the $t_{2g}$ orbitals. Hence the $^5T_{2g}$ ground state configuration possesses
an additional threefold orbital degeneracy to the $S=2$ spin degeneracy. The middle panel of Fig.~\ref{fig:excitations} shows the first excited $^5E_g$ configuration \cite{JoyVasudevanFePS3-Optical-1992} where the electron with the anti-aligned spin is put into the excited $e_{g}$
orbitals, leading to a two-fold orbital degeneracy. Since the contribution of the Coulomb energy is identical, the excitation energy is given by the splitting $\Delta E_{et} = E_{e_{g}} - E_{t_{2g}}$. 
For the second excitation only
one contributing Slater determinant is displayed in the right panel of Fig.~\ref{fig:excitations}. This many-body state is a superposition of several Slater determinants. In the depicted
configuration, relative to the ground state
(Fig.~\ref{fig:excitations} left panel) one electron of the $e_{g}$ orbital is placed
into the ground state $t_{2g}$ orbitals with opposite spin. There is an energy gain of $\Delta E_{et}$
but a simultaneous loss of Hund's energy $J$. 
Furthermore, the pair fluctuation term in 
$H_\text{imp}$ induces a coupling between the two double occupied $t_{2g}$ orbitals and the empty $e_{g}$ orbital: This excited state is
a superposition of different Slater determinants also involving virtual
contributions from populating one $e_{g}$ orbital with two electrons. 
Therefore, the excitation energy of this $^3T_{1g}$ state depends on $\Delta E_{et}$ and $J$ but also to some degree on $U$. 

It has been suggested \cite{NITSCHKE2025100019} that the
configurations shown in Fig.~\ref{fig:excitations} 
are the three relevant 3$d$ shell configurations for understanding
the \ac{trARPES} experiments with pump laser energies of $\hbar\omega=1.1,1.8$eV.
Although the transition of the ground state to one of the two excited states is dipole-forbidden, the final state can be reached by a $d-p$ dipole transition
of an electron to a neighboring S $p$ orbital with a consecutive $p-d$ hopping back to 3$d$ shell as outlined in Sec.~\ref{ddtheory}. In NiPS$_3$ it is believed \cite{He:2024yb} that the  $d-p$ dipole transition can lead to a Hund's exciton which can propagate in a magnetic background.

\begin{figure}[tb!]
    \centering
    \hspace*{0cm}\includegraphics[width=0.4\textwidth]{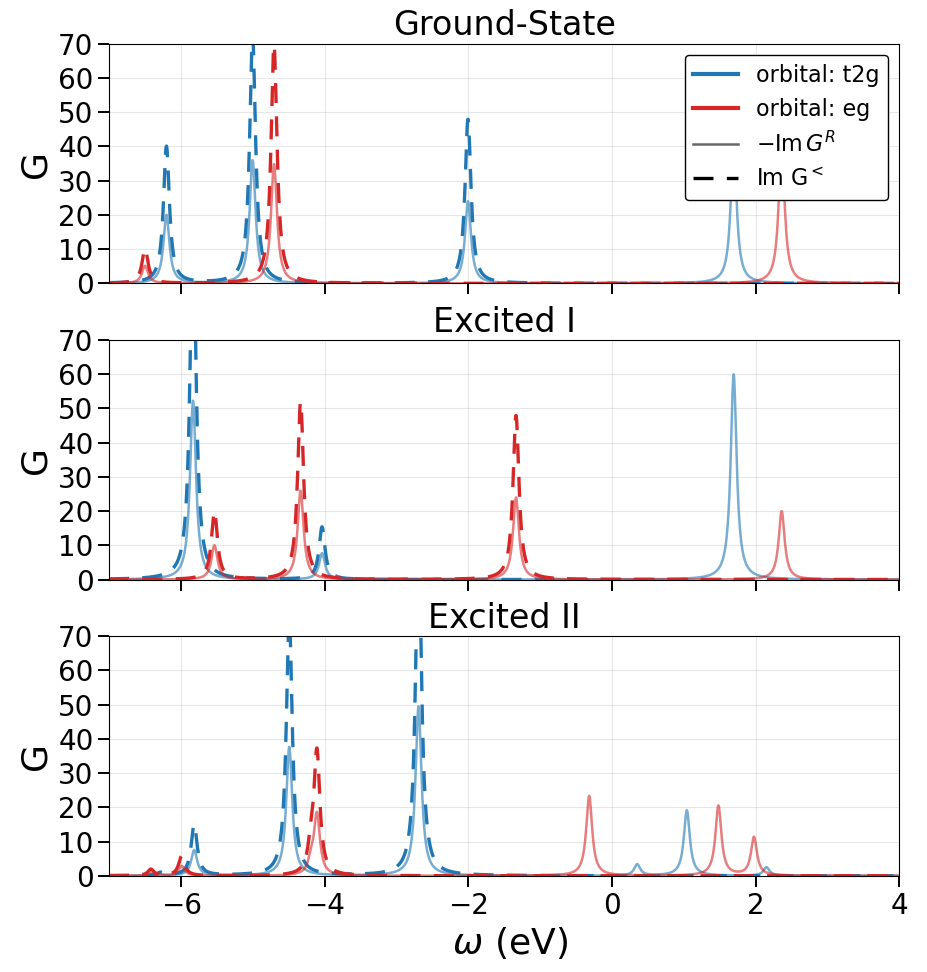}
    \caption{Retarded and lesser atomic Green's functions (spin-summed), for the ground-state, first excited ($\simeq0.7$~eV) and second excited ($\simeq2.0$~eV) paramagnetic atomic configurations ($U=5.50$, $J=0.60$ and $\Delta E\simeq0.7$~eV). The parameters considered in the evaluation are frequency grid of $\Delta\omega\simeq0.01$~eV, Lorentzian broadening $\eta=0.05$ and relative temperature $\beta=600$.} 
    \label{fig:param_atomic}
\end{figure}

In summary, the U, J and $\Delta E$ values selected here give  excitation energies of the order of $\simeq0.7$~eV and $\simeq2.0$~eV in the local atomic problem. We report in Fig.~\ref{fig:param_atomic} the atomic retarded and lesser Green's functions, for the ground-state atomic configuration and for the two first atomic excitation configurations, $^5E_{g}$ and $^3T_{1g}$. Here, we combine the contributions from the $t_{2g}$ and $e_g$ orbitals, and sum over the spin channels.

After an exact diagonalization of $H_\text{imp}$ the real-time definitions reported in Sec.~\ref{sec:II.B} are used to analytically obtain the Fourier transformation. We also
check the spectral sum-rule of the retarded Green's function. We notice that the first excited peaks of the lesser Green's functions correspond to the ground-state spectral function  when shifted by the excitation energy $\simeq0.7$~eV.

\subsection{\label{sec:III-C} Understanding equilibrium \ac{ARPES} data}

Although the main focus of this paper is to present
and apply our simplified \ac{NEQ} approach to pump-probe setups such as \ac{trARPES}
for correlated insulators, in this section we set the stage 
by gauging the method in equilibrium. We used the values $U=5.50$~eV and $J=0.60$~eV and the \ac{DFT}
energy difference between the $t_{2g}$ and $e_{g}$ orbitals of  $0.7$eV for the local 3$d$ problem. With our Hubbard I approach we adjusted the double counting corrections on the Fe d orbitals and the Fermi energy such that charge neutrality is observed as well as the model predicts a filling of $\sim6$~$e^-$ in the Fe $d$ shell. 

\subsubsection{Density of states and correlated band structure}

\begin{figure}[tb!]
    \centering
\includegraphics[width=0.4\textwidth]{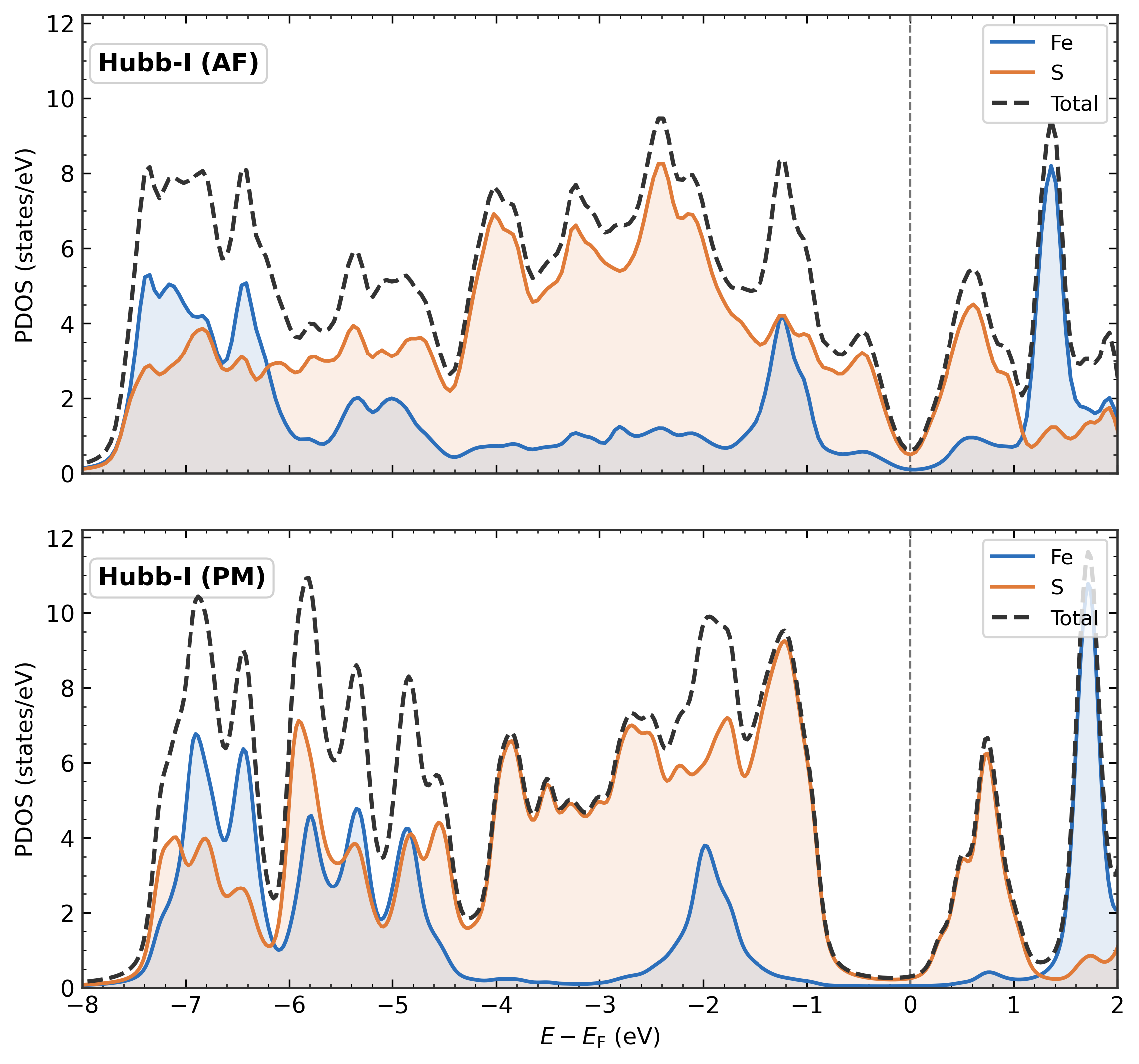}
    \caption{Projected density of states obtained from Hubbard I, in the antiferromagnetic (top) and paramagnetic phase (bottom). Color scheme: Blue Fe $d$ orbitals projection; Orange: S $p$ orbitals projection. The dotted line corresponds to the total of the two projections.}
    \label{fig:pdos}
\end{figure}
\begin{figure}[htb!]
    \centering
    \hspace*{0cm}\includegraphics[width=0.4\textwidth]{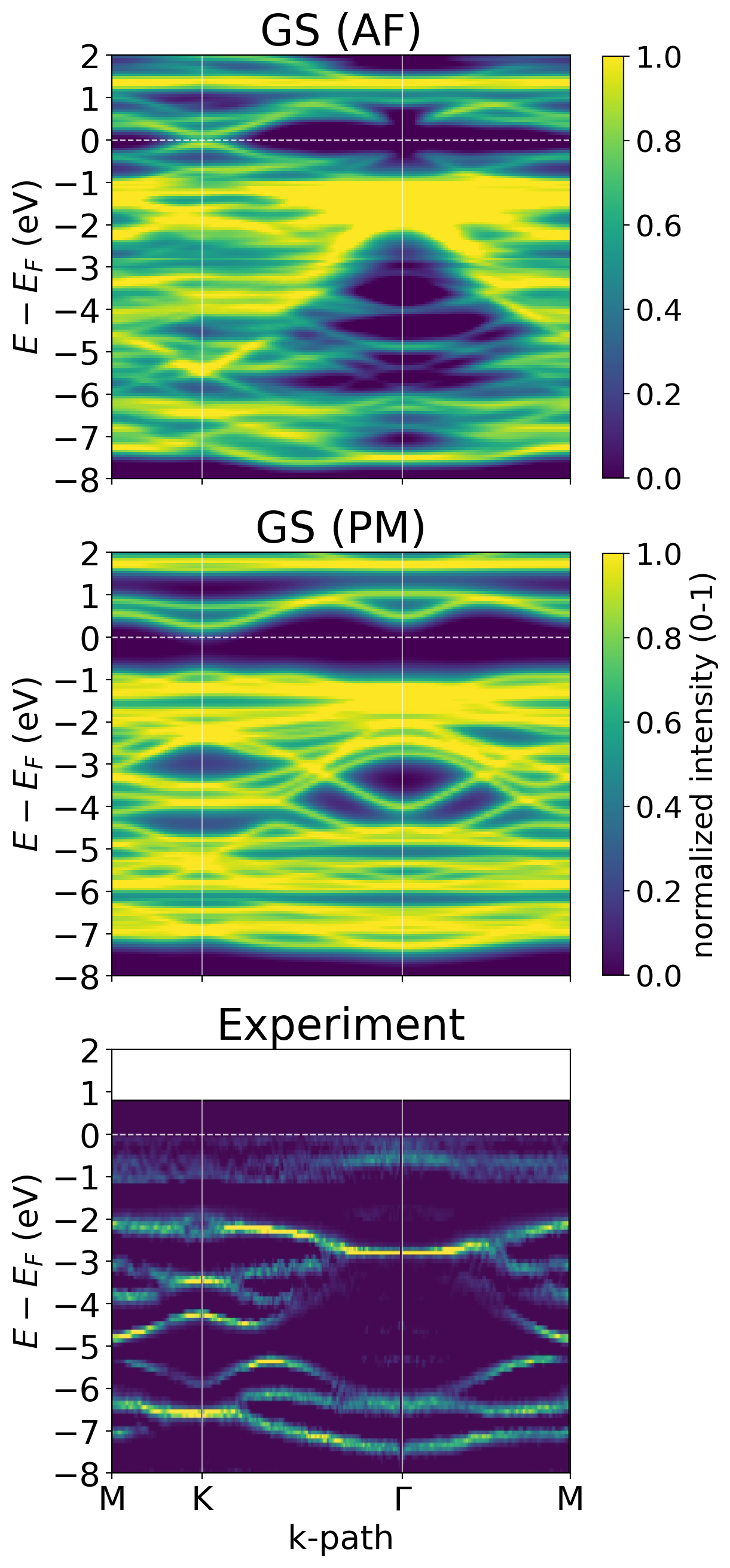}
    \caption{Band structure obtained from \ac{DFT} + Hubbard I calculations, in the antiferromagnetic (top) and paramagnetic phases (center). The experimental results obtained from ARPES measurements (curvature-ARPES \cite{10.1063/1.3585113}) of the paramagnetic phase are reported (bottom). To increase the contrast in the calculated image, we select the data in the intensity window from $0.1$ to $0.9$.}
    \label{fig:bands}
\end{figure}

FePS$_3$ is found in a paramagnetic phase at high temperatures and in a zig-zag antiferromagnetic order where the spins align in an out-of-plane moment orientations below $T_N \approx 120K$ \cite{AF-FePS3-2025}.
We combined the \ac{DFT} +  Hubbard I scheme, as outlined above. The  projected density of states for the  \ac{AF} and paramagnetic (PM) phase are shown in Fig.~\ref{fig:pdos}.
We find a qualitative agreement between the two phases concerning the the $d$ and $p$ orbital disposition. The antiferromagnetic ordering is associated with a reduction of the valence-conduction gap. Recently  blue shifts of the Hubbard band gap of the d-orbitals 
in correlated materials \cite{Hafez-Torbati-MBS-2021,PhysRevB.106.205117} was reported upon entering the \ac{AF} phase. Inspecting the \ac{PDOS} contribution of the Fe $d$ orbital clearly shows
a shift of the $d$ spectral weight to lower energies in our calculations  as depicted in  Fig.~\ref{fig:pdos}.
Note however, that the model calculations in Refs.~\cite{Hafez-Torbati-MBS-2021,PhysRevB.106.205117} use an effective single-band Hubbard model augmented with auxiliary Hund's rules spins while we take into account all five $d$ orbital and in particular the $d-p$ hybridization to the S $p$ orbitals. Although a spectral weight blue shift in the $d$ \ac{PDOS} spectrum is observed, the physical valence-conduction gap shows a red-shift when entering the \ac{AF} phase. 

We found that the occupation of the Fe d orbitals is of $\simeq3.10$~$e^-$ for spin channel in the paramagnetic phase, while it is of $\simeq4.89$~$e^-$ for the majority spin channel and of $\simeq1.60$~$e^-$ for the minority spin channel in the antiferromagnetic phase. We measured a conduction-valence band gap of $\simeq0.05$~eV in the AF phase, and of $\simeq1.10$~eV in the PM phase. The corresponding band structures for the paramagnetic and the antiferromagnetic phase are shown in the two top panels in  Fig.~\ref{fig:bands}. We augmented the results  with experimental results extracted from ARPES data for the paramagnetic phase shown in the lower panel of Fig.~\ref{fig:bands} (note that the ARPES data only show the occupied bands).

\subsubsection{\ac{ARPES} and \ac{trARPES} k-maps band structure}

\begin{figure}[htb!]
    \centering
\includegraphics[width=0.3\textwidth]{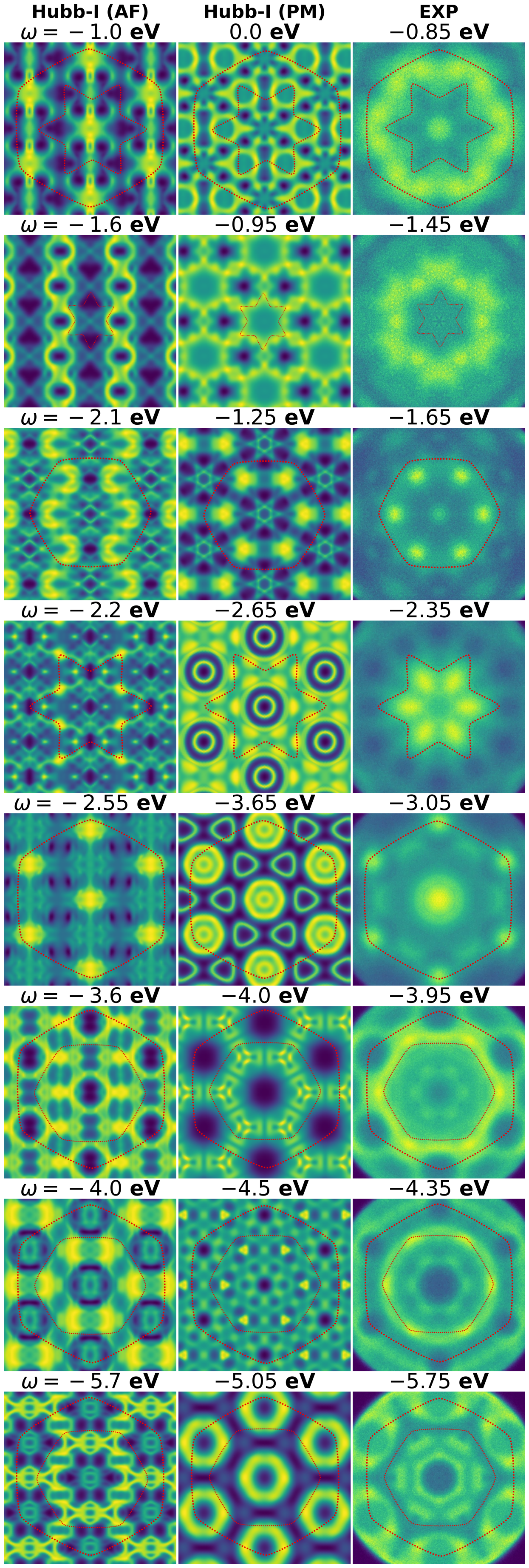}
    \caption{K-maps (1st and 2nd Brillouin zone) obtained from \ac{DFT} + Hubbard I calculations, in the antiferromagnetic phase (left), paramagnetic phase (center), and from experiments (right). 
    The color scale is from 0 (dark blue) to 1 (bright yellow) (in arbitrary units), and is related to the degree of electronic occupation in the Brillouin zone: 0 empty and 1 filled. The $\omega$ is the energy cut evaluated with respect to the top valence band (an $\Delta E=0.1$~eV energy interval has been considered around each cut and averaged). An integration over $k_z$ has been performed. The dotted line serves as a guide to the eye. A logarithmic color scale has been applied.}
    \label{fig:kmaps1}
\end{figure}

\ac{ARPES} is  an experimental procedure where a light pulse excites electrons in the material into an unbound state, then these electrons propagate to the surface, leave the material and their kinetic properties are determined with angular resolution. Setting aside  questions of surface states and finite penetration depth of the light, we assume that that there is a 1:1 correspondence of the reconstructed k-maps in \ac{ARPES} for a fixed initial electron energy and the projection of the k-dependent lesser \ac{GF} into the $k_x-k_y$-plane for a fixed energy $\omega$. Since FePS$_3$ is a 2d van der Waals material, and the $k_z$ is not conserved in the \ac{ARPES} experiment, we perform a integration over $k_z$ to eliminate the very small dependency of the band structure (lesser \ac{GF}) on $k_z$.

In Figure~\ref{fig:kmaps1} the intensity k-maps for the valence states obtained from the bulk lesser Green's function are shown for different electron energies $\omega$ in eV. The left-column depicts the calculated results for the \ac{AF} phase, the middle column depicts our results for the paramagnetic phase. We augmented the calculation with experimental data taken in the paramagnetic phase shown as right column. We note a significant similarity between the calculated paramagnetic k-maps and the experimental ones at comparable energy planes (middle and right column). Clearly, the k-maps differ in the different energy planes, and not exact 1:1 correspondence between theory and experiment is expected. However, the $C_6$ symmetry is visible in the experimental and the calculated k-maps, as indicated by the thin red lines.

Remember that we are not taking into account the surface sensitivity of the measurements, which  could be included into by replacing the bulk by a layered material. Therefore we selected the energy windows of our calculation such that they match the experimental data the best: they follow monotonically the experimental cuts with a reasonable energy correspondence.

The antiferromagnetic k-maps reported in the left column of Fig.~\ref{fig:kmaps1}, instead, show a breaking of the $C_6$ symmetry. This is visible at the energies $\omega=-1.0,-1.6,-2.1,-2.2$~eV. We selected matching k-maps between the PM and the \ac{AF} phase calculated for the same parameters. The energy difference between the two phases k-maps, is mainly related to the different position of the top valence energy level. 

The numerical difference in the calculation of the two magnetic phases is the use of a magnetic super-cell and a very small local magnetic field applied to the Fe $d$ orbitals leading to a spin polarized ground state.

In more advanced mean-field calculations, one could calculate the effective $d-d$ exchange interaction for the tight-binding model and plug the values into an appropriate Heisenberg model. This would allow to determine the local spin polarization self-consistently below $T_N$ which evolves continuously from zero at $T_N$ to the maximal local polarization at $T=0$. Our calculations assume $T=0$, use the zig-zag antiferromagnetic order as an input for the exact diagonalization of the Fe 3$d$ Hamiltonian and solve the Dyson equation for the lattice as outlined above to illustrate the change in the \ac{AF} phase. 

We already noted above that using the parameters of the PM phase have problems in opening the band gap in the \ac{AF} phase: we attribute that to the large $d-p$ hybridization due to the extended nature of the constructed Wannier orbitals. We would like to emphasise that the main focus of this paper is actually the \ac{NEQ} dynamics accessed experimentally with \ac{trARPES}. Therefore, we do not optimize the calculation to describe the PM to \ac{AF} phase transition as well as the evolution of the insulator gap with temperature.

\subsection{\ac{NEQ} or \ac{trARPES} in the paramagnetic phase: Excited State Projection (ESP)}
\label{sec:simple-pop-inversion}

\subsubsection{\label{sec:neq-dos} \ac{NEQ} density of state}

\begin{figure}[tb!]
    \centering
    \hspace*{0cm}\includegraphics[width=0.4\textwidth]{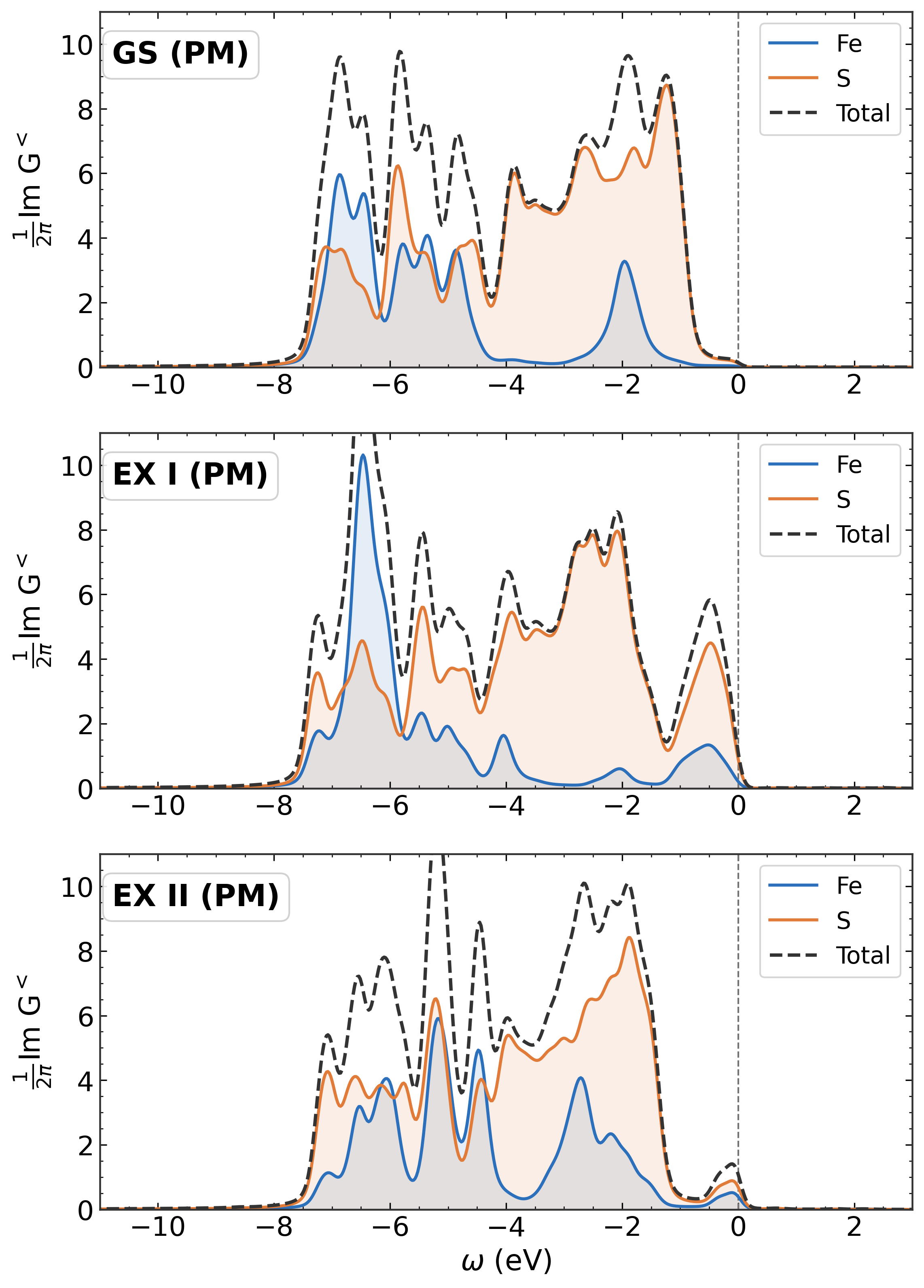}
    \caption{Imaginary part of the lesser Green's function in the paramagnetic phase for the ground state (top), first excited state (center) and second excited state atomic configuration, projected on the different orbital contributions. Here the IAP-ESP method is used for the NEQ quantities. The two excited configurations reported are related to the two atomic excited configurations at $\simeq0.7$~eV and $\simeq2.0$~eV, obtained considering in the local problem the parameters $U=5.50$~eV, $J=0.60$~eV and $\Delta E\simeq0.7$~eV. Color scheme: Blue Fe $d$ orbitals projection; Orange: S $p$ orbitals projection. The dotted line corresponds to the total of the two projections.}
    \label{fig:glesser}
\end{figure}

To set the stage for the investigation of the full local dynamics under the influence
of a light pump pulse, we only consider an instantaneous population inversion of the $d$ orbitals in this section.

The local lesser Green's function of the lattice is calculated from the Hubbard I \acs{GF} using the lattice Dyson equations as outlined in Sec.~\ref{sec:3$d$-dynamics}. The atomic lesser and retarded \acs{GF} entering the Dyson equations of the lattice describe the removal of a single electron leaving the 3$d$ shell in a $N=5$ configuration. Hereby, the local atom density operator is given by the projector onto the ground state multiplets, or  the laser frequency matching exciting states in the configuration space $d^6$ depicted in Fig.~\ref{fig:excitations} and described in Sec.~\ref{sec:comput}. The atomic lesser and retarded \acs{GF} are plotted Fig.~\ref{fig:param_atomic} and the change in the atomic lesser and retarded orbital \ac{GF} calculated with respect to the different density operators were already discussed in Ref.~\cite{NITSCHKE2025100019} and  reported for completeness in Sec.~\ref{sec:3$d$-dynamics}.

We note that the equilibrium spectral peaks located at $\omega\simeq-2.0,-5.0,-5.5,-7.0$~eV in the lattice lesser \ac{GF} are broadened by the lattice dispersion as shown in the upper panel of Fig.~\ref{fig:glesser}. The spectrum is shifted by the excitation energy $\simeq0.7$~eV for the first excited configuration depicted in the central panel of  Fig.~\ref{fig:glesser}.
It appears that the total spectral weight below the chemical potential is conserved so a decrease in the height of the peaks is associated with an increase in their width. At $\omega \simeq-7$~eV the contribution of the different peaks result in a significantly higher peak compared to the PM equilibrum phase. We note that the $d$ peaks in the lattice spectra are additionally modified by $d-p$ hybridization effects and not simply shifted of the excitation energy as suggested from the atomic picture in Sec.~\ref{sec:3$d$-dynamics}. The hybridization between $d$ and $p$ orbitals is not only responsible for the dipole forbidden transition (Sec.~\ref{ddtheory}), but is also modifying the effective bands in the excited configuration.

When the atomic 3$d$ system is in the second excited configuration schematically shown in Fig.~\ref{fig:excitations} the picture completely changes as seen in the bottom panel of Fig.~\ref{fig:glesser}. The lattice spectral function shows Fe $d$ and S $p$ orbital contributions which peak at the same location and is an indication of strong $d-p$ mixing.

Lastly, we note in Fig.~\ref{fig:glesser} as the excitation induces an insulator to metal transition (IMT): the excited electrons, populating now the conduction bands, can access different energy states with no energy cost. For completeness, we report in Appendix.~\ref{subsec:ex_bands} the band structure for the excited states configurations in Fig.\ \ref{fig:bands_ex}. In this regard, we emphasize that the excitation does not merely induce a shift of the chemical potential, but rather modifies the band structure and redistributes the spectral weight. Moreover, we note that the first atomic excitation falls within the conduction–valence band gap, which underscores that this excited state originates from a many-body description that goes beyond the single-particle picture.

\subsubsection{\ac{trARPES} k-maps for FePS$_3$}

\begin{figure}[tb!]
   \centering
\includegraphics[width=0.3\textwidth]{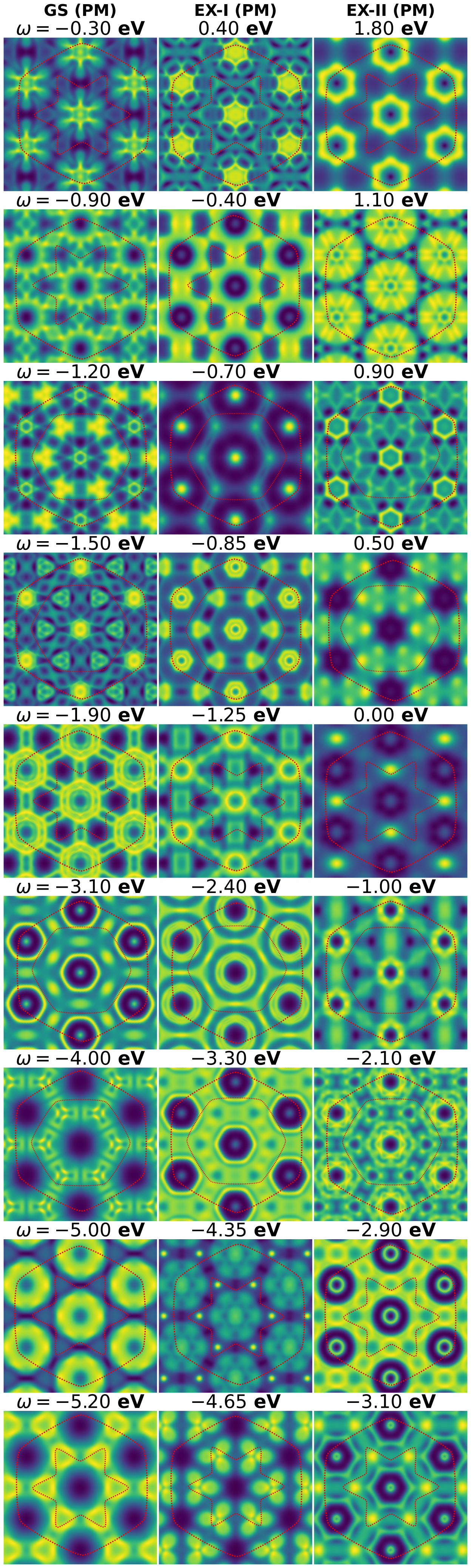}
    \caption{K-maps obtained from a \ac{DFT} + Hubbard I calculation in the paramagnetic phase, taking into account the IAP-ESP method for the first (center) and second (right) atomic excitation, in comparison with the mirrored valence k-maps (left). Color scheme and legend as the one reported in Fig.\ref{fig:kmaps1}.}
    \label{fig:exc1}
\end{figure}

After presenting the k-summed lattice lesser \ac{GF} in the previous section \ref{sec:neq-dos}, we use $G^<(\vec{k},\omega)$ to determine the \ac{trARPES} measurements of the k-maps. The results are depicted in  Fig.~\ref{fig:exc1}. We report in the left column the ground state configuration and a few selected valence energy cuts, in the center column the first excited configuration and energy cuts at energies shifted by $\simeq0.5-0.7$~eV with respect to the selected valence energy cuts, and on the right column the second excited configuration and energy cuts at $\simeq2.0$~eV energy difference with respect to the selected valence energy cuts.

The calculated k-maps shown in the left and the central column of Fig.~\ref{fig:exc1} are very similar between each other \footnote{The same is true if the local LDA
tight-binding model is replace by an effective model with $\Delta E_{et}=1.1eV$ which is closer to the experimental
excitation energy \cite{JoyVasudevanFePS3-Optical-1992}}. This is in qualitative agreement with the experimental  observation in \ac{trARPES} data of Ref.\ \cite{NITSCHKE2025100019}. A better agreement is achieved considering the entire time evolution, as reported in  Sec.\ \ref{sec:III.E} below.

The energy cuts that show a major mirroring in the first excitation configuration tends to correspond to $d$ peaks in the $k$-summed lesser Green's function depicted in Fig.~\ref{fig:glesser}. This is expected as the excitation is involving mainly the $d$ orbitals. We note, however, that the resemblance between the ground state and first excitation k-maps is not 1 to 1. We attribute the differences to the hybridization effects between the $d$ and $p$ orbitals as discussed in Sec.\ \ref{sec:neq-dos}.

In contrary, the k-maps calculated for a locally excitation state of $ \omega \simeq 2$~eV plotted in the right column of Fig.~\ref{fig:exc1} show very different characteristics compared to the ground state k-maps in the PM phase. At $\omega\simeq-1.50$ and $\simeq-1.90$~eV one can clearly see the shift of the peaks to different k-values and the disappearance of the $\vec{k}=(0,0)$ peak. Also the maps at other energies are noticeable different. This result is also in line with the
experimental findings in \cite{NITSCHKE2025100019}.

While the ground states as well as the first excited state on the Fe 3$d^6$ subspace can be written as a single Slater determinant leading to very similar atomic lesser \ac{GF} after shifting the frequency axis by the excitation energy, the second excited state on the Fe 3$d^6$ subspace involves a linear combination of several Slater determinants being a true many-body excitation. Consequently, the fingerprint of the local atomic \acs{GF} is significantly different leading to different lattice \acs{GF}.

Note that due to the fact that the $p$ orbitals remain unmodified by the excitation, some k-maps between the ground state and the excited state, are similar, even at the same energy cut.

In this section we presented a proof of principle study for the calculation of \ac{trARPES} k-maps at a low numerical cost. We (i) used the simplified Hubbard I approach to obtain an approximation for the lattice \acs{GF} (justified in an insulating phase) and (ii) considered a population inversion evolution of the local density matrix, assuming that the laser energies are selected such that essentially local effective $d-d$ transitions are driven by the laser pulse \cite{JoyVasudevanFePS3-Optical-1992} (since the $p-d$ transfer is fast due to the strong hybridization).

\subsection{\ac{NEQ} or \ac{trARPES} in the paramagnetic phase: Excited State Dynamics (ESD)}

\label{sec:III.E}

\subsubsection{Effective transfer matrix elements}

We project the effective $d-d$ transition vector $\bm{d}_{\alpha\beta}(\omega)$,  Eq.~\eqref{eq:effective_dft}, onto the polarization of the vector potential, $\bm{n}_A$,
and averaged the scalar product over the energy interval around the excitation energies  $\omega_0\pm \Delta$ (where $\Delta\simeq0.05$~eV) 
\begin{align}
\label{eq:effective_dipole}
    d_{\alpha\beta}(\omega_0)=\tfrac{1}{2\Delta}\int_{\omega_0-\Delta}^{\omega_0+\Delta}\,\bm{n}_A\cdot\bm{d}_{\alpha\beta}(\omega)\, d\omega.
\end{align}
where $\omega_0$ is the driving frequency of the laser pulse. Note that the electric field shares the orientation $\bm{n}_A$: $\bm{E}\propto \bm{n}_A $ since $\bm{E}(t)=-\partial_t \bm{A}(t)$.

\begin{table}[tb!]
    \centering
    \begin{minipage}{0.60\linewidth}
        \centering
         \begin{tabular}{lcc}
    \toprule
         & $t_{2g}$ & $e_g$ \\
    \midrule
    $t_{2g}$ & 0.0897 & 0.1170 \\
    $e_g$    & 0.1181 & 0.0808 \\
    \bottomrule
  \end{tabular}
        \caption{Mean strength of the effective $d-d$ transition matrix $d^i_{\alpha\beta}(\omega_0)$ (in units of e\AA) at $\omega_0\simeq0.7$~eV (the modulus is averaged over the three directions $\bm{n}_A=\hat{x},\hat{y},\hat{z}$ and Fe atoms, and divided by the resonance frequency $\omega_0$).}
        \label{tab:dipole_066eV}
    \end{minipage}
    \hfill
    \begin{minipage}{0.60\linewidth}
        \centering
        \begin{tabular}{lcc}
    \toprule
         & $t_{2g}$ & $e_g$ \\
    \midrule
    $t_{2g}$ & 0.0123 & 0.0501 \\
    $e_g$    & 0.0501 & 0.0233 \\
    \bottomrule
  \end{tabular}
        \caption{
        Mean strength of the effective $d-d$ transition matrix $d^i_{\alpha\beta}(\omega_0)$ (in units of  e\AA) at $\omega_0\simeq2.0$~eV (the modulus is averaged over the three directions $\bm{n}_A=\hat{x},\hat{y},\hat{z}$ and Fe atoms, and divided by the resonance frequency $\omega_0$).}
        \label{tab:dipole_2eV}
    \end{minipage}
\end{table}

Averaging $|d^i_{\alpha\beta}(\omega_0)|$ over the two Fe sites, the different orbitals and polarization directions $\bm{n}_A$, we obtain the tables \ref{tab:dipole_066eV} and  \ref{tab:dipole_2eV} for the two excitation energies $\omega_0\simeq0.7$~eV and $\omega_0\simeq2.0$~eV. We note that the effective transfer matrix elements for the lower excitation energy are larger than the ones for the higher laser pumping energy.
Due to the location of the $p$ orbitals in the PDOS, as reported in Fig.~\ref{fig:pdos}, the virtual excitations via the $p$ orbitals are slightly suppressed for $\omega_0\simeq2.0$~eV compared to $\omega_0\simeq0.7$~eV.
This averaged transition matrix element $d^i_{\alpha\beta}(\omega_0)$  enters Eq.\ \eqref{eq:pulse_1} as a Hermitian matrix $\bar d^i_{\alpha\beta}(\omega_0)= d^i_{\alpha\beta}(\omega_0)+(d^i)^*_{\beta\alpha}(\omega_0)$ for each of the two Fe sites $i$.

To this end, we substitute the  matrix elements $d^i_{\alpha\beta}(\omega)$ into the 
pulse Hamiltonian, Eq.\ \eqref{eq:pulse_1}, replacing the excitation energy value $\omega_0$ in the $d(\omega)$ term. We consider the complete expression stated in Eq.\ \eqref{eq:effective_dft} and not the averaged quantity (over the orbital sets or the polarization directions).

We stress that a unitary excitation acting on an initially incoherent (equally populated) ground-state manifold cannot produce a complete population inversion
since the excited state has a lower degeneracy than the ground state. In our case, the ratio between the degeneracy of the excited state (degeneracy $=10$) ($\omega_0\simeq0.7$~eV) and that of the ground state (degeneracy $=15$) is $2/3$, which sets an upper limit of approximately $67\%$ on the filling of the excited state. When driving the d-subsystem in resonance, the Rabi oscillation frequency is given by the modules of the transfer matrix element. Therefore, different Rabi frequencies yield interference effects, significantly reducing this value.  A singular-value decomposition (SVD) of the transfer matrix provides insight into the number of excitation channels and into how effectively the excited state is filled. In particular, a larger number of non-zero singular values, especially if of comparable magnitude, indicates the presence of multiple excitation channels, whereas a larger number of vanishing singular values indicates the presence of multiple dark states. Considering the $\hat{z}$ projection, we obtain from the Hermitian part of the transfer matrix the following singular values $(0.299,0.298,0.068,0.067,0.000)$. These values underline two main channels of excitation, which relate to the specific orbital transitions $d_{xz}\rightarrow d_{z^2}$ and $d_{xz}\rightarrow d_{x^2-y^2}$, as can be observed in the population dynamics in panel (c) of Fig.~\ref{fig:linblad_evolution}.

\subsubsection{Density Matrix Evolution}
\label{sec:DM-evolution}

We integrate the von-Neumann equation for the density matrix in the subspace of $N=6$ electrons for each Fe site using the matrix elements $d^i_{\alpha\beta}(\omega_0)$, Eq.\ \eqref{eq:effective_dipole}, in the  pulse Hamiltonian $\hat{H}_\text{pulse}(t)$, and a Runge-Kutta method (RK4) \cite{press1992numerical}.
Note that $\hat{H}_\text{pulse}(t)$ as well as the Lindblad dissipators are particle number conserving, and, therefore, the density matrix is block-diagonal in the number of electrons $N$ on the d-shell.

We selected  the following parameters for the dynamics: linear polarization, i.e., $\bm{n}_A\propto\hat{z}$ , $\gamma_{decay}\simeq5.3\times10^{-4}$~eV, $\gamma_{therm}\simeq 3.5\times10^{-5}$~eV, $\gamma_{deph}\simeq4.0\times10^{-3}$eV, $\sigma\simeq40$~fs, $A\simeq$~0.16Vfs/\AA, and the total simulation time $395$~fs. The RK4 time step $\delta t=0.02$~fs was used, and the pulse center was set at $t_p=140$~fs. The amplitude is selected in order to maximize the excited population, looking at the area of the pulse function considering the channel with the strongest dipole contribution.

\begin{figure}[tb!]
   \centering
\includegraphics[width=0.45\textwidth]{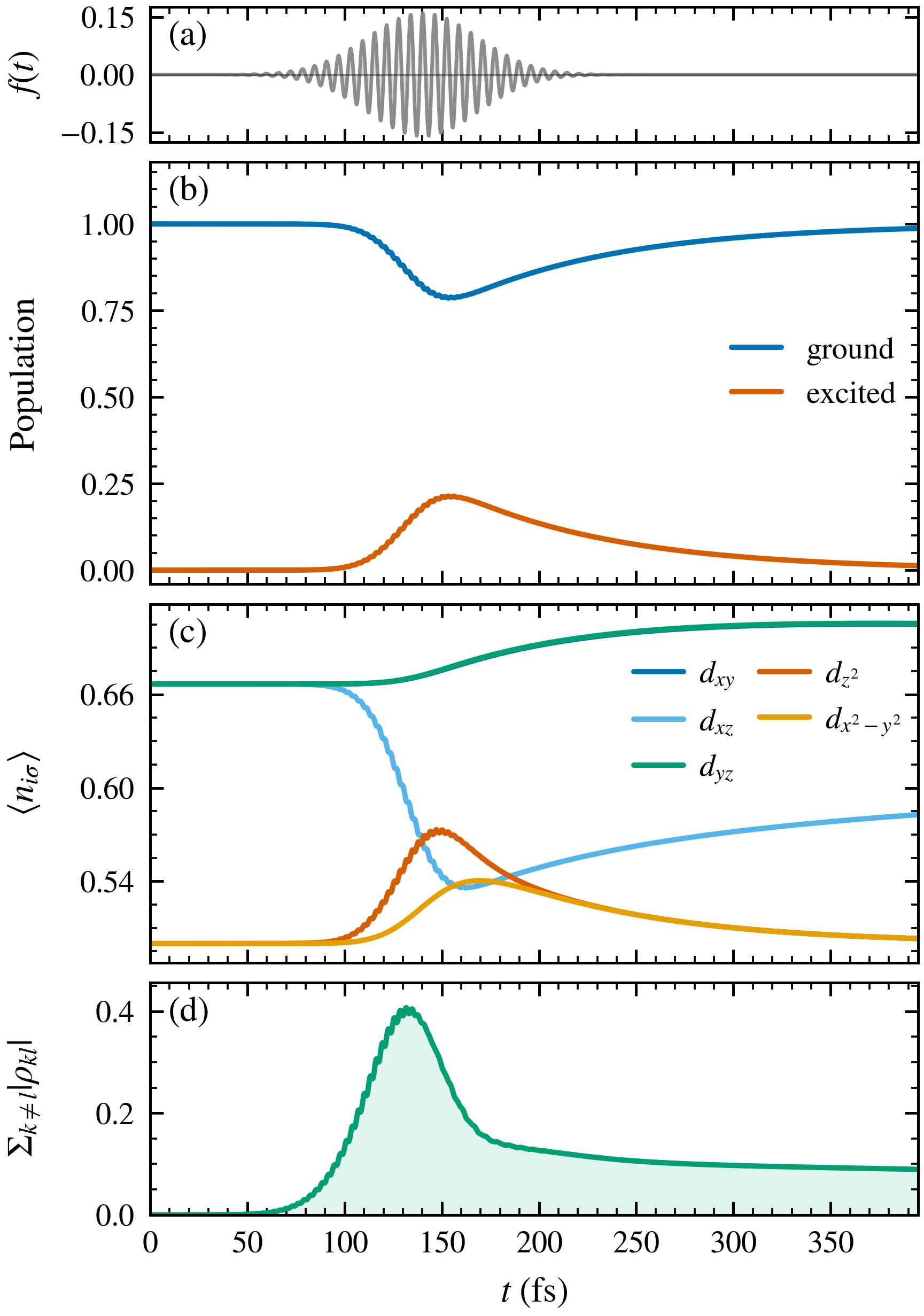}
    \caption{Density matrix dynamics taking into account an external pulse at resonance with the first atomic excitation. Decoherence and dissipative channels are taken into account. In panel (a) we report the external pulse, in panel (b) the population of the ground state and the first excited state, in panel (d) the population of the d orbitals, and in panel (c) the out-of-diagonal terms of the density matrix. The trace of the density matrix is equal to one during the entire evolution. The thermalized states is equivalent to the starting ground state, and it is reached on longer time scales $1$~ps than the one reported.}
    \label{fig:linblad_evolution}
\end{figure}

In Fig.~\ref{fig:linblad_evolution} we report the dynamics of the density matrix for the atomic Fe site subject to a single laser pulse in resonance with the first excitation energy at $\omega_0 \simeq0.7$~eV. Fig.~\ref{fig:linblad_evolution}(a) depicts the pulse shape on the simulated time interval, while Fig.~\ref{fig:linblad_evolution}(c) shows the population dynamics of the lower $t_{2g}$ orbitals vs the upper $e_g$ orbitals for linear polarized light in $\hat z$ direction. Note that in the ground state  multiplets the $t_{2g}$ orbitals are occupied with four electrons and two are located in the  $e_g$ orbitals according to Hund's rule, while in the first excited multiplet around $\omega_0\approx 0.7eV$ above the ground states, one of the $t_{2g}$ electrons is placed into one excited  $e_{g}$ orbital with the same spin degeneracy. In our case, the major laser pump pulse induced transition takes an electron from the $d_{xz}$ orbital and places it into the $d_{z^2}$ orbitals as shown in Fig.~\ref{fig:linblad_evolution}(c). With this configuration, no population inversion could be achieved. The slow decay after the pulse at $t_p=140$~fs is related to the three Lindblad decay rates $\gamma_{decay}$, $\gamma_{therm}$ and $\gamma_{deph}$. In particular, the dephasing is on a time scale of $1/(\gamma_{deph})\simeq165$~fs, the decay is on a time scale of $1/(N_{gs}\gamma_{decay})\simeq83$~fs and the thermalization is on a time scale of $1/(N_{gs}\gamma_{therm})\simeq1.27$~ps.

The second pump frequency of $\omega_0\simeq2.0$~eV corresponds to the $^5T_{2} \to\, ^3T_{1g}$ spectroscopy transition \cite{NITSCHKE2025100019}, and, therefore, is spin forbidden. Since the total spin is conserved by the Hamiltonian $H_0$ and also commutes with $\hat{H}_\text{pulse}(t)$, we cannot describe this transition with our effective $d-d$ transfer matrix approach. Either the spin-orbit interaction is included on the single-particle level in the DFT and in the \ac{TB} model, or the local cluster 
must be extended to the local $p$ ligands, and  spin-exchange processes between the $p$ and the $d$ subsystem must be taken into account. In this case, a $d$ subsystem spin-flip is always associated with a spin-flip in the $p$ subsystem. This exchange mechanism is the same which causes the \ac{AF} magnetic ordering at low temperatures. We, however, leave this extension of the theory to a future work.

\begin{figure}[tb!]
    \centering
    \hspace*{0cm}\includegraphics[width=0.4\textwidth]{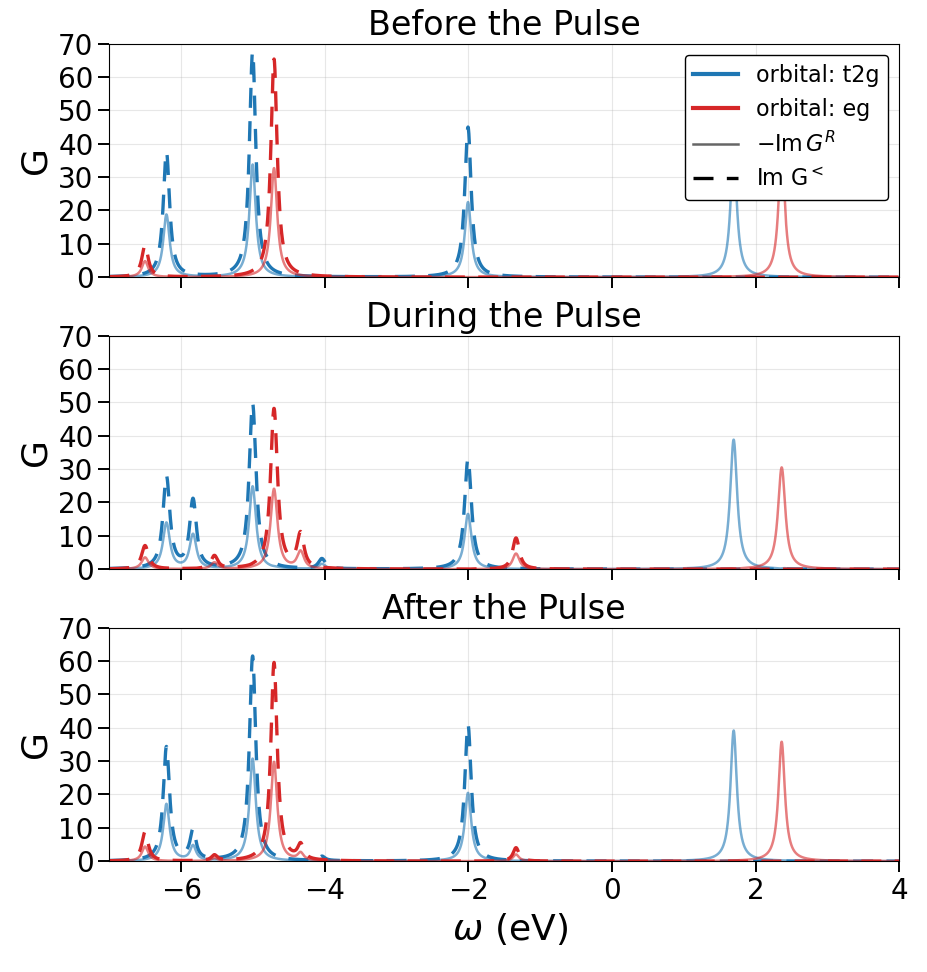}
    \caption{Time evolution of the retarded and lesser atomic Green's functions (spin-summed) for the first excited ($\simeq0.7$~eV) paramagnetic atomic configurations ($U=5.50$, $J=0.60$). In particular, the following time steps $T\simeq-66$~fs, $T\simeq155$~fs and $T\simeq240$~fs in the density matrix evolution are considered, for respectively the top, the central and the bottom panel. The parameters considered in the evaluation are frequency grid of $\Delta\omega\simeq 2$~meV, Lorentzian broadening $\eta=50$meV and relative temperature $\beta=600/$eV.} 
    \label{fig:time_evolution_atomic}
\end{figure}

At this point we focus on three particular instants of the time evolution: before the pulse, shortly after the center of the pulse, $T\approx t_p$ and at a later time, in the decoherence-dissipation regime. For each of the sections we report the retarded and lesser atomic Green's functions (spin-summed) in Fig.~\ref{fig:time_evolution_atomic}.

We observe that the $e_g$ states are partially populated after the pulse, and then during the decoherence-dissipation path depleted in accordance with Fig.~\ref{fig:linblad_evolution}(c). The originally sharp local excitation peaks acquire a frequency broadening, i.e. satellites features, due to the time evolution: The eigenstates of $H_0$ are not sharp and well defined due to the laser pumping of the system as illustrated in the middle and the lower panel of Fig.~\ref{fig:time_evolution_atomic}.

\subsubsection{NEQ density of state}

\begin{figure}[tb!]
    \centering
    \hspace*{0cm}\includegraphics[width=0.4\textwidth]{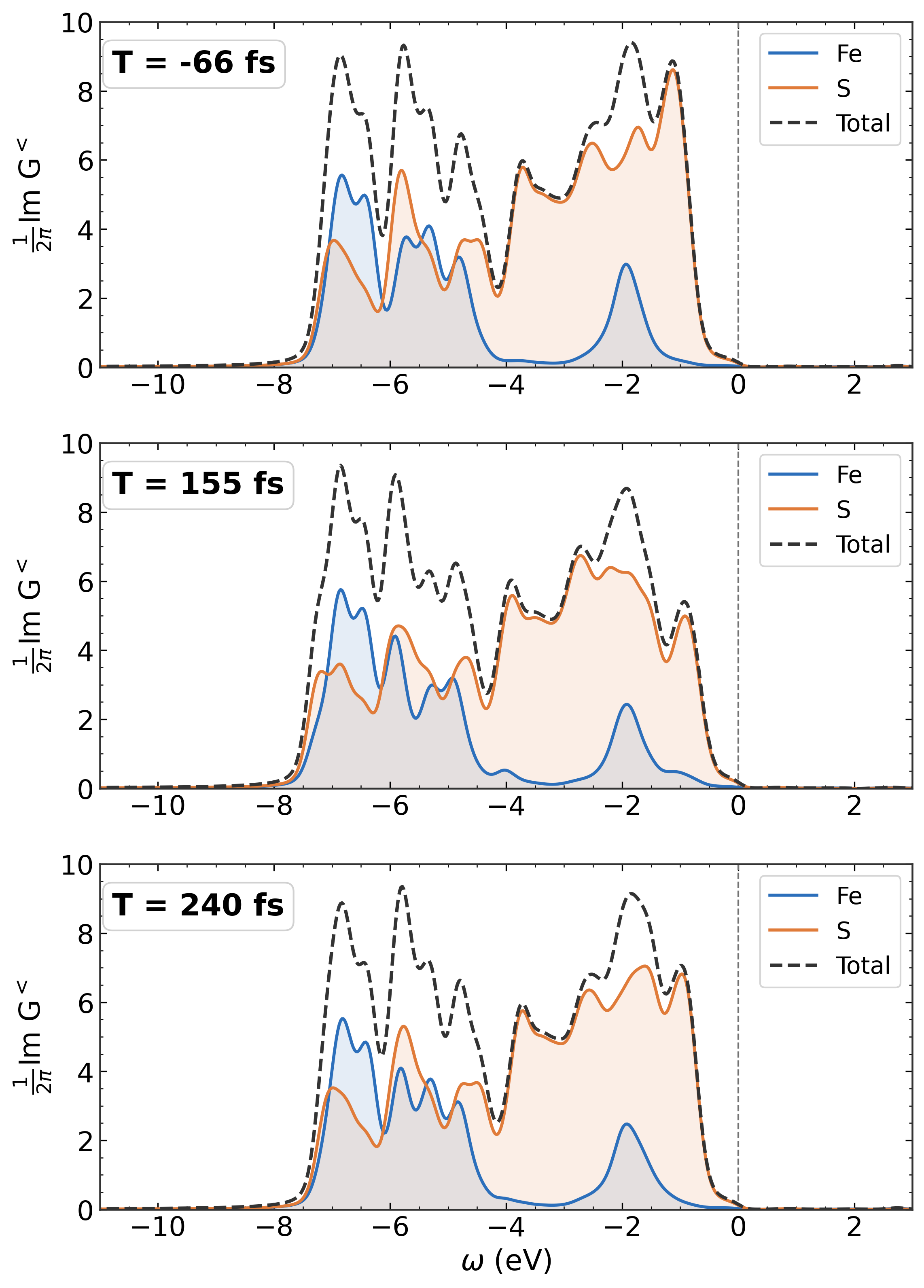}
    \caption{Imaginary part of the lesser Green's function in the paramagnetic phase for the three different time steps $T=-66$~fs, $T=155$~fs and $T=240$~fs in the evolution of the density matrix trough the first excited atomic configuration ($\simeq0.7$~eV). Color scheme: Blue Fe $d$ orbitals projection; Orange: S $p$ orbitals projection. The dotted line corresponds to the total of the two projections.}
    \label{fig:glesser_evol}
\end{figure}

Using the local atomic self-energies as input, we derive the lattice lesser \acp{GF} at the three characteristic times: prior, at and after the laser pump pulse in the decoherence regime. The results of the imaginary part of the local lesser \ac{GF} for Fe (d-subsystem) S (p-subsystem) are shown in Fig.~\ref{fig:glesser_evol} for the three Wigner times $T=-66,155,240$fs. Clearly visible is an additional small peak located around $\omega\approx -1$eV induced by the light pulse, and the depletion of the equilibrium peak at $\omega\approx -2$eV. This can be understood already from the atomic dynamics depicted in Fig.\ \ref{fig:time_evolution_atomic}. Simultaneously, the spectral distribution of the  S (p-subsystem) is changing as function of time.

Negative values of the diagonal \acp{GF} can appear \cite{FreericksManyBody2019} when considering $T$ slices in the transient of the pulse. In particular, during the pulse, the lesser Green's function and the spectral function show negative values at both $\omega\simeq2.0$ and $\omega\simeq-1.6(7)$~eV, i.e. the equilibrium positions of the top valence and bottom conduction $d$ states.

These values are of the order of $20\%$ of the corresponding average value over the surrounding frequency interval $\pm 0.2$~eV. The fraction of affected k-points, at the above frequencies ($\omega\simeq2.0$ and $\omega\simeq-1.6(7)$~eV), are at most $\sim 1\%$ of the total k-points for the lesser Green's function, and instead of the order of $\sim 60\%$ of the total for the spectral function. Taking into account that the two quantities are evaluated on different grids (a $40\times40\times40$ k-point-mesh with $\Delta\omega = 0.1$~eV for the lesser Green's function (having averaged over few frequencies), a $40\times40\times1$ k-point-mesh with $\Delta\omega = 0.03$~eV for the spectral function), which affects both the resolution of narrow features and the number of frequencies entering the local average.

Note that the Wigner transforms at the average time $T$ is not observable, and a finite probe pulse width has to be taken into account \cite{FreericksPRL2009}. Once the Wigner transforms are convoluted with Gaussian pulse width as stated in Eq.~\eqref{eq:convoluted}, the negativity is completely cured ($\sigma_T\simeq 5$~fs and $\sigma_{\omega}\simeq 0.07$~eV, $dT\simeq 2$~fs, $\Delta T\simeq\pm17$~fs, $\Delta\omega\simeq\pm0.43$~eV and $d\omega\simeq0.02$~eV). The regularization of the lesser self-energy and the artificial $\eta$ broadening have probably some role in regularizing the Wigner transforms.

We checked the total occupation of the unit cell as well as the occupation of each subsystem separately. While the atomic dynamics is conserving by construction, this is not obvious when using the atomic self-energy as correlation term in the Keldysh equation of motion, since our approximation is not $\Phi[G]$ derivable. The total filling as the charge redistribution between the two subsystems are preserved as a function of time in the order of 1\% (numerical discretisation errors), and the spectral weight is always conserved.

\subsubsection{trARPES k-maps for FePS$_3$}

\begin{figure}[htb!]
    \centering
    \hspace*{0cm}\includegraphics[width=0.2\textwidth]{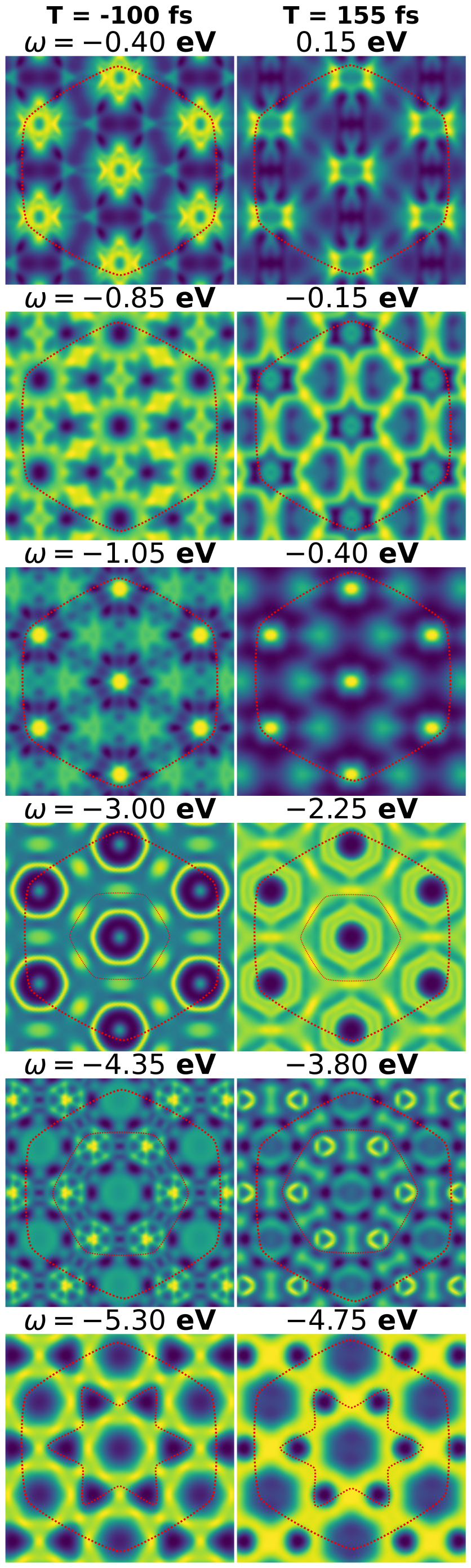}
    \caption{K-maps obtained from a \ac{DFT} + Hubbard I calculation in the paramagnetic phase, taking into account the IAP-ESD method for the first excited atomic configuration ($\simeq0.7$~eV). Energy slices are evaluated with respect to the top valence ($\simeq-1.0$~eV in Fig.~\ref{fig:glesser_evol}). Color scheme and legend as the one reported in the preceding Fig.\ \ref{fig:kmaps1}.}
    \label{fig:kmaps_evolution}
\end{figure}

We report the time evolution for the different k-maps at some relevant energies
in Fig~\ref{fig:kmaps_evolution}. It seems that the time evolved k-maps, i.e., the partial filling of the conduction bands, reproduce a mirroring of the valence bands as observed in experimental data \cite{NITSCHKE2025100019} more effectively than a naive population inversion scheme presented in Sec.\ \ref{sec:simple-pop-inversion}. So we can conclude that a realistic dynamic simulation of the data is significantly better in reproducing the experimental findings than a projection method approach.

\subsubsection{Summary of the section}

In this section we presented the calculation of trARPES k-maps, considering the entire local evolution as a response to an external laser pulse. Although we only used the simplifying Hubbard I assumption to extract an estimate for the local self-energy, the results from this time evolution approach seem more promising than the projection method approach.

\section{Conclusions}
We have introduced an instantaneous approximation for the non-equilibrium dynamics of correlated multiorbital systems. The approach is motivated by a separation of time scales in which the Green's function decays rapidly with respect to the relative time $\tau$, while the pump-induced state evolves more slowly with respect to the average time $T$. Under this assumption, the non-equilibrium problem can be expressed in terms of an instantaneous, frequency-dependent self-energy $\Sigma(T,\omega,\vec{k})$. By combining this approximation with a one-shot DMFT construction at the Hubbard I level, we incorporate the explicitly calculated dynamics of the local correlated shell into the momentum-dependent lattice Green's function without solving a fully self-consistent non-equilibrium impurity problem.

We have also derived an effective local coupling for nominally dipole-forbidden $d-d$ transitions. The underlying microscopic process consists of a dipole-allowed transition between a transition-metal $d$ orbital and a ligand $p$ orbital, followed by $p-d$ hopping that restores the charge of the correlated shell while leaving it in an excited multiplet state. When this ligand-mediated process is fast compared with the local dynamics, it can be represented by an effective instantaneous transition matrix element acting within the $d$ shell. Beyond this limit, the same process gives rise to an energy-dependent vertex correction to the optical response. This construction provides a microscopic justification for describing the pump-induced dynamics within the correlated $d$ subspace.
As a proof of principle, we applied the method to paramagnetic and antiferromagnetic FePS$_3$. Starting from a \ac{DFT} derived tight-binding model, we treated the local Fe 3$d$ correlations within the Hubbard I approximation and reconstructed the momentum-resolved spectra through the lattice Dyson equation. For the antiferromagnetic phase, a supercell unfolding procedure was used to represent the calculated spectra in the Brillouin zone of the primitive cell, thereby facilitating comparison with experimental momentum maps.

The calculations reproduce the main qualitative features observed in recent trARPES experiments on FePS$_3$. Following excitation of the first $d-d$ transition, the calculated momentum maps remain similar to their equilibrium counterparts. Excitation of the second $d-d$ transition instead produces substantially stronger modifications of the momentum-dependent spectral weight. These results support an interpretation in which the experimentally observed transient electronic structure is governed, to a significant extent, by the population of different local Fe multiplet states and their embedding into the dispersive valence-band structure.

The present treatment is intentionally approximate. It neglects the self-consistent feedback of the lattice on the local non-equilibrium dynamics and does not include surface contributions. Consequently, a one-to-one quantitative correspondence with the measured momentum maps is not expected. Nevertheless, the ability of the approach to capture their principal excitation-dependent features indicates that the instantaneous approximation provides a useful and computationally economical framework for interpreting pump-driven local excitations in correlated materials. Its extension to self-consistent impurity schemes offers a route toward a more quantitative description of time-resolved photoemission experiments.

\begin{acknowledgments}
MM gratefully acknowledges fruitful discussions with Anna Sidochenko. MC and LS acknowledge financial support from the Deutsche
Forschungsgemeinschaft (DFG) through project dd2D (CI 157/11-1, Project number: 555818086). FBA acknowledge financial support  financial support by the DFG via Grant No. AN 275/10-1.
\end{acknowledgments}

\appendix
\section{\label{subsec:wann}Wannierisation Procedure}

\begin{figure}[b]
    \centering
    \hspace*{0cm}\includegraphics[width=0.5\textwidth]{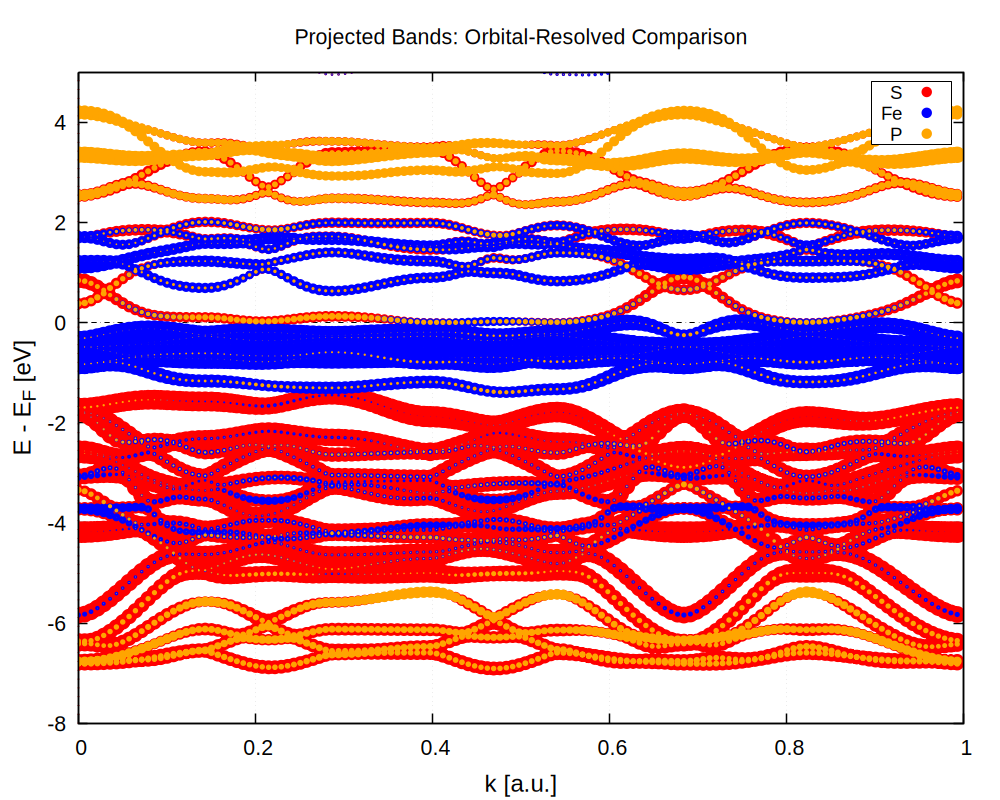}
    \caption{Band structure of the \ac{DFT} calculation for paramagnetic FePS$_3$ projected on the different atomic contributions: Red: S atoms; Blue: Fe atoms; and Orange: P atoms.}
    \label{fig:projbands}
\end{figure}

The band structure of the \ac{DFT} calculation projected on the different atomic contributions is reported in Fig.~\ref{fig:projbands}. The metallic $d$ bands at the Fermi energy plotted in blue are clearly visible: Charge neutrality and an independent electron picture only leaves the \ac{LDA} (\ac{DFT}) the option to position the only partially filled $d$ bands around the chemical potential. The goal is to generate a Mott-Hubbard/charge transfer insulator by including correlation effects which typically leads to fractional spectral weight of bands in the case of a Mott insulator. While for a strongly correlated metal, a full \ac{DMFT} calculation is needed, already a Hubbard I approximation leads to reasonably accurate description of a Mott insulator away from the critical coupling $U_c$ at which the metal-insulator phase transition occurs. 

The challenge to obtain an insulator using  a tight-binding model extracted from the \ac{DFT} calculation
in a single cycle \ac{DMFT} calculation which is equivalent to the Hubbard I approximation is connected to the $p$ bands close the Fermi energy which hybridize the Fe $d$ orbitals. In a full electronic structure calculation using a combination of \ac{DFT} and \ac{DMFT} \cite{RevModPhys.78.865} the Kohn-Sham equations of the conventional \ac{LDA} are modified by a feedback loop from the \ac{DMFT}. In FePS$_3$ we expect that
the first \ac{DMFT} step generates corrected Fe $d$ bands and pushes the Fe $d$ orbital energy to much lower energies, away from the Fermi energy. This energy shift would separate the S $p$ bands and the Fe $d$ bands avoiding binding and anti-binding $d-p$ states inside the conduction-valence gap. A shift of these bands at higher energies at the tight-binding level is not sufficient, being already hybridized at the Wannier level.

\begin{figure}[tb!]
    \centering
    \hspace*{0cm}\includegraphics[width=0.48\textwidth]{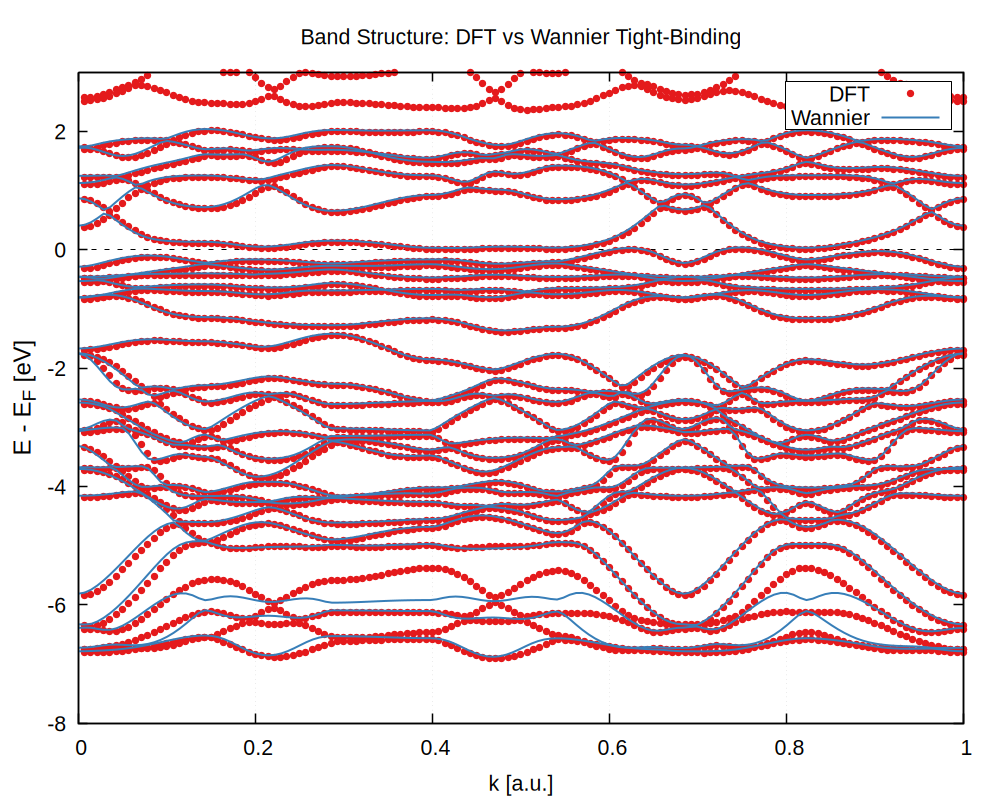}
    \caption{Band structures respectively of the \ac{DFT} calculation (Red dots) and of the Wannier tight-binding calculation (Blue lines)
    for for paramagnetic FePS$_3$.}
    \label{fig:bands_wannier}
\end{figure}

The band structure obtained from the Wannier tight-binding model is compared to the original \ac{DFT} (\ac{LDA}) dispersion in Fig.~\ref{fig:bands_wannier}.
%, in comparison with the \ac{DFT} (LDA) one.
In the construction of our tight-binding model we consider as initial projectors only Fe $d$ orbitals and S $p$ orbitals. 
The tight-binding band structure is in very good agreement with the \ac{DFT} results. Only the $p$ bands located about $\approx 6$~eV below the Fermi energy are not reproduced exactly. These bands are strongly influenced by the neglected P $p$ orbitals. 

Enlarging the model, including P $p$ orbitals, gives Wannier functions with stronger localization and an excellent agreement with the \ac{LDA} band structure, but as underlined in the main text and above does not solve the intrinsic problem of the significantly overestimated $d-p$ hopping parameters. Although the point group symmetry for each atom in the unit cell is preserved at the \ac{DFT} level, is not completely preserved in the calculated k-maps. This effect is traced back to the construction of the tight-binding model, where no symmetry constrains are applied when localizing the different Wannier orbitals.

\begin{figure}[t] 
    \centering
    \includegraphics[width=0.4\textwidth]{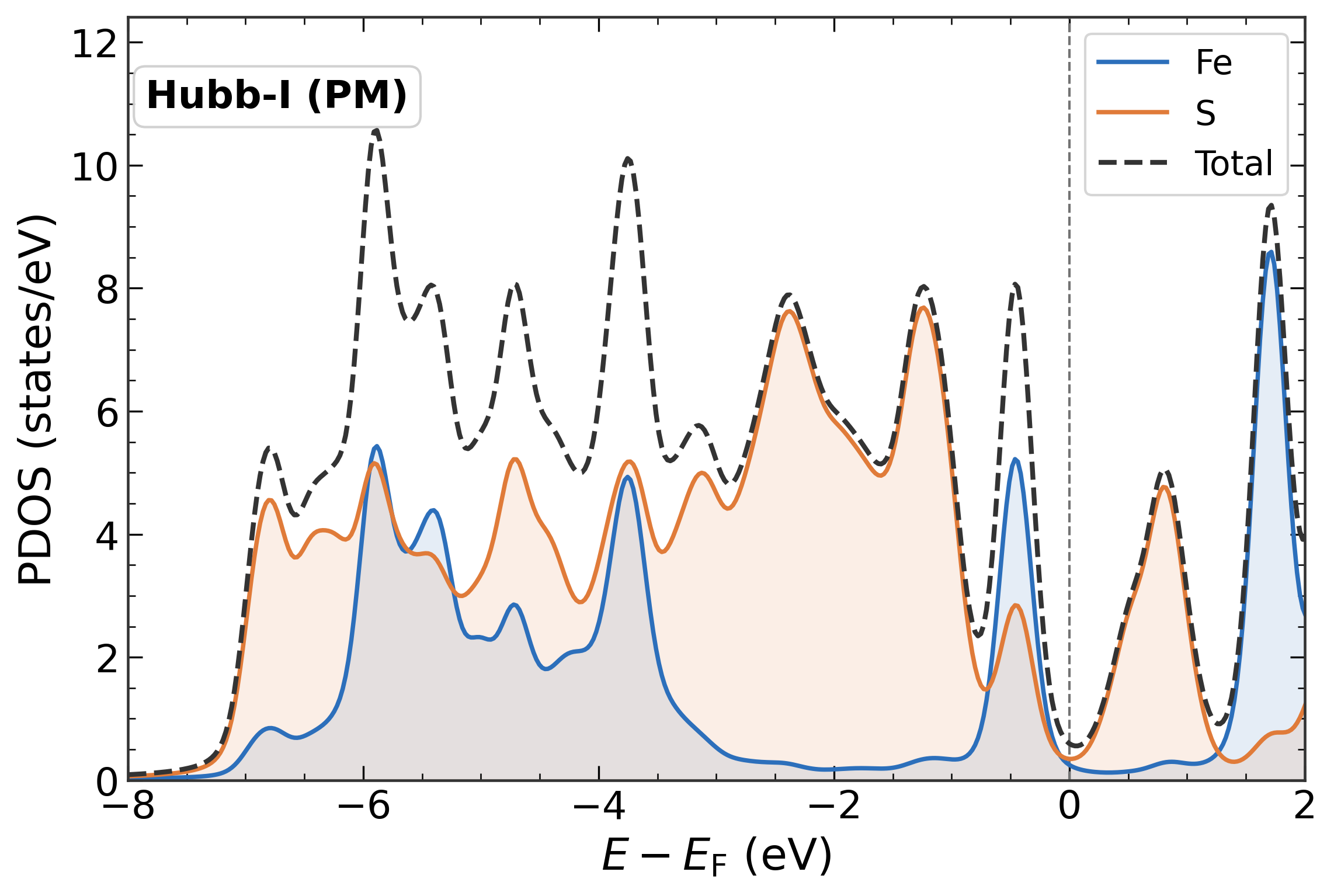}
    \caption{\ac{PDOS} obtained from Hubbard I + $\Lambda$ in the paramagnetic phase. Color scheme: Blue Fe $d$ orbitals projection; Orange: S $p$ orbitals projection. The dotted line corresponds to the total of the two projections.}
    \label{fig:ws_pdos}
\end{figure}

\begin{figure}[htb!]
%[hb!]
    \centering
    %\hspace*{0cm}
    \includegraphics[width=0.4\textwidth]{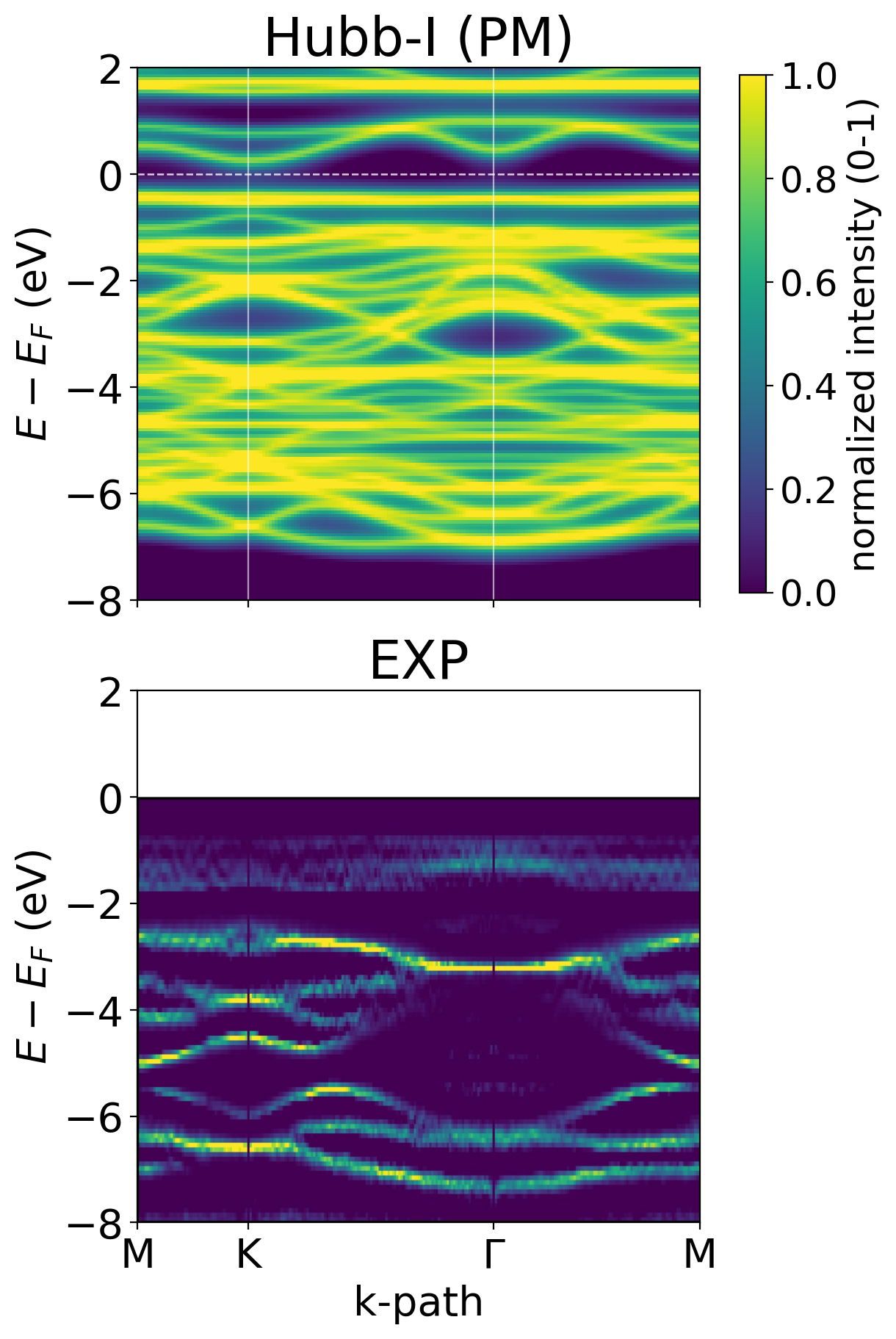}
    \caption{Upper panel: The equilibrium paramagnetic spectral function $\rho(\omega,\vec{k})$ as a color coded
    map in $\omega$-k plane obtained from \ac{DFT} + Hubbard I + $\Lambda$ calculations. Lower panel: 
    The experimental results obtained from ARPES measurements (curvature-ARPES \cite{10.1063/1.3585113}) of the paramagnetic phase as already shown in Fig.\ \ref{fig:bands} for comparison.}
    \label{fig:ws_bands}
\end{figure}
\section{Hubbard I + $\Lambda$}
\label{subsec:lambda}
\begin{figure}[htb!]
    \centering
\includegraphics[height=0.7\textheight]{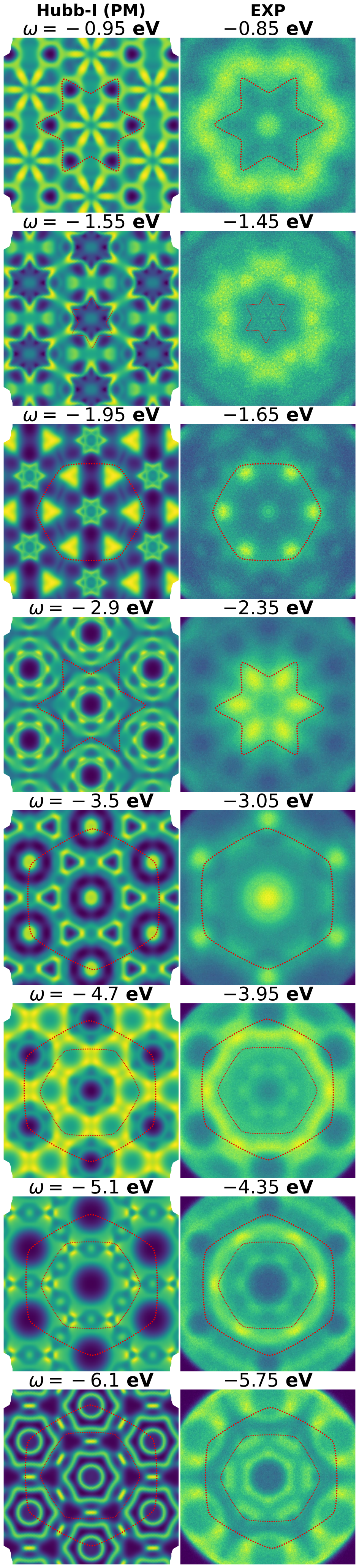}
    \caption{K-maps (1st and 2nd Brillouin zone) obtained from \ac{DFT} + Hubbard I + $\Lambda$ calculations, in the paramagnetic phase (right), and from experiments (left). The color scale is from 0 (dark blue) to 1 (bright yellow) (in arbitrary units), and is related to the degree of electronic occupation in the Brillouin zone: 0 empty and 1 filled. A logarithmic color scale is considered. The $\omega$ is the energy cut evaluated with respect to the top valence band (an $\Delta E=0.1$~eV energy interval has been considered around each cut and averaged). An integration over $k_z$ has been performed. The dotted line serves as a guide to the eye.}
    \label{fig:ws_kmaps}
\end{figure}

\begin{figure}[htb!]
   \centering
\includegraphics[height=0.6\textheight]{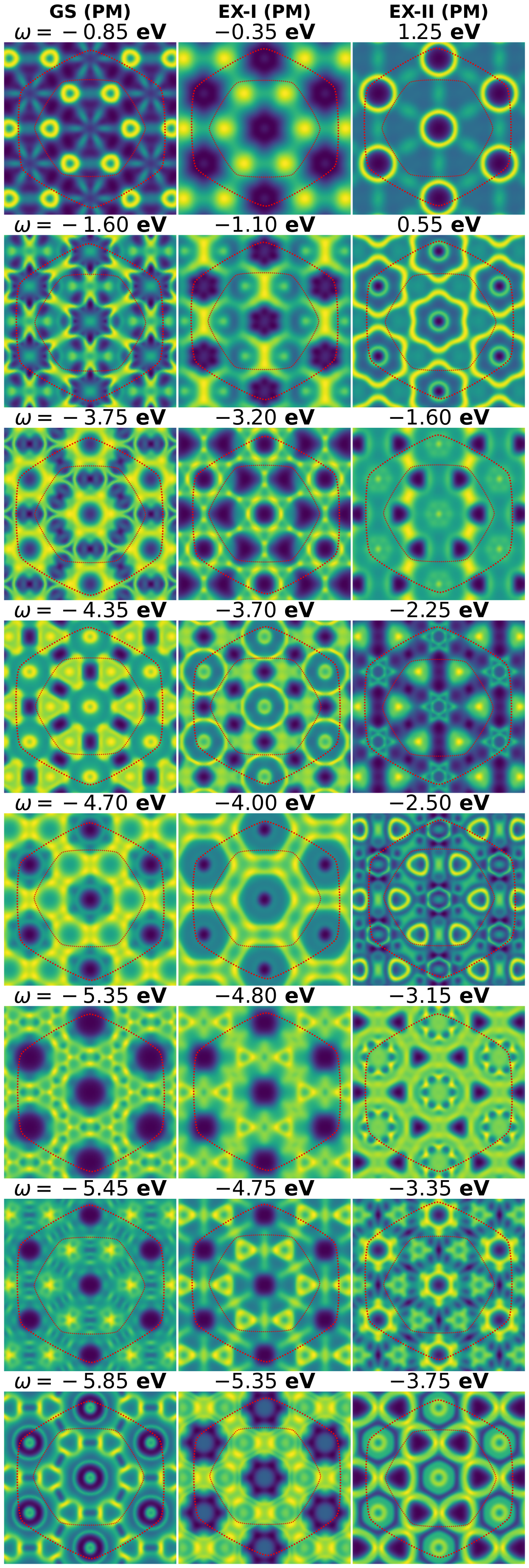}
    \caption{k-maps obtained from a \ac{DFT} + Hubbard I +$\Lambda$ calculation in the paramagnetic phase, taking into account the IAP-ESP method for the first (center) and second (right) atomic excitation, in comparison with the mirrored valence k-maps (left). Color scheme and legend as the one reported in Fig.\ref{fig:kmaps1}.}
    \label{fig:ws_exc1}
\end{figure}

In Appendix \ref{subsec:wann} and in the main text, we pointed out that in self-consistent many-body calculation, the Fe $d$ bands should be pushed down to lower energies and as a consequence, the $d-p$ hopping will be slightly reduced. 
From a many-body perspective, a Mott insulator can also be obtained by an increase in the effective $U$. 
As an alternative, we can maintain the previously used literature values $U=4.15$~eV and $J=0.65$~eV \cite{NITSCHKE2025100019} and artificially re-scale the overestimated $p-d$ hopping by a scaling factor $\Lambda=0.8$. 
Instead of the empirical energy difference between the $t_{2g}$ and $e_{g}$ orbitals of $\simeq1.1$~eV, we used also here the \ac{DFT} value of $\simeq0.7$~eV changing the local $d-d$ excitation energies to $\simeq0.7$~eV and $\simeq2.2$~eV. 
Note that we obtained analogue results considering the empirical energy difference, i.\ e.\ 
replacing the \ac{LDA} $t_{2g}-e_{g}$ splitting by $\Delta E_{et}\approx 1.1$~eV.
The resulting \ac{PDOS} of the paramagnetic phase is shown in Fig.~\ref{fig:ws_pdos}. The occupation of the $Fe$ $d$ orbitals is of $\simeq3.15$~$e^-$ for spin channel in the paramagnetic phase, which is close to the experimentally reported value of  $\approx 6.3$ $e^{-}$  \cite{AF-FePS3-2025}. The spectral function $\rho(\omega,\vec{k})$ corresponds to the renormalized many-body band structure and is plotted in the upper panel of Fig.~\ref{fig:ws_bands},
and we added  experimental results  extracted from an \ac{ARPES} measurement  from  Fig.\ \ref{fig:bands} for comparison as lower panel.
We used the equilibrium data shown in the Figs.\ \ref{fig:ws_pdos} and \ref{fig:bands} to generate the \ac{ARPES} k-maps in the Hubbard I + $\Lambda$ approximation. The resulting k-maps are presented in Fig.~\ref{fig:ws_kmaps} left column in the same style as in the main text. We augmented the figure with the experimental data from Fig.\ \ref{fig:kmaps1} of the main text as right column.
The calculated results are in very good agreement with the experimental \ac{ARPES}  k-maps. The spectral function, however, shows more deviations from the experimental \ac{ARPES} data as in the Hubbard I approximation presented in the main text. 
We performed the same calculation for the first an second excitation as outlined in the main text. In Fig.~\ref{fig:ws_exc1} the resulting k-maps are shown using the \ac{DFT} + Hubbard I +$\Lambda$ approximation.
\section{\label{subsec:ex_bands}Excited Band Structure}
\begin{figure}[htb!]
    \centering
    \hspace*{0cm}\includegraphics[width=0.30\textwidth]{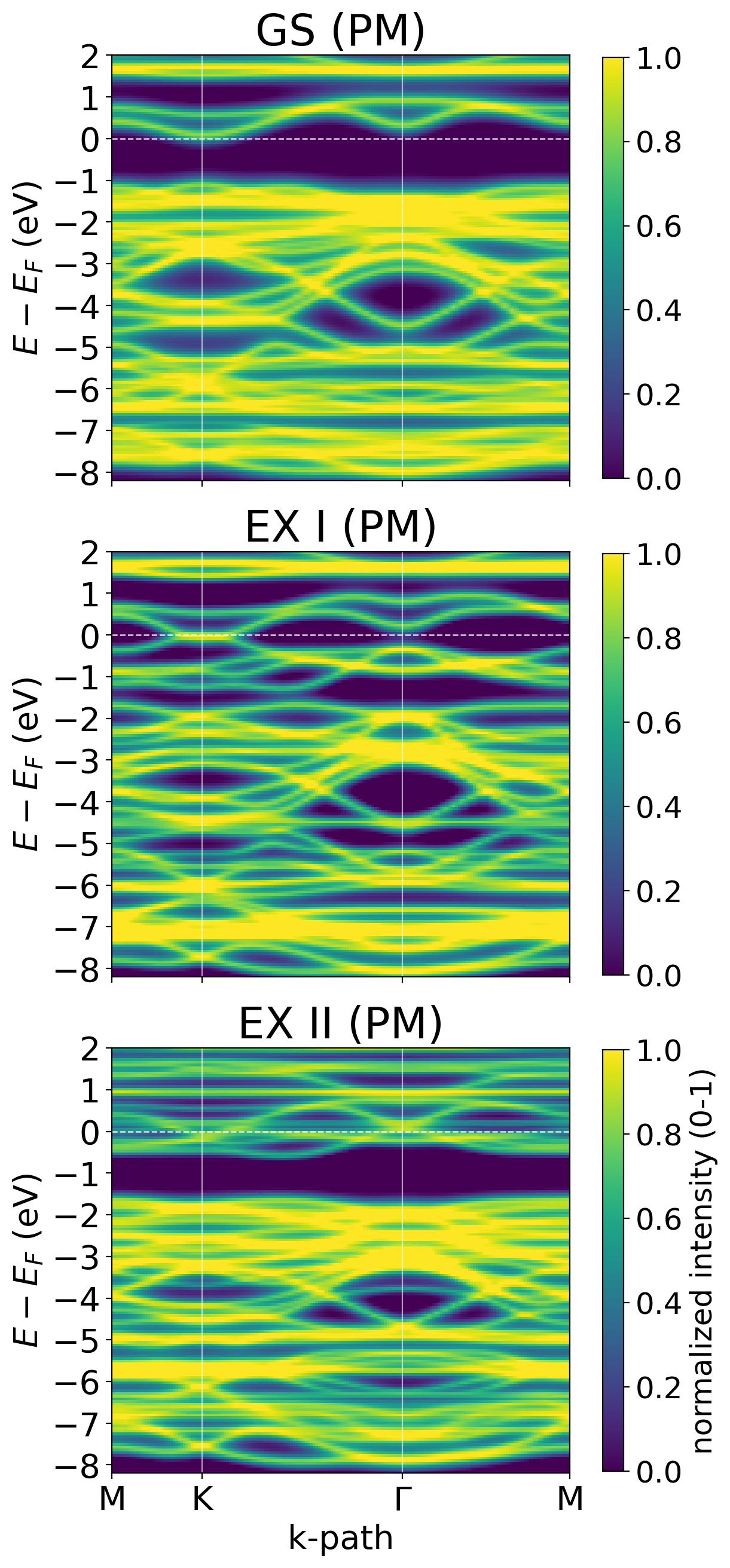}
    \caption{Band structure obtained from \ac{DFT} + Hubbard I calculations in the paramagnetic phase, taking into account the IAP-ESP method for the first (center) and second (bottom) atomic excitations, in comparison with the ground state result (top). To increase the contrast in the images, we select the data in the intensity window from $0.1$ to $0.9$}
    \label{fig:bands_ex}
\end{figure}
\begin{figure}[htb!]
    \centering
    \hspace*{0cm}\includegraphics[width=0.30\textwidth]{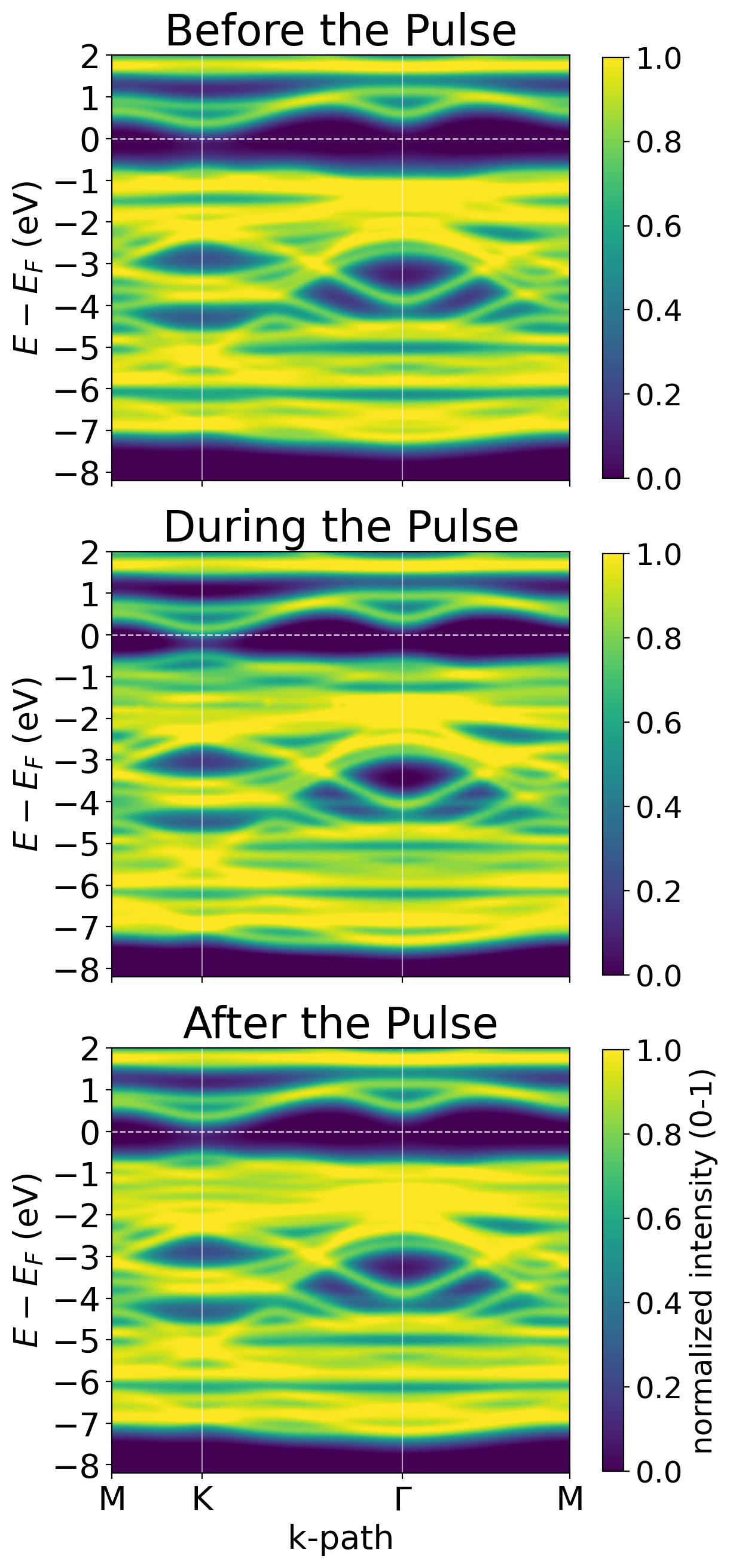}
    \caption{Band structure obtained from \ac{DFT} + Hubbard I calculations in the paramagnetic phase, taking into account the IAP approximation in the time evolution method for the first atomic excitation (IAP-ESD), considering three different times steps $T=-66$~fs, $T=155$~fs and $T=240$~fs, respectively before, during and after the pulse. To increase the contrast in the images, we select the data in the intensity window from $0.1$ to $0.9$.}
    \label{fig:bands_ex_T}
\end{figure}
In Fig.~\ref{fig:bands_ex} we report  the band structure obtained considering the out-of-equilibrium retarded Green's function from the IAP approximation in the excited state projection method. The excitation induces an insulator to metal transition (IMT), as the lesser Green's function is already underlining in Fig.~\ref{fig:glesser} in Sec.~\ref{sec:neq-dos}. In particular, it is clear as the excitation is not merely inducing a shift of the chemical potential, but rather modifying the band structure and redistributing the spectral weight. 
In Fig.~\ref{fig:bands_ex_T} we plot the band structure obtained considering the out-of-equilibrium retarded Green's function from the IAP approximation in the time evolution method for the first excitation. In this case, we can follow directly the reduction of the valence-conduction band gap with the incoming pulse.
\section{Effective $d-d$ Transition Matrix Elements}
We explicitly state the effective $d-d$ transition matrix elements obtained from the Eq.~\eqref{eq:effective_dipole} at the first excitation energy $\omega_0\simeq0.7$~eV 
in the Tables~\ref{tab:dipole_fe_x},\ref{tab:dipole_fe_y} and \ref{tab:dipole_fe_z}.
\begin{table}[tbh!]
    \begin{minipage}{1.0\linewidth}
        \centering
        \resizebox{\linewidth}{!}{%
        \begin{tabular}{c|ccccc}
                            & $d_{xy}$ & $d_{xz}$ & $d_{yz}$ & $d_{z^2}$ & $d_{x^2-y^2}$ \\
            \hline
            $d_{xy}$        & $-0.002+0.000i$ & $0.001-0.002i$ & $-0.042+0.184i$ & $-0.024-0.115i$ & $0.003-0.003i$ \\
            $d_{xz}$        & $-0.003+0.001i$ & $-0.002+0.000i$ & $0.069+0.220i$ & $0.204-0.209i$ & $-0.006-0.001i$ \\
            $d_{yz}$        & $0.042-0.185i$ & $-0.069-0.210i$ & $-0.002+0.001i$ & $-0.005+0.004i$ & $-0.126+0.047i$ \\
            $d_{z^2}$       & $0.020+0.116i$ & $-0.217+0.214i$ & $0.007-0.006i$ & $0.000+0.000i$ & $0.051-0.618i$ \\
            $d_{x^2-y^2}$   & $-0.008+0.002i$ & $0.002-0.002i$ & $0.133-0.052i$ & $-0.068+0.594i$ & $-0.002+0.002i$
        \end{tabular}
        }
        \caption{$x$ component of the effective $d-d$ transition matrix elements  $d^x_{\alpha\beta}(\omega_0)$ (e\AA/fs), averaged over $\Delta=0.05$~eV and averaged over the two Fe sites.}
        \label{tab:dipole_fe_x}
    \end{minipage}
\end{table}
\begin{table}[htb!]
          \begin{minipage}{1.0\linewidth}
        \centering
        \resizebox{\linewidth}{!}{%
        \begin{tabular}{c|ccccc}
                            & $d_{xy}$ & $d_{xz}$ & $d_{yz}$ & $d_{z^2}$ & $d_{x^2-y^2}$ \\
            \hline
            $d_{xy}$        & $0.005-0.003i$ & $0.000-0.013i$ & $-0.047+0.210i$ & $0.003-0.069i$ & $-0.007+0.005i$ \\
            $d_{xz}$        & $-0.001+0.019i$ & $0.002+0.001i$ & $-0.079+0.370i$ & $0.035+0.300i$ & $0.008+0.006i$ \\
            $d_{yz}$        & $0.046-0.213i$ & $0.074-0.365i$ & $0.003-0.002i$ & $-0.002+0.000i$ & $-0.047-0.147i$ \\
            $d_{z^2}$       & $-0.008+0.073i$ & $-0.036-0.305i$ & $-0.004+0.000i$ & $0.000+0.007i$ & $0.007+0.086i$ \\
            $d_{x^2-y^2}$   & $0.010-0.011i$ & $-0.010+0.004i$ & $0.045+0.149i$ & $-0.006-0.086i$ & $0.009-0.007i$
        \end{tabular}
         }
        \caption{$y$ component of the effective $d-d$ transition matrix elements  $d^y_{\alpha\beta}(\omega_0)$ (e\AA/fs), averaged over $\Delta=0.05$~eV and averaged over the two Fe sites.}
        \label{tab:dipole_fe_y}
    \end{minipage}
\end{table}
\begin{table}[htb!]
    \begin{minipage}{1.0\linewidth}
        \centering
        \resizebox{\linewidth}{!}{%
        \begin{tabular}{c|ccccc}
                            & $d_{xy}$ & $d_{xz}$ & $d_{yz}$ & $d_{z^2}$ & $d_{x^2-y^2}$ \\
            \hline
            $d_{xy}$        & $0.000+0.000i$ & $-0.002+0.000i$ & $-0.013-0.033i$ & $-0.009-0.008i$ & $-0.003+0.004i$ \\
            $d_{xz}$        & $0.000+0.000i$ & $0.000-0.001i$ & $-0.007+0.021i$ & $0.015-0.147i$ & $0.001+0.006i$ \\
            $d_{yz}$        & $0.014+0.033i$ & $0.007-0.021i$ & $0.000+0.000i$ & $0.005-0.003i$ & $0.003-0.004i$ \\
            $d_{z^2}$       & $0.009+0.008i$ & $-0.014+0.147i$ & $-0.005+0.002i$ & $-0.001+0.001i$ & $0.003+0.013i$ \\
            $d_{x^2-y^2}$   & $0.002-0.005i$ & $-0.001-0.005i$ & $-0.003+0.003i$ & $-0.002-0.015i$ & $0.000+0.001i$
        \end{tabular}
        }
        \caption{$z$ component of the effective $d-d$ transition matrix elements  $d^z_{\alpha\beta}(\omega_0)$ (e\AA/fs), averaged over $\Delta=0.05$~eV and averaged over the two Fe sites.}
        \label{tab:dipole_fe_z}
     \end{minipage}
\end{table}
\pagebreak

\bibliography{bibliography}% Produces the bibliography via BibTeX.

@article{Peierls:1933ec,
	author = {Peierls, R.},
	date = {1933/11/01},
	doi = {10.1007/BF01342591},
	id = {Peierls1933},
	isbn = {0044-3328},
	journal = {Zeitschrift f{\"u}r Physik},
	number = {11},
	pages = {763--791},
	title = {{Zur Theorie des Diamagnetismus von Leitungselektronen}},
	url = {https://doi.org/10.1007/BF01342591},
	volume = {80},
	year = {1933}}

@article{PhysRevB.101.205140,
	author = {Li, Jiajun and Golez, Denis and Mazza, Giacomo and Millis, Andrew J. and Georges, Antoine and Eckstein, Martin},
	doi = {10.1103/PhysRevB.101.205140},
	issue = {20},
	journal = {Phys. Rev. B},
	month = {May},
	numpages = {16},
	pages = {205140},
	publisher = {American Physical Society},
	title = {Electromagnetic coupling in tight-binding models for strongly correlated light and matter},
	url = {https://link.aps.org/doi/10.1103/PhysRevB.101.205140},
	volume = {101},
	year = {2020}}

@article{LuttingerPeierls1951,
	author = {Luttinger, J. M.},
	doi = {10.1103/PhysRev.84.814},
	issue = {4},
	journal = {Phys. Rev.},
	month = {Nov},
	numpages = {0},
	pages = {814--817},
	publisher = {American Physical Society},
	title = {The Effect of a Magnetic Field on Electrons in a Periodic Potential},
	url = {https://link.aps.org/doi/10.1103/PhysRev.84.814},
	volume = {84},
	year = {1951}}

@article{TomczakEffPeierls2009,
	author = {Tomczak, Jan M. and Biermann, Silke},
	doi = {10.1103/PhysRevB.80.085117},
	issue = {8},
	journal = {Phys. Rev. B},
	month = {Aug},
	numpages = {13},
	pages = {085117},
	publisher = {American Physical Society},
	title = {{}Optical properties of correlated materials: Generalized Peierls approach and its application to ${\text{VO}}_{2}$}}

@article{He:2024yb,
	author = {He, W. and Shen, Y. and Wohlfeld, K. and Sears, J. and Li, J. and Pelliciari, J. and Walicki, M. and Johnston, S. and Baldini, E. and Bisogni, V. and Mitrano, M. and Dean, M. P. M.},
	date = {2024/04/25},
	doi = {10.1038/s41467-024-47852-x},
	id = {He2024},
	isbn = {2041-1723},
	journal = {Nature Communications},
	number = {1},
	pages = {3496},
	title = {{Magnetically propagating Hund's exciton in van der Waals antiferromagnet NiPS3}},
	url = {https://doi.org/10.1038/s41467-024-47852-x},
	volume = {15},
	year = {2024}}

@article{PhysRevB.106.205117,
  title = {{Simplified approach to the magnetic blue shift of Mott gaps}},
  author = {Hafez-Torbati, Mohsen and Anders, Frithjof B. and Uhrig, G\"otz S.},
  journal = {Phys. Rev. B},
  volume = {106},
  issue = {20},
  pages = {205117},
  numpages = {16},
  year = {2022},
  month = {Nov},
  publisher = {American Physical Society},
  doi = {10.1103/PhysRevB.106.205117},
  url = {https://link.aps.org/doi/10.1103/PhysRevB.106.205117}
}

@article{Hafez-Torbati-MBS-2021,
	author = {Hafez-Torbati, Mohsen and Bossini, Davide and Anders, Frithjof B. and Uhrig, G\"otz S.},
	doi = {10.1103/PhysRevResearch.3.043232},
	issue = {4},
	journal = {Phys. Rev. Res.},
	month = {Dec},
	numpages = {13},
	pages = {043232},
	publisher = {American Physical Society},
	title = {{Magnetic blue shift of Mott gaps enhanced by double exchange}},
	url = {https://link.aps.org/doi/10.1103/PhysRevResearch.3.043232},
	volume = {3},
	year = {2021}}

@article{Bulla1999,
	author = {Bulla, R.},
	doi = {10.1103/PhysRevLett.83.136},
	issue = {1},
	journal = {Phys. Rev. Lett.},
	month = {Jul},
	numpages = {0},
	pages = {136--139},
	publisher = {American Physical Society},
	title = {{Zero Temperature Metal-Insulator Transition in the Infinite-Dimensional Hubbard Model}},
	url = {https://link.aps.org/doi/10.1103/PhysRevLett.83.136},
	volume = {83},
	year = {1999}}

@article{AF-FePS3-2025,
	author = {Wei, Yuan and Tseng, Yi and Elnaggar, Hebatalla and Zhang, Wenliang and Asmara, Teguh Citra and Paris, Eugenio and Domaine, Gabriele and Strocov, Vladimir N. and Testa, Luc and Favre, Virgile and Di Luca, Mario and Banerjee, Mitali and Wildes, Andrew R. and de Groot, Frank M. F. and R{\o}nnow, Henrik M. and Schmitt, Thorsten},
	date = {2025/06/17},
	doi = {10.1038/s41535-025-00777-0},
	id = {Wei2025},
	isbn = {2397-4648},
	journal = {npj Quantum Materials},
	number = {1},
	pages = {61},
	title = {{Spin-orbital excitations encoding the magnetic phase transition in the van der Waals antiferromagnet FePS3}},
	url = {https://doi.org/10.1038/s41535-025-00777-0},
	volume = {10},
	year = {2025}}

@article{RevModPhys.68.13,
  title = {Dynamical mean-field theory of strongly correlated fermion systems and the limit of infinite dimensions},
  author = {Georges, Antoine and Kotliar, Gabriel and Krauth, Werner and Rozenberg, Marcelo J.},
  journal = {Rev. Mod. Phys.},
  volume = {68},
  issue = {1},
  pages = {13--125},
  numpages = {0},
  year = {1996},
  month = {Jan},
  publisher = {American Physical Society},
  doi = {10.1103/RevModPhys.68.13},
  furl = {https://link.aps.org/doi/10.1103/RevModPhys.68.13}
}

@article{RevModPhys.78.865,
  title = {Electronic structure calculations with dynamical mean-field theory},
  author = {Kotliar, G. and Savrasov, S. Y. and Haule, K. and Oudovenko, V. S. and Parcollet, O. and Marianetti, C. A.},
  journal = {Rev. Mod. Phys.},
  volume = {78},
  issue = {3},
  pages = {865--951},
  numpages = {0},
  year = {2006},
  month = {Aug},
  publisher = {American Physical Society},
  doi = {10.1103/RevModPhys.78.865},
  furl = {https://link.aps.org/doi/10.1103/RevModPhys.78.865}
}

@article{castellanos2022van,
  title={{Van der Waals heterostructures}},
  author={Castellanos-Gomez, Andres and Duan, Xiangfeng and Fei, Zhe and Gutierrez, Humberto Rodriguez and Huang, Yuan and Huang, Xinyu and Quereda, Jorge and Qian, Qi and Sutter, Eli and Sutter, Peter},
  journal={Nature Reviews Methods Primers},
  volume={2},
  number={1},
  pages={58},
  year={2022},
  publisher={Nature Publishing Group UK London}
}

@article{RevModPhys.96.015003,
  title = {{Time-resolved ARPES studies of quantum materials}},
  author = {Boschini, Fabio and Zonno, Marta and Damascelli, Andrea},
  journal = {Rev. Mod. Phys.},
  volume = {96},
  issue = {1},
  pages = {015003},
  numpages = {56},
  year = {2024},
  month = {Feb},
  publisher = {American Physical Society},
  doi = {10.1103/RevModPhys.96.015003},
  furl = {https://link.aps.org/doi/10.1103/RevModPhys.96.015003}
}

@book{stefanucci2013nonequilibrium,
  title={Nonequilibrium many-body theory of quantum systems: a modern introduction},
  author={Stefanucci, Gianluca and Van Leeuwen, Robert},
  year={2013},
  publisher={Cambridge University Press}
}

@article{nitschke2023valence,
  title={{Valence band electronic structure of the van der Waals antiferromagnet FePS3}},
  author={Nitschke, Jonah Elias and Esteras, Dorye L and Gutnikov, Michael and Schiller, Karl and Ma{\~n}as-Valero, Samuel and Coronado, Eugenio and Stupar, Matija and Zamborlini, Giovanni and Ponzoni, Stefano and Baldov{\'\i}, Jos{\'e} J and others},
  journal={Materials Today Electronics},
  volume={6},
  pages={100061},
  year={2023},
  publisher={Elsevier}
}

@article{NITSCHKE2025100019,
title = {{Tracing the ultrafast buildup and decay of d-d transitions in FePS3}},
journal = {Newton},
volume = {1},
number = {2},
pages = {100019},
year = {2025},
issn = {2950-6360},
doi = {https://doi.org/10.1016/j.newton.2025.100019},
furl = {https://www.sciencedirect.com/science/article/pii/S2950636025000118},
author = {Jonah Elias Nitschke and Lasse Sternemann and Michael Gutnikov and Karl Schiller and Eugenio Coronado and Alan Omar and Giovanni Zamborlini and Clara Saraceno and Matija Stupar and Alberto M. Ruiz and Dorye L. Esteras and José J. Baldoví and Frithjof Anders and Mirko Cinchetti}
}

@article{giannozzi2017advanced,
  title={{Advanced capabilities for materials modelling with Quantum ESPRESSO}},
  author={Giannozzi, Paolo and Andreussi, Oliviero and Brumme, Thomas and Bunau, Oana and Buongiorno Nardelli, M and Calandra, Matteo and Car, Roberto and Cavazzoni, Carlo and Ceresoli, Davide and Cococcioni, Matteo and others},
  journal={Journal of physics: Condensed matter},
  volume={29},
  number={46},
  pages={465901},
  year={2017},
  publisher={IOP Publishing}
}

@article{giannozzi2009quantum,
  title={{QUANTUM ESPRESSO: a modular and open-source software project for quantum simulations of materials}},
  author={Giannozzi, Paolo and Baroni, Stefano and Bonini, Nicola and Calandra, Matteo and Car, Roberto and Cavazzoni, Carlo and Ceresoli, Davide and Chiarotti, Guido L and Cococcioni, Matteo and Dabo, Ismaila and others},
  journal={Journal of physics: Condensed matter},
  volume={21},
  number={39},
  pages={395502},
  year={2009}
}

@article{PhysRevLett.115.136402,
  title = {Spin Signature of Nonlocal Correlation Binding in Metal-Organic Frameworks},
  author = {Thonhauser, T. and Zuluaga, S. and Arter, C. A. and Berland, K. and Schr\"oder, E. and Hyldgaard, P.},
  journal = {Phys. Rev. Lett.},
  volume = {115},
  issue = {13},
  pages = {136402},
  numpages = {6},
  year = {2015},
  month = {Sep},
  publisher = {American Physical Society},
  doi = {10.1103/PhysRevLett.115.136402},
  url = {https://link.aps.org/doi/10.1103/PhysRevLett.115.136402}
}

@article{PhysRevB.13.5188,
  title = {{Special points for Brillouin-zone integrations}},
  author = {Monkhorst, Hendrik J. and Pack, James D.},
  journal = {Phys. Rev. B},
  volume = {13},
  issue = {12},
  pages = {5188--5192},
  numpages = {0},
  year = {1976},
  month = {Jun},
  publisher = {American Physical Society},
  doi = {10.1103/PhysRevB.13.5188},
  url = {https://link.aps.org/doi/10.1103/PhysRevB.13.5188}
}

@article{RevModPhys.84.1419,
  title = {{Maximally localized Wannier functions: Theory and applications}},
  author = {Marzari, Nicola and Mostofi, Arash A. and Yates, Jonathan R. and Souza, Ivo and Vanderbilt, David},
  journal = {Rev. Mod. Phys.},
  volume = {84},
  issue = {4},
  pages = {1419--1475},
  numpages = {0},
  year = {2012},
  month = {Oct},
  publisher = {American Physical Society},
  doi = {10.1103/RevModPhys.84.1419},
  url = {https://link.aps.org/doi/10.1103/RevModPhys.84.1419}
}

@article{FreericksPRL2009,
	author = {Freericks, J. K. and Krishnamurthy, H. R. and Pruschke, Th.},
	doi = {10.1103/PhysRevLett.102.136401},
	issue = {13},
	journal = {Phys. Rev. Lett.},
	month = {Mar},
	numpages = {4},
	pages = {136401},
	publisher = {American Physical Society},
	title = {Theoretical Description of Time-Resolved Photoemission Spectroscopy: Application to Pump-Probe Experiments},
	url = {https://link.aps.org/doi/10.1103/PhysRevLett.102.136401},
	volume = {102},
	year = {2009}}

@book{FreericksManyBody2019,
	author = {James K. Freericks},
	editor = {E. Pavarini and E. Koch and S. Zhang},
	publisher = {Forschungszentrum J{\"u}lich},
	title = {{Many-Body Methods for Real Materials Modeling and Simulation}},
	volume = {9},
	year = {2019}}

@article{PARCOLLET2015398,
title = {{TRIQS: A toolbox for research on interacting quantum systems}},
journal = {Computer Physics Communications},
volume = {196},
pages = {398-415},
year = {2015},
issn = {0010-4655},
doi = {https://doi.org/10.1016/j.cpc.2015.04.023},
url = {https://www.sciencedirect.com/science/article/pii/S0010465515001666},
author = {Olivier Parcollet and Michel Ferrero and Thomas Ayral and Hartmut Hafermann and Igor Krivenko and Laura Messio and Priyanka Seth}}

@article{OUVRARD19851181,
title = {{Structural determination of some MPS3 layered phases (M = Mn, Fe, Co, Ni and Cd)}},
journal = {Materials Research Bulletin},
volume = {20},
number = {10},
pages = {1181-1189},
year = {1985},
issn = {0025-5408},
doi = {https://doi.org/10.1016/0025-5408(85)90092-3},
url = {https://www.sciencedirect.com/science/article/pii/0025540885900923},
author = {G. Ouvrard and R. Brec and J. Rouxel}}

@book{Kamenev_2011, place={Cambridge}, title={Field Theory of Non-Equilibrium Systems}, publisher={Cambridge University Press}, author={Kamenev, Alex}, year={2011}}

@article{PhysRevB.57.6884,
  title = {{Ab initio calculations of quasiparticle band structure in correlated systems: LDA++ approach}},
  author = {Lichtenstein, A. I. and Katsnelson, M. I.},
  journal = {Phys. Rev. B},
  volume = {57},
  issue = {12},
  pages = {6884--6895},
  numpages = {0},
  year = {1998},
  month = {Mar},
  publisher = {American Physical Society},
  doi = {10.1103/PhysRevB.57.6884},
  url = {https://link.aps.org/doi/10.1103/PhysRevB.57.6884}
}

@article{JoyVasudevanFePS3-Optical-1992,
	author = {Joy, P. A. and Vasudevan, S.},
	doi = {10.1103/PhysRevB.46.5134},
	issue = {9},
	journal = {Phys. Rev. B},
	month = {Sep},
	numpages = {0},
	pages = {5134--5141},
	publisher = {American Physical Society},
	title = {{Optical-absorption spectra of the layered transition-metal thiophosphates M${\mathrm{PS}}_{3}$ (M=Mn, Fe, and Ni)}},
	url = {https://link.aps.org/doi/10.1103/PhysRevB.46.5134},
	volume = {46},
	year = {1992}}

@article{optical-conductivity-Hubbard-2010,
	author = {M. I. Katsnelson and A. I. Lichtenstein},
	journal = {J. Phys.: Condens. Matter},
	pages = {382201},
	title = {Theory of optically forbidden d--d transitions in strongly correlated crystals},
	volume = {22},
	year = {2021}}

@article{FreericksPumpProbe09,
	author = {J. K. Freericks and H. R. Krishnamurthy and Th. Pruschke},
	doi = {10.1103/PhysRevLett.102.136401},
	eid = {136401},
	journal = {Phys. Rev. Lett.},
	number = {13},
	numpages = {4},
	pages = {136401},
	publisher = {APS},
	title = {Theoretical Description of Time-Resolved Photoemission Spectroscopy: Application to Pump-Probe Experiments},
	url = {http://link.aps.org/abstract/PRL/v102/e136401},
	volume = {102},
	year = {2009}}

@article{PhysRevB.106.L180409,
  title = {{Antiferromagnetic fluctuations and orbital-selective Mott transition in the van der Waals ferromagnet ${\text{Fe}}_{3\ensuremath{-}x}{\text{GeTe}}_{2}$}},
  author = {Bai, Xiaojian and Lechermann, Frank and Liu, Yaohua and Cheng, Yongqiang and Kolesnikov, Alexander I. and Ye, Feng and Williams, Travis J. and Chi, Songxue and Hong, Tao and Granroth, Garrett E. and May, Andrew F. and Calder, Stuart},
  journal = {Phys. Rev. B},
  volume = {106},
  issue = {18},
  pages = {L180409},
  numpages = {6},
  year = {2022},
  month = {Nov},
  publisher = {American Physical Society},
  doi = {10.1103/PhysRevB.106.L180409},
  url = {https://link.aps.org/doi/10.1103/PhysRevB.106.L180409}
}

@article{Keldysh65,
	author = {L. V. Keldysh},
	journal = {Sov. Phys. JETP},
	pages = {1018},
	volume = {20},
	year = {1965}}

@article{RevModPhys.58.323,
  title = {Quantum field-theoretical methods in transport theory of metals},
  author = {Rammer, J. and Smith, H.},
  journal = {Rev. Mod. Phys.},
  volume = {58},
  issue = {2},
  pages = {323--359},
  numpages = {0},
  year = {1986},
  month = {Apr},
  publisher = {American Physical Society},
  doi = {10.1103/RevModPhys.58.323},
  url = {https://link.aps.org/doi/10.1103/RevModPhys.58.323}
}

@article{RevModPhys.86.779,
	author = {Aoki, Hideo and Tsuji, Naoto and Eckstein, Martin and Kollar, Marcus and Oka, Takashi and Werner, Philipp},
	doi = {10.1103/RevModPhys.86.779},
	issue = {2},
	journal = {Rev. Mod. Phys.},
	month = {Jun},
	numpages = {59},
	pages = {779--837},
	publisher = {American Physical Society},
	title = {Nonequilibrium dynamical mean-field theory and its applications},
	url = {https://link.aps.org/doi/10.1103/RevModPhys.86.779},
	volume = {86},
	year = {2014}}

@article{KAROLAK201011,
title = {{Double counting in LDA+DMFT—The example of NiO}},
journal = {Journal of Electron Spectroscopy and Related Phenomena},
volume = {181},
number = {1},
pages = {11-15},
year = {2010},
RRRRnote = {Proceedings of International Workshop on Strong Correlations and Angle-Resolved Photoemission Spectroscopy 2009},
issn = {0368-2048},
doi = {https://doi.org/10.1016/j.elspec.2010.05.021},
url = {https://www.sciencedirect.com/science/article/pii/S0368204810001222},
author = {M. Karolak and G. Ulm and T. Wehling and V. Mazurenko and A. Poteryaev and A. Lichtenstein}
}

@article{PhysRevLett.87.067205,
  title = {Finite-Temperature Magnetism of Transition Metals: An ab initio Dynamical Mean-Field Theory},
  author = {Lichtenstein, A. I. and Katsnelson, M. I. and Kotliar, G.},
  journal = {Phys. Rev. Lett.},
  volume = {87},
  issue = {6},
  pages = {067205},
  numpages = {4},
  year = {2001},
  month = {Jul},
  publisher = {American Physical Society},
  doi = {10.1103/PhysRevLett.87.067205},
  url = {https://link.aps.org/doi/10.1103/PhysRevLett.87.067205}
}

@article{FinkelsteinVanVleck1940,
    author = {Finkelstein, R. and Van Vleck, J. H.},
    title = {{On the Energy Levels of Chrome Alum}},
    journal = {The Journal of Chemical Physics},
    volume = {8},
    number = {10},
    pages = {790-797},
    year = {1940},
    month = {10},
    issn = {0021-9606},
    doi = {10.1063/1.1750581},
    url = {https://doi.org/10.1063/1.1750581}
}

@article{10.1063/1.3585113,
    author = {Zhang, P. and Richard, P. and Qian, T. and Xu, Y.-M. and Dai, X. and Ding, H.},
    title = {A precise method for visualizing dispersive features in image plots},
    journal = {Review of Scientific Instruments},
    volume = {82},
    number = {4},
    pages = {043712},
    year = {2011},
    month = {04},
    issn = {0034-6748},
    doi = {10.1063/1.3585113},
    url = {https://doi.org/10.1063/1.3585113}
}

@book{breuer2002theory,
  title={The theory of open quantum systems},
  author={Breuer, Heinz-Peter and Petruccione, Francesco},
  year={2002},
  publisher={OUP Oxford}
}

@book{press1992numerical,
  title={Numerical recipes in C},
  author={Press, William H and Teukolsky, Saul A and Vetterling, William T and Flannery, Brian P and others},
  volume={2},
  number={1},
  year={1992},
  publisher={Cambridge university press, Cambridge}
}

@article{PhysRevB.105.L041108,
  title = {Orbital-selective Mott phase and non-Fermi liquid in ${\mathrm{FePS}}_{3}$},
  author = {Kim, Minsung and Kim, Heung-Sik and Haule, Kristjan and Vanderbilt, David},
  journal = {Phys. Rev. B},
  volume = {105},
  issue = {4},
  pages = {L041108},
  numpages = {5},
  year = {2022},
  month = {Jan},
  publisher = {American Physical Society},
  doi = {10.1103/PhysRevB.105.L041108},
  url = {https://link.aps.org/doi/10.1103/PhysRevB.105.L041108}
}

@article{PhysRevB.85.205133,
  title = {Dipole matrix element approach versus Peierls approximation for optical conductivity},
  author = {Wissgott, P. and Kune{\ifmmode \check{s}\else \v{s}\fi{}}, J. and Toschi, A. and Held, K.},
  journal = {Phys. Rev. B},
  volume = {85},
  issue = {20},
  pages = {205133},
  numpages = {9},
  year = {2012},
  month = {May},
  publisher = {American Physical Society},
  doi = {10.1103/PhysRevB.85.205133},
  url = {https://link.aps.org/doi/10.1103/PhysRevB.85.205133}
}

\end{document}